\documentclass[12pt]{article}%
\usepackage{amssymb}
\usepackage[singlespacing]{setspace}
\usepackage{amsmath}
\usepackage{amsfonts}
\usepackage{graphicx}
\usepackage{cite}%
\allowdisplaybreaks
\begin{document}

\author{\vspace{0.16in}Hartmut Wachter\thanks{e-mail:
Hartmut.Wachter@physik.uni-muenchen.de}\\Max-Planck-Institute\\for Mathematics in the Sciences\\Inselstr. 22, D-04103 Leipzig\\\hspace{0.4in}\\Arnold-Sommerfeld-Center\\Ludwig-Maximilians-Universit\"{a}t\\Theresienstr. 37, D-80333 M\"{u}nchen}
\title{Analysis on q-deformed quantum spaces}
\date{}
\maketitle

\begin{abstract}
A q-deformed version of classical analysis is given to quantum spaces of
physical importance, i.e. Manin plane, q-deformed Euclidean space in three or
four dimensions, and q-deformed Minkowski space. The subject is presented in a
rather complete and selfcontained way. All relevant notions are introduced and
explained in detail. The different possibilities to realize the objects of
q-deformed analysis are discussed and their elementary properties are studied.
In this manner attention is focused on star products, q-deformed tensor
products, q-deformed translations, q-deformed partial derivatives, dual
pairings, q-deformed exponentials, and q-deformed integration. The main
concern of this work is to show that these objects fit together in a
consistent framework, which is suitable to formulate physical theories on
quantum spaces. \newpage

\end{abstract}
\tableofcontents

\section{Introduction}

\subsection{General motivation for deforming spacetime}

Relativistic quantum field theory is not a fundamental theory, since its
formalism leads to divergencies. In some cases like that of quantum
electrodynamics one is able to overcome the difficulties with the divergencies
by applying the so-called renormalization procedure due to Feynman, Schwinger,
and Tomonaga \cite{Schw}. Unfortunately, this procedure is not successful if
we want to deal with quantum gravity. Despite the fact that gravitation is a
rather weak interaction we are not able to treat it perturbatively. The reason
for this lies in the fact that transition amplitudes of n-th order to the
gravitation constant diverge like a momentum integral of the general form
\cite{Wei}
\begin{equation}
\int p^{2n-1}dp, \label{ImInt}%
\end{equation}
leaving us with an infinite number of ultraviolet divergent Feynman diagrams
that cannot be removed by redefining finitely many physical parameters.

It is surely legitimate to ask for the reason for these fundamental
difficulties. It is commonplace that the problems with the divergences in
relativistic quantum field theory result from an incomplete description of
spacetime at very small distances \cite{Schw}. Niels Bohr and Werner
Heisenberg have been the first who suggested that quantum field theories
should be formulated on a spacetime lattice \cite{Cass, Heis}. Such a
spacetime lattice would imply the existence of a smallest distance $a$ with
the consequence that plane-waves of wave-length smaller than twice the lattice
spacing could not propagate. In accordance with the relationship between
wave-length $\lambda$ and momentum $p$ of a plane-wave, i.e.
\begin{equation}
\lambda\geq\lambda_{\min}=2a\quad\Rightarrow\quad\frac{1}{\lambda}\sim p\leq
p_{\max}\sim\frac{1}{2a},
\end{equation}
it follows then that physical momentum space would be bounded. Hence, the
domain of all momentum integrals in Eq. (\ref{ImInt}) would be bounded as well
with the consequence that momentum integrals should take on finite values.

\subsection{q-Deformation of symmetries as an attempt to get a more detailed
description of nature}

Discrete spacetime structures in general do not respect classical Poincar\'{e}
symmetry. A possible way out of this difficulty is to modify not only
spacetime but also its corresponding symmetries. How are we to accomplish
this? First of all let us recall that classical spacetime symmetries are
usually described by Lie groups. If we realize that Lie groups are manifolds
the Gelfand-Naimark\textsf{\ }theorem tells us that Lie groups can be
naturally embedded in the category of algebras\textsf{\ }\cite{GeNe}. The
utility of this interrelation lies in formulating the geometrical structure of
Lie groups in terms of a Hopf structure \cite{Hopf}. The point is that during
the last two decades generic methods have been discovered for continuously
deforming matrix groups and Lie algebras within the category of Hopf algebras.
It is this development which finally led to the arrival of quantum groups and
quantum spaces \cite{Ku83, Wor87, Dri85, Jim85, Drin86, RFT90, Tak90}.

From a physical point of view the most realistic and interesting deformations
are given by q-deformed versions of Minkowski space and Euclidean spaces as
well as their corresponding symmetries, i.e. respectively Lorentz symmetry and
rotational symmetry \cite{CSSW90, PW90, SWZ91, Maj91, LWW97}. Further studies
even allowed to establish differential calculi on these q-deformed quantum
spaces \cite{WZ91, CSW91, Song92, OSWZ92} representing nothing other than
q-analogs of translational symmetry. In this sense we can say that
q-deformations of the complete Euclidean and Poincar\'{e} symmetries are now
available \cite{Maj93-2}. Finally, Julius Wess and his coworkers were able to
show that q-deformation of spaces and symmetries can indeed lead to the wanted
discretizations of the spectra of spacetime observables \cite{FLW96, Wes97,
CHMW99, CW98}, which nourishes the hope that q-deformation might give a new
method to regularize quantum field theories \cite{GKP96, MajReg, Oec99, Blo03}.

\subsection{Foundations of q-deformed analysis}

In order to formulate quantum field theories on quantum spaces it is necessary
to provide us with some essential tools of a q-deformed analysis
\cite{Wess00}. The main question is how to define these new tools, which
should be q-analogs of classical notions. Towards this end the considerations
of Shahn Majid have proved very useful \cite{Maj91Kat, Maj94Kat, Maj93-Int,
Maj94-10}. The key idea of this approach is that all the quantum spaces to a
given quantum symmetry form a braided tensor category. Consequently,
operations and objects concerning quantum spaces must rely on this framework
of a braided tensor category, in order to guarantee their well-defined
behavior under quantum group transformations. This so-called principle of
covariance can be seen as the essential guideline for constructing a
consistent theory.

In our previous work \cite{WW01, BW01, Wac02, Wac03, Wac04, Wac05, MSW04,
SW04} we have applied these general considerations, exposed in Refs.
\cite{Maj93-Int, Maj94-10, Maj93-6, Maj93-5, Maj95}, to quantum spaces of
physical importance, i.e. q-deformed quantum plane, q-deformed Euclidean space
with three or four dimensions, and q-deformed Minkowski space. In this manner,
we obtained explicit expressions for computing star products, operator
representations of partial derivatives and symmetry generators, q-integrals,
q-exponentials, q-translations, and braided products. With this toolbox of
essential elements of q-analysis we should be able to develop a q-deformed
version of quantum field theory along the same line of reasonings as in the
classical case. To achieve this we have to gain further insight into the
properties of the new objects. In other words, we need a set of consistent
rules telling us how to perform calculations with the new objects.

In this respect we have to be aware of one\ remarkable difference between
q-deformed objects and their classical counterparts. For each of our
q-deformed objects we can find different realizations, which in the undeformed
case become identical. The reason for this lies in the fact that the braided
tensor category in which a deformed object lives is not uniquely determined.
This becomes quite clear, if one realizes that each braided category is
characterized by a so-called braiding mapping $\Psi$. Its inverse $\Psi^{-1}$
gives an equally good braiding which, in general, leads to a different braided
category. But we have to deal with both categories, since they are linked via
the operation of conjugation \cite{OZ92, Maj94star, Maj95star}.

\subsection{Intention and content of this work}

Important developments in theoretical physics are often accompanied by the
arrival of new mathematical techniques. Classical mechanics, for example, is
deeply connected to classical analysis. It is also well-known that general
relativity and quantum mechanics are heavily based on ideas of Riemannian
geometry and functional analysis, respectively. As it was pointed out above,
the mathematics of quantum groups and quantum spaces can provide a framework
for a more detailed description of spacetime at very small distances. If this
is the case the formalism should be prepared in a manner that allows to
formulate and evaluate physical theories on quantum spaces in a rather
systematical way.

In this article we develop basic ideas of q-analysis on quantum spaces of
physical importance in a rather complete and self-contained way. We introduce
and explain all relevant notions in detail and discuss the different
possibilities for realizing the objects of q-analysis.\ Furthermore, we give a
review of elementary properties and derive useful calculational rules. We show
how the various objects fit together in a consistent framework. It is also our
aim to introduce a convenient notation, which will prove useful in further
investigations. In doing so, we hope to provide the foundations for performing
concrete calculations, as they are necessary in formulating and evaluating
field theories on quantum spaces.

In particular, we intend to proceed as follows. In Section \ref{Sec1} we deal
with star products, q-deformed tensor products, and q-deformed translations.
In Section \ref{SecPart} we concern ourselves with q-deformed partial
derivatives. Section \ref{SecExp} is devoted to the discussion of dual
pairings and q-deformed exponentials. In Section \ref{SecInt} we introduce a
powerful concept of integration on q-deformed quantum spaces. Section
\ref{SecCon} closes our considerations by a short conclusion. For reference
and for the purpose of introducing consistent and convenient notation, we
provide a review of key notation and results in Appendix \ref{AppQuan}. Last
but not least Appendix \ref{MinRech} contains a discussion of the details
about\ integration on q-deformed Minkowski space.

\section{Star products, q-tensor products, and q-translations\label{Sec1}}

This section is devoted to star products, q-deformed tensor products, and
q-deformed translations. Our considerations apply to quantum spaces like the
Manin plane, the q-deformed Euclidean space in three or four dimensions, and
the q-deformed Minkowski space.

\subsection{Definition of star products and q-deformed tensor products}

For our purposes, it is at first sufficient to consider a quantum space as an
algebra of formal power series in non-commuting coordinates $X^{1}%
,X^{2},\ldots,X^{n},$ i.e.%

\begin{equation}
\mathcal{A}_{q}=\mathbb{C}\left[  \left[  X_{1},\ldots,X_{n}\right]  \right]
/\mathcal{I},
\end{equation}
where $\mathcal{I}$ denotes the ideal generated by the relations of the
non-commuting coordinates. The term 'coordinate', however, is a little bit
misleading, since the generators of quantum spaces cannot only play the role
of coordinates but also that of partial derivatives or differentials.

The two-dimensional Manin plane is one of the simplest examples for a quantum
space \cite{Man88}. It consists of all the power series in two coordinates
$X^{1}$ and $X^{2}$ being subject to the commutation relations
\begin{equation}
X^{1}X^{2}=qX^{2}X^{1},\quad q>1.
\end{equation}
We can think of $q$ as a deformation parameter measuring the coupling among
different spatial degrees of freedom. In the classical case, i.e. if $q$
becomes $1$ we regain commutative coordinates.

Next, we would like to focus our attention on the question how to perform
calculations on an algebra of quantum space coordinates. This can be
accomplished by a kind of pullback transforming operations on non-commutative
coordinate algebras to those on commutative ones. For this to become more
clear, we have to realize that the non-commutative algebras\ we are dealing
with satisfy the \textit{Poincar\'{e}-Birkhoff-Witt property}. It tells us
that the dimension of a subspace of homogenous polynomials should be the same
as for commuting coordinates. This property is the deeper reason why monomials
of a given normal ordering constitute a basis of $\mathcal{A}_{q}$. Due to
this fact, we can establish a vector space isomorphism between $\mathcal{A}%
_{q}$ and a commutative algebra $\mathcal{A}$ generated by ordinary
coordinates $x^{1},x^{2},\ldots,x^{n}$:
\begin{align}
\mathcal{W}  &  :\mathcal{A}\longrightarrow\mathcal{A}_{q},\nonumber\\
\mathcal{W}((x^{1})^{i_{1}}\ldots(x^{n})^{i_{n}})  &  \equiv(X^{1})^{i_{1}%
}\ldots(X^{n})^{i_{n}}. \label{AlgIsoN}%
\end{align}

This vector space isomorphism can even be extended to an algebra isomorphism
by introducing a non-commutative product in $\mathcal{A}$, the so-called
\textit{star product} \cite{BFF78, Moy49, MSSW00}. This product is defined via
the relation
\begin{equation}
\mathcal{W}(f\circledast g)=\mathcal{W}(f)\cdot\mathcal{W}(g), \label{StarDef}%
\end{equation}
being tantamount to%
\begin{equation}
f\circledast g\equiv\mathcal{W}^{-1}\left(  \mathcal{W}\left(  f\right)
\cdot\mathcal{W}\left(  g\right)  \right)  ,
\end{equation}
where $f$ and $g$ are formal power series in $\mathcal{A}$. In the case of the
Manin plane, the star product is of the well-known form%
\begin{align}
&  f(x^{i})\circledast g(x^{j})=q^{-\hat{n}_{x^{2}}\hat{n}_{y^{1}}}\left.
\left[  f(x^{i})\,g(y^{j})\right]  \right\vert _{y\rightarrow x}\nonumber\\
&  =f(x^{i})\,g(x^{j})+O(h),\quad\text{with\textsf{\quad}}q=e^{h},
\label{Star2-dimN}%
\end{align}
where we have introduced the operators%
\begin{equation}
\hat{n}_{x^{i}}\equiv x^{i}\frac{\partial}{\partial x^{i}},\quad i=1,2.
\end{equation}
From the last equality in (\ref{Star2-dimN}) we see that star products on
quantum spaces lead to modifications of commutative products. Evidently, this
modifications vanish in the classical limit $q$ $\rightarrow$ $1$.

Next, we would like to deal with tensor products of quantum spaces. To this
end, we have to specify the commutation relations between generators of
distinct quantum spaces. These relations are completely determined by the
requirement of being consistent with the action of a Hopf algebra
$\mathcal{H}$ describing the symmetry of the quantum spaces. Making for the
commutation relations between two quantum space generators $X^{i}%
\in\mathcal{A}_{q}$ and $Y^{j}\in\mathcal{A}_{q}^{\prime}$ an ansatz of the
form%
\begin{equation}
X^{i}Y^{j}=k\,C_{kl}^{ij}\,Y^{k}X^{l},\quad k,C_{kl}^{ij}\in\mathbb{C},
\end{equation}
the coefficients $C_{kl}^{ij}$ have to be determined in such a way that the
following condition of covariance is satisfied:%
\begin{equation}
(h_{(1)}\triangleright X^{i})(h_{(2)}\triangleright Y^{j})=k\,C_{kl}%
^{ij}\,(h_{(1)}\triangleright Y^{k})(h_{(2)}\triangleright X^{l}),
\end{equation}
where $h$ denotes an arbitrary element of $\mathcal{H}$. Notice that the
coproduct of $h$ is written in the so-called Sweedler notation, i.e.
$\Delta(h)=h_{(1)}\otimes h_{(2)}$, and repeated indices are to be summed.

It is well-known that for a quasitriangular Hopf algebra $\mathcal{H}$ the
last identity is fulfilled if we substitute for the coefficients $C_{kl}^{ij}$
a suitable representation of the universal R-matrix $\mathcal{R=R}%
_{[1]}\otimes\mathcal{R}_{[2]}\in\mathcal{H}\otimes\mathcal{H}$. To be more
specific, we have%
\begin{align}
X^{i}Y^{j}  &  =(\mathcal{R}_{[2]}\triangleright Y^{j})\,(\mathcal{R}%
_{[1]}\triangleright X^{i})\nonumber\\
&  =(Y^{j}\triangleleft\mathcal{R}_{[2]})\,(X^{i}\triangleleft\mathcal{R}%
_{[1]})=k\,\hat{R}_{kl}^{ij}\,Y^{k}X^{l}. \label{VerRN}%
\end{align}
Alternatively, we can take a representation of the transposed inverse of
$\mathcal{R}$, i.e. $\tau(\mathcal{R}^{-1})=\mathcal{R}_{[2]}^{-1}%
\otimes\mathcal{R}_{[1]}^{-1},$ giving us%
\begin{align}
X^{i}Y^{j}  &  =(\mathcal{R}_{[1]}^{-1}\triangleright Y^{j})\,(\mathcal{R}%
_{[2]}^{-1}\triangleright X^{i})\nonumber\\
&  =(Y^{j}\triangleleft\mathcal{R}_{[1]}^{-1})\,(X^{i}\triangleleft
\mathcal{R}_{[2]}^{-1})=k^{-1}(\hat{R}^{-1})_{kl}^{ij}\,Y^{k}X^{l}.
\label{VerRInN}%
\end{align}
To understand the second identity in (\ref{VerRN}) and (\ref{VerRInN}), one
has to realize that%
\begin{align}
(S^{-1}\otimes S^{-1})\circ\mathcal{R}  &  =\mathcal{R},\\
(S^{-1}\otimes S^{-1})\circ\mathcal{R}^{-1}  &  =\mathcal{R}^{-1},\nonumber
\end{align}
and
\begin{equation}
S^{-1}(h)\triangleright a=a\triangleleft h,\quad h\in\mathcal{H},\quad
a\in\mathcal{A}_{q},
\end{equation}
where $S^{-1}$ denotes the inverse of the antipode of $\mathcal{H}$.

The identities in (\ref{VerRN}) and (\ref{VerRInN}) extend to arbitrary
quantum space elements. Especially, we have for the commutation relations of a
single quantum space generator $X^{i}$ with an arbitrary element $g$ of
another quantum space%
\begin{align}
X^{i}a  &  =(\mathcal{R}_{[2]}\triangleright a)\,(\mathcal{R}_{[1]}%
\triangleright X^{i})=\big ((\mathcal{\bar{L}}_{x}\mathcal{)}_{j}%
^{i}\triangleright a\big )\,X^{j},\nonumber\\
aX^{i}  &  =(X^{i}\triangleleft\mathcal{R}_{[2]})\,(a\triangleleft
\mathcal{R}_{[1]})=X^{j}\big (a\triangleleft(\mathcal{L}_{x})_{j}^{i}\big ),
\label{LGXconN}%
\end{align}
or%
\begin{align}
X^{i}a  &  =(\mathcal{R}_{[1]}^{-1}\triangleright a)\,(\mathcal{R}_{[2]}%
^{-1}\triangleright X^{i})=\big ((\mathcal{L}_{x}\mathcal{)}_{j}%
^{i}\triangleright a\big )\,X^{j},\nonumber\\
aX^{i}  &  =(X^{i}\triangleleft\mathcal{R}_{[1]}^{-1})\,(a\triangleleft
\mathcal{R}_{[2]}^{-1})=X^{j}\big (a\triangleleft(\mathcal{\bar{L}}_{x}%
)_{j}^{i}\big ), \label{LGXN}%
\end{align}
where we introduced the so-called L-matrix together with its conjugate. It
should be obvious that the entries of the L-matrices live in the quantum
algebra $\mathcal{H}$. As an example, we give the non-vanishing entries of the
L-matrices in the case of the Manin plane \cite{MSW04}:
\begin{align}
(\mathcal{L}_{a})_{1}^{1}  &  =\Lambda(a)\tau^{-1/4},\nonumber\\
(\mathcal{L}_{a})_{1}^{2}  &  =-q\lambda\Lambda(a)\tau^{-1/4}T^{+},\nonumber\\
(\mathcal{L}_{a})_{2}^{2}  &  =\Lambda(a)\tau^{1/4}, \label{LMa2dimN}%
\end{align}
and likewise
\begin{align}
(\mathcal{\bar{L}}_{a})_{1}^{1}  &  =\Lambda^{-1}(a)\tau^{1/4},\nonumber\\
(\mathcal{\bar{L}}_{a})_{2}^{1}  &  =q^{-1}\lambda\Lambda^{-1}(a)\tau
^{-1/4}T^{-},\nonumber\\
(\mathcal{\bar{L}}_{a})_{2}^{2}  &  =\Lambda^{-1}(a)\tau^{-1/4}.
\label{LMa2dimKonNN}%
\end{align}
Notice that $\tau^{\pm1/4}$ and $T^{\pm}$ denote generators of the quantum
algebra $U_{q}(su_{2})$ and $\Lambda(a)$ stands for a unitary scaling
operator. The constant $\lambda$ is given by $\lambda\equiv q-q^{-1}$. The
L-matrices for q-deformed Euclidean space in three or four dimensions and
those for q-deformed Minkowski space can easily be read off from the results
in Refs. \cite{BW01} and \cite{MSW04}.

By virtue of the algebra isomorphism $\mathcal{W}$ we are able to introduce
the following operations:%
\begin{align}
f(x^{i})\,\odot_{L}\,g(y^{j})  &  \equiv\mathcal{W}^{-1}\big (\mathcal{R}%
_{[1]}^{-1}\triangleright\mathcal{W}(g)\big)\otimes\mathcal{W}^{-1}%
\big (\mathcal{R}_{[2]}^{-1}\triangleright\mathcal{W}(f)\big),\nonumber\\
f(x^{i})\,\odot_{\bar{R}}\,g(y^{j})  &  \equiv\mathcal{W}^{-1}%
\big (\mathcal{W}(g)\triangleleft\mathcal{R}_{[1]}^{-1}\big)\otimes
\mathcal{W}^{-1}\big (\mathcal{W}(f)\triangleleft\mathcal{R}_{[2]}^{-1}\big),
\label{BraidDef1}%
\end{align}
and
\begin{align}
f(x^{i})\,\odot_{\bar{L}}\,g(y^{j})  &  \equiv\mathcal{W}^{-1}%
\big (\mathcal{R}_{[2]}\triangleright\mathcal{W}(g)\big)\otimes\mathcal{W}%
^{-1}\big (\mathcal{R}_{[1]}\triangleright\mathcal{W}(f)\big),\nonumber\\
f(x^{i})\,\odot_{R}\,g(y^{j})  &  \equiv\mathcal{W}^{-1}\big (\mathcal{W}%
(g)\triangleleft\mathcal{R}_{[2]}\big)\otimes\mathcal{W}^{-1}\big (\mathcal{W}%
(f)\triangleleft\mathcal{R}_{[1]}\big), \label{BraidDef2N}%
\end{align}
where $f$ and $g$ denote formal power series in the commutative coordinate
algebras $\mathcal{A}_{x}$ and $\mathcal{A}_{y}$, respectively. The operations
in (\ref{BraidDef1}) and (\ref{BraidDef2N}) are referred to as \textit{braided
products.} Using the explicit form of the identities in (\ref{LGXconN}) and
(\ref{LGXN}) we derived in Ref. \cite{Wac05}\ commutation relations between
normal ordered monomials. With these results at hand we are able to write down
explicit formulae for computing braided products. As an example we give the
expression we obtained for the braided product of the Manin plane
\cite{Wac05}:%
\begin{align}
&  f(x^{1},x^{2})\,\odot_{L}\,g(y^{1},y^{2})\nonumber\\
&  =\sum_{i=0}^{\infty}q^{i^{2}}(-\lambda)^{i}\,\frac{(y^{2})^{i}\otimes
(x^{1})^{i}}{[[i]]_{q^{-2}}!}\,q^{-\hat{n}_{y^{1}}\otimes\hat{n}_{x^{2}}%
-2\hat{n}_{y^{2}}\otimes\hat{n}_{x^{1}}-2\hat{n}_{y^{1}}\otimes\hat{n}_{x^{1}%
}-2\hat{n}_{y^{2}}\otimes\hat{n}_{x^{1}}}\nonumber\\
&  \qquad\times(D_{q^{-2}}^{1})^{i}g(q^{-i}y^{1},q^{-2i}y^{2})\otimes
(D_{q^{-2}}^{2})^{i}f(q^{-2i}x^{1},q^{-i}x^{2}).
\end{align}
In the above formulae we introduced the so-called\ \textit{Jackson
derivatives}, which are defined by \cite{Jack08}
\begin{equation}
D_{q^{a}}^{i}f(x^{i})=\frac{f(x^{i})-f(q^{a}x^{i})}{(1-q^{a})x^{i}},\quad
a\in\mathbb{C}.
\end{equation}
It should be mentioned that the equalities in (\ref{VerRN}) and (\ref{VerRInN}%
) imply that
\begin{align}
f(x^{i})\,\odot_{L}\,g(y^{j})  &  =f(x^{i})\,\odot_{\bar{R}}\,g(y^{j}%
),\nonumber\\
f(x^{i})\,\odot_{\bar{L}}\,g(y^{j})  &  =f(x^{i})\,\odot_{R}\,g(y^{j}).
\label{BraidIdN}%
\end{align}
Thus, it would be sufficient to deal with $\odot_{L}$ and $\odot_{\bar{L}},$ only.

Now, we are in a position to introduce the tensor product of quantum spaces.
It is equipped with a multiplication being determined by%
\begin{equation}
(a\otimes a^{\prime})(b\otimes b^{\prime})=\big(a(\mathcal{R}_{[2]}%
\triangleright b)\big)\otimes\big((\mathcal{R}_{[1]}\triangleright a^{\prime
})b^{\prime}\big),
\end{equation}
or
\begin{equation}
(a\otimes a^{\prime})(b\otimes b^{\prime})=\big(a(\mathcal{R}_{[1]}%
^{-1}\triangleright b)\big)\otimes\big((\mathcal{R}_{[2]}^{-1}\triangleright
a^{\prime})b^{\prime}\big),
\end{equation}
where $a,b\in\mathcal{A}_{q}$ and $a^{\prime},b^{\prime}\in\mathcal{A}%
_{q}^{\prime}.$ We see that multiplication on a tensor product requires to
know the commutation relations between the elements of the two tensor factors.
Essentially for us is the fact that the algebra isomorphism $\mathcal{W}$
allows us to represent the tensor product of quantum spaces on a tensor
product of commutative coordinate algebras. To be more specific, this can be
achieved by extending the braided products in (\ref{BraidDef1}) and
(\ref{BraidDef2N}) as follows:%
\begin{align}
&  \big(f(x^{i})\otimes f^{\prime}(y^{j})\big)\,\odot_{L}\,\big(g(x^{k}%
)\otimes g^{\prime}(y^{l})\big)\nonumber\\
&  \equiv\big[f\overset{x}{\circledast}\mathcal{W}^{-1}\big (\mathcal{R}%
_{[1]}^{-1}\triangleright\mathcal{W}(g)\big)\big]\otimes\big[\mathcal{W}%
^{-1}\big (\mathcal{R}_{[2]}^{-1}\triangleright\mathcal{W}(f^{\prime
})\big)\overset{y}{\circledast}g^{\prime}\big], \label{BraTens1}%
\end{align}
and similarly%
\begin{align}
&  \big(f(x^{i})\otimes f^{\prime}(y^{j})\big)\,\odot_{\bar{L}}\,\big(g(x^{k}%
)\otimes g^{\prime}(y^{l})\big)\nonumber\\
&  \equiv\big[f\overset{x}{\circledast}\mathcal{W}^{-1}\big (\mathcal{R}%
_{[2]}\triangleright\mathcal{W}(g)\big)\big]\otimes\big[\mathcal{W}%
^{-1}\big (\mathcal{R}_{[1]}\triangleright\mathcal{W}(f^{\prime}%
)\big)\overset{y}{\circledast}g^{\prime}\big], \label{BraTens2}%
\end{align}
where $f,g\in\mathcal{A}_{x}$ and $f^{\prime},g^{\prime}\in\mathcal{A}_{y}.$

\subsection{Definition and basic properties of q-deformed translations}

Before introducing q-translations, it is useful to make contact with the
notion of a \textit{cross product algebra }\cite{KS97}. It is well-known that
we can combine a Hopf algebra $\mathcal{H}$ with its representation space
$\mathcal{A}_{q}$ to form a left cross product algebra $\mathcal{A}_{q}%
\rtimes\mathcal{H}$ built on $\mathcal{A}_{q}\otimes\mathcal{H}$ with product%
\begin{equation}
(a\otimes h)(b\otimes g)=a(h_{(1)}\triangleright b)\otimes h_{(2)}g,\quad
a,b\in\mathcal{A}_{q},\mathcal{\quad}h,g\in\mathcal{H}. \label{LefCrosPro}%
\end{equation}
There is also a right-handed version of this notion called a right cross
product algebra $\mathcal{H}\ltimes\mathcal{A}$ and built on $\mathcal{H}%
\otimes\mathcal{A}$ with product%
\begin{equation}
(h\otimes a)(g\otimes b)=hg_{(1)}\otimes(a\triangleleft g_{(2)})b.
\label{RigCrosPro}%
\end{equation}
When $\mathcal{A}_{q}$ is a q-deformed quantum space the cross product
algebras have a Hopf structure. On quantum space generators the corresponding
coproduct, antipode, and counit take the form \cite{OSWZ92, Maj93-2}%
\begin{align}
\Delta_{\bar{L}}(X^{i})  &  =X^{i}\otimes1+(\mathcal{\bar{L}}_{x})_{j}%
^{i}\otimes X^{j},\nonumber\\
\Delta_{L}(X^{i})  &  =X^{i}\otimes1+(\mathcal{L}_{x})_{j}^{i}\otimes
X^{j},\label{HopfStrucN}\\[0.16in]
S_{\bar{L}}(X^{i})  &  =-S(\mathcal{\bar{L}}_{x})_{j}^{i}\,X^{j},\nonumber\\
S_{L}(X^{i})  &  =-S(\mathcal{L}_{x})_{j}^{i}\,X^{j},\label{SExplN}\\[0.16in]
\varepsilon_{\bar{L}}(X^{i})  &  =\varepsilon_{L}(X^{i})=0.
\end{align}
There are opposite Hopf structures related to the above ones via%
\begin{equation}
\Delta_{\bar{R}/R}=\tau\circ\Delta_{\bar{L}/L},\qquad S_{\bar{R}/R}=S_{\bar
{L}/L}^{-1},\qquad\varepsilon_{\bar{R}/R}=\varepsilon_{\bar{L}/L},
\label{RightHopf}%
\end{equation}
where $\tau$ shall\ denote the usual transposition of tensor factors. Notice
that for the antipodes $S_{R}$ and $S_{\bar{R}}$ it holds%
\begin{align}
S_{\bar{R}}(X^{i})  &  =-X^{j}S^{-1}(\mathcal{\bar{L}}_{x})_{j}^{i}%
,\nonumber\\
S_{R}(X^{i})  &  =-X^{j}S^{-1}(\mathcal{L}_{x})_{j}^{i}, \label{S-1ExplN}%
\end{align}
which is a direct consequence of the Hopf algebra axiom%
\begin{equation}
a_{(2)}S^{-1}(a_{(1)})=\varepsilon(a).
\end{equation}

An essential observation is that coproducts of coordinates imply their
translations \cite{Maj93-5, Maj93-2, Maj-93/3, Me95, Wac04, SW04}. This can be
seen as follows. The coproduct on coordinates is an algebra homomorphism. If
the coordinates constitute a module coalgebra then the algebra structure of
the coordinates $X^{i}$ is carried over to their coproduct $\Delta_{A}%
(X^{i}).$ More formally, we have
\begin{equation}
\Delta_{A}(X^{i}X^{j})=\Delta_{A}(X^{i})\Delta_{A}(X^{j})\quad\text{and\quad
}\Delta_{A}(h\rhd X^{i})=\Delta(h)\rhd\Delta_{A}(X^{i}). \label{Trans}%
\end{equation}
Due to this fact we can think of (\ref{HopfStrucN}) as nothing other than an
addition law for q-deformed vector components.

To proceed any further we have to realize that our algebra morphism
$\mathcal{W}^{-1}$ can be extended by
\begin{align}
\mathcal{W}_{L}^{-1}  &  :\mathcal{A}_{q}\rtimes\mathcal{H}\longrightarrow
\mathcal{A},\nonumber\\
\mathcal{W}_{L}^{-1}((X^{1})^{i_{1}}\ldots(X^{n})^{i_{n}}\otimes h)  &
\equiv\mathcal{W}^{-1}((X^{1})^{i_{1}}\ldots(X^{n})^{i_{n}})\,\varepsilon(h),
\label{ExtAlgIsoL}%
\end{align}
or
\begin{align}
\mathcal{W}_{R}^{-1}  &  :\mathcal{H}\ltimes\mathcal{A}_{q}\longrightarrow
\mathcal{A},\nonumber\\
\mathcal{W}_{R}^{-1}(h\otimes(X^{1})^{i_{1}}\ldots(X^{n})^{i_{n}})  &
\equiv\varepsilon(h)\,\mathcal{W}^{-1}((X^{1})^{i_{1}}\ldots(X^{n})^{i_{n}}),
\label{ExtAlgIsoR}%
\end{align}
with $\varepsilon$ being the counit of the Hopf algebra $\mathcal{H}.$ The
meaning of the definitions in (\ref{ExtAlgIsoL}) and (\ref{ExtAlgIsoR})
becomes more clear when we assume the existence of a vacuum state $|0\rangle$
with%
\begin{align}
\langle0|(X^{1})^{i_{1}}\ldots(X^{n})^{i_{n}}|0\rangle &  =\mathcal{W}%
^{-1}((X^{1})^{i_{1}}\ldots(X^{n})^{i_{n}}),\nonumber\\
\langle0|0\rangle &  =1.
\end{align}
Furthermore, we require for the vacuum state to be invariant under symmetry
operations, i.e.
\begin{equation}
h|0\rangle=\varepsilon(h)|0\rangle.
\end{equation}
Combining (\ref{ExtAlgIsoL}) and (\ref{ExtAlgIsoR}) then yields
\begin{align}
&  \langle0|(X^{1})^{i_{1}}\ldots(X^{n})^{i_{n}}h|0\rangle=\langle
0|(X^{1})^{i_{1}}\ldots(X^{n})^{i_{n}}|0\rangle\varepsilon(h)\nonumber\\
&  =\mathcal{W}^{-1}((X^{1})^{i_{1}}\ldots(X^{n})^{i_{n}})\varepsilon
(h)=\mathcal{W}_{L}^{-1}((X^{1})^{i_{1}}\ldots(X^{n})^{i_{n}}\otimes h).
\end{align}
Similar considerations hold for $\mathcal{W}_{R}^{-1}.$

With these mappings at hand we are able to introduce q-deformed translations:%
\begin{align}
f(x^{i}\oplus_{L/\bar{L}}y^{j})  &  \equiv((\mathcal{W}_{L}^{-1}%
\otimes\mathcal{W}_{L}^{-1})\circ\Delta_{L/\bar{L}})(\mathcal{W}%
(f)),\nonumber\\
f(x^{i}\oplus_{R/\bar{R}}y^{j})  &  \equiv((\mathcal{W}_{R}^{-1}%
\otimes\mathcal{W}_{R}^{-1})\circ\Delta_{R/\bar{R}})(\mathcal{W}%
(f)),\label{DefCoProN}\\[0.16in]
f(\ominus_{L/\bar{L}}\,x^{i})  &  \equiv(\mathcal{W}_{R}^{-1}\circ
S_{L/\bar{L}})(\mathcal{W}(f)),\nonumber\\
f(\ominus_{R/\bar{R}}\,x^{i})  &  \equiv(\mathcal{W}_{L}^{-1}\circ
S_{R/\bar{R}})(\mathcal{W}(f)). \label{DefAntiN}%
\end{align}
The following formulae referring to the Manin plane shall serve as concrete
examples:%
\begin{align}
f(x^{i}\oplus_{L}y^{j})  &  =\sum_{k_{1},k_{2}=0}^{\infty}\frac{(x^{1}%
)^{k_{1}}(x^{2})^{k_{2}}}{[[k_{1}]]_{q^{-2}}![[k_{2}]]_{q^{-2}}!}%
\,\big ((D_{q^{-2}}^{1})^{k_{1}}(D_{q^{-2}}^{2})^{k_{2}}f\big )(q^{-k_{2}%
}y^{1}),\label{CoForm2dim}\\[0.16in]
f(\ominus_{L}\,x^{i})  &  =q^{-(\hat{n}_{x^{1}})^{2}-(\hat{n}_{x^{2}}%
)^{2}-2\hat{n}_{x^{1}}\hat{n}_{x^{2}}}\,f(-qx^{1},-qx^{2}).
\end{align}
It should be mentioned that Eq. (\ref{CoForm2dim}) can be seen as q-deformed
version of the Taylor rule in two dimensions.

It is useful to notice that the considerations so far provide us with nothing
other than a realization of a braided Hopf algebra on an algebra of commuting
coordinates. For this reason we take another point of view which is provided
by category theory. A category is a collection of objects $X,Y,Z,\ldots$
together with a set Mor$(X,Y)$ of morphisms between two objects $X,Y$. The
composition of morphisms has similar properties as the composition of maps. We
are interested in tensor categories. These categories have a product, denoted
$\otimes$ and called the tensor product. It admits several 'natural'
properties such as associativity and existence of a unit object. For a more
formal treatment we refer to Refs. \cite{Maj91Kat, Maj94Kat, Maj95, MaL74}. If
the action of a quasitriangular Hopf algebra $\mathcal{H}$ on the tensor
product of two quantum spaces $X$ and $Y$ is defined by
\begin{equation}
h\triangleright(v\otimes w)=(h_{(1)}\triangleright v)\otimes(h_{(2)}%
\triangleright w)\in X\otimes Y,\quad h\in\mathcal{H},
\end{equation}
then the representations (quantum spaces) of the given Hopf algebra
$\mathcal{H}$ are the objects of a tensor category.

In this tensor category exist a number of morphisms of particular importance
that are covariant with respect to the Hopf algebra action. First of all, for
any pair of objects $X,Y$ there is an isomorphism $\Psi_{X,Y}:X\otimes
Y\rightarrow Y\otimes X$ such that $(g\otimes f)\circ\Psi_{X,Y}=\Psi
_{X^{\prime},Y^{\prime}}\circ(f\otimes g)$ for arbitrary morphisms $f\in$
Mor$(X,X^{\prime})$ and $g\in$ Mor$(Y,Y^{\prime})$. In addition to this one
requires the hexagon axiom to hold. The hexagon axiom is the validity of the
two conditions
\begin{equation}
\Psi_{X,Z}\circ\Psi_{Y,Z}=\Psi_{X\otimes Y,Z},\quad\Psi_{X,Z}\circ\Psi
_{X,Y}=\Psi_{X,Y\otimes Z}.
\end{equation}
A tensor category equipped with such mappings $\Psi_{X,Y}$ for each pair of
objects $X,Y$ is called a \textit{braided tensor category}. The mappings
$\Psi_{X,Y}$ as a whole are often referred to as the braiding of the tensor
category. Furthermore, for any quantum space algebra $X$ in this category
there are morphisms $\Delta:X\rightarrow X\otimes X,$ $S:X\rightarrow X,$ and
$\varepsilon:X\rightarrow\mathbb{C}$ forming a \textit{braided Hopf algebra},
i.e. $\Delta,$ $S,$ and $\varepsilon$ obey the usual axioms of a Hopf algebra,
but now as morphisms in the braided category. At this point, it should be
noted that the inverse mappings $\Psi_{X,Y}^{-1}$ give a braiding as well, to
which we can assign another braided Hopf structure denoted in the following by
$\bar{\Delta}:X\rightarrow X\otimes X,$ $\bar{S}:X\rightarrow X,$ and
$\bar{\varepsilon}:X\rightarrow\mathbb{C}$. For further details we recommend
Refs.\ \cite{Maj95}, \cite{Maj95star}, and \cite{ChDe96}.

In what follows it is important to realize that braided products and
q-deformed translations are linked to the braiding mappings and the Hopf
structures of a braided Hopf algebra, respectively. This way the axioms of a
braided Hopf algebra carry over to the corresponding operations on commutative
algebras. To this end, we first collect the axioms of a braided Hopf algebra,
which are given by (see for example Ref. \cite{Maj93-Int})%
\begin{align}
m\circ(S\otimes S)\circ\Psi^{-1}  &  =m\circ\Psi^{-1}\circ(S\otimes S)=S\circ
m,\nonumber\\
m\circ(\bar{S}\otimes\bar{S})\circ\Psi &  =m\circ\Psi\circ(\bar{S}\otimes
\bar{S})=\bar{S}\circ m,\label{Hopf1AnfN}\\[0.1in]
m\circ(S^{-1}\otimes S^{-1})\circ\Psi &  =m\circ\Psi\circ(S^{-1}\otimes
S^{-1})=S^{-1}\circ m,\nonumber\\
m\circ(\bar{S}^{-1}\otimes\bar{S}^{-1})\circ\Psi^{-1}  &  =m\circ\Psi
^{-1}\circ(\bar{S}^{-1}\otimes\bar{S}^{-1})=\bar{S}^{-1}\circ m,
\label{Hopf1EndN}%
\end{align}
and
\begin{align}
\Delta\circ S  &  =(S\otimes S)\circ(\Psi\circ\Delta),\nonumber\\
\bar{\Delta}\circ\bar{S}  &  =(\bar{S}\otimes\bar{S})\circ(\Psi^{-1}\circ
\bar{\Delta}),\label{Hopf2AnfN}\\[0.1in]
(\Psi\circ\Delta)\circ S^{-1}  &  =(S^{-1}\otimes S^{-1})\circ\Delta
,\nonumber\\
(\Psi^{-1}\circ\bar{\Delta})\circ\bar{S}^{-1}  &  =(\bar{S}^{-1}\otimes\bar
{S}^{-1})\circ\bar{\Delta}, \label{Hopf2EndN}%
\end{align}
where $m$ denotes\ multiplication in the quantum space algebra. Furthermore,
there are the axioms
\begin{align}
m\circ(S\otimes\text{id})\circ\Delta &  =m\circ(\text{id}\otimes S)\circ
\Delta=\varepsilon,\nonumber\\
m\circ(\bar{S}\otimes\text{id})\circ\bar{\Delta}  &  =m\circ(\text{id}%
\otimes\bar{S})\circ\bar{\Delta}=\bar{\varepsilon}, \label{HopfVerAnfN}%
\\[0.1in]
m\circ(S^{-1}\otimes\text{id})\circ(\Psi\circ\Delta)  &  =m\circ
(\text{id}\otimes S^{-1})\circ(\Psi\circ\Delta)=\varepsilon,\nonumber\\
m\circ(\bar{S}^{-1}\otimes\text{id})\circ(\Psi^{-1}\circ\bar{\Delta})  &
=m\circ(\text{id}\otimes\bar{S}^{-1})\circ(\Psi^{-1}\circ\bar{\Delta}%
)=\bar{\varepsilon}. \label{HopfVerEndN}%
\end{align}

It is our aim to reformulate the above axioms in terms of the new operations
on commutative coordinate algebras. For this to achieve we need a kind of
dictionary which contains the correspondence between the set of morphisms in
the braided category and the new operations on commutative algebras:%
\begin{align}
\Delta &  \rightarrow f(x^{i}\oplus_{L}y^{j}),\quad & S  &  \rightarrow
f(\ominus_{L}\,x^{i}),\nonumber\\
\bar{\Delta}  &  \rightarrow f(x^{i}\oplus_{\bar{L}}y^{j}),\quad & \bar{S}  &
\rightarrow f(\ominus_{\bar{L}}\,x^{i}),\label{UebReg1}\\[0.1in]
\Psi\circ\Delta &  \rightarrow f(x^{i}\oplus_{R}y^{j}),\quad & S^{-1}  &
\rightarrow f(\ominus_{R}\,x^{i}),\nonumber\\
\Psi^{-1}\circ\bar{\Delta}  &  \rightarrow f(x^{i}\oplus_{\bar{R}}y^{j}),\quad
& \bar{S}^{-1}  &  \rightarrow f(\ominus_{\bar{R}}\,x^{i}), \label{UebReg2}%
\\[0.1in]
\Psi &  \rightarrow f(x^{i})\,\odot_{\bar{L}}\,g(y^{j}),\quad & \Psi^{-1}  &
\rightarrow f(x^{i})\,\odot_{L}\,g(y^{j}),\nonumber\\
m  &  \rightarrow f\circledast g. &  &
\end{align}
In addition to this, we should mention that%
\begin{align}
f(x^{i}\oplus_{L}y^{j})  &  =f(x^{i}\oplus_{\bar{R}}y^{j}),\nonumber\\
f(x^{i}\oplus_{\bar{L}}y^{j})  &  =f(x^{i}\oplus_{R}y^{j}),
\end{align}
and%
\begin{align}
f(\ominus_{L}\,x^{i})  &  =f(\ominus_{\bar{R}}\,x^{i}),\nonumber\\
f(\ominus_{\bar{L}}\,x^{i})  &  =f(\ominus_{R}\,x^{i}).
\end{align}
These identifications follow from arguments similar to those leading to
(\ref{BraidIdN}) or from a direct inspection of the explicit formulae for
q-deformed translations \cite{Wac04}.

Now, we are ready to rewrite the axioms in (\ref{Hopf1AnfN}) and
(\ref{Hopf1EndN}) as%
\begin{align}
f(\ominus_{L}\,x^{i})\odot_{L}g(\ominus_{L}\,x^{j})  &  =(f\circledast
g)(\ominus_{L}\,x^{i}),\nonumber\\
f(\ominus_{\bar{L}}\,x^{i})\odot_{\bar{L}}g(\ominus_{\bar{L}}\,x^{j})  &
=(f\circledast g)(\ominus_{\bar{L}}\,x^{i}),\label{HopfKomAnfN}\\[0.1in]
f(\ominus_{R}\,x^{i})\odot_{R}g(\ominus_{R}\,x^{j})  &  =(f\circledast
g)(\ominus_{R}\,x^{i}),\nonumber\\
f(\ominus_{\bar{R}}\,x^{i})\odot_{\bar{R}}g(\ominus_{\bar{R}}\,x^{j})  &
=(f\circledast g)(\ominus_{\bar{R}}\,x^{i}). \label{HopfKomEndN}%
\end{align}
Notice that we take the convention that tensor factors which are addressed by
the same coordinates have to be multiplied via the star product. This means,
for example,%
\begin{equation}
f(x^{i})\odot_{\bar{L}}g(x^{j})=(\mathcal{R}_{[2]}\triangleright
g(x^{j}))\overset{x}{\circledast}(\mathcal{R}_{[1]}\triangleright f(x^{i})).
\end{equation}
Likewise we find from the axioms in (\ref{Hopf2AnfN}) and (\ref{Hopf2EndN})
that%
\begin{align}
f(\ominus_{L}(x^{i}\oplus_{L}y^{j}))  &  =f((\ominus_{L}\,x^{i})\oplus
_{R}(\ominus_{L}\,y^{j})),\nonumber\\
f(\ominus_{\bar{L}}(x^{i}\oplus_{\bar{L}}y^{j}))  &  =f((\ominus_{\bar{L}%
}\,x^{i})\oplus_{\bar{R}}(\ominus_{\bar{L}}\,y^{j})),\\[0.1in]
f(\ominus_{R}(x^{i}\oplus_{R}y^{j}))  &  =f((\ominus_{R}\,x^{i})\oplus
_{L}(\ominus_{R}\,y^{j})),\nonumber\\
f(\ominus_{\bar{R}}(x^{i}\oplus_{\bar{R}}y^{j}))  &  =f((\ominus_{\bar{R}%
}\,x^{i})\oplus_{\bar{L}}(\ominus_{\bar{R}}\,y^{j})).
\end{align}
Furthermore, we are able to translate the axioms in (\ref{HopfVerAnfN}) and
(\ref{HopfVerEndN}) into the identities%
\begin{align}
f((\ominus_{L}\,x^{i})\oplus_{L}x^{j})  &  =f(x^{i}\oplus_{L}(\ominus
_{L}\,x^{j}))=f(0),\nonumber\\
f((\ominus_{\bar{L}}\,x^{i})\oplus_{\bar{L}}x^{j})  &  =f(x^{i}\oplus_{\bar
{L}}(\ominus_{\bar{L}}\,x^{j}))=f(0),\label{qAddN}\\[0.16in]
f((\ominus_{R}\,x^{i})\oplus_{R}x^{j})  &  =f(x^{i}\oplus_{R}(\ominus
_{R}\,x^{j}))=f(0),\nonumber\\
f((\ominus_{\bar{R}}\,x^{i})\oplus_{\bar{R}}x^{j})  &  =f(x^{i}\oplus_{\bar
{R}}(\ominus_{\bar{R}}\,x^{j}))=f(0), \label{qAdd2N}%
\end{align}
where%
\begin{equation}
f(0)\equiv\varepsilon(\mathcal{W}(f))=\left.  f(x^{i})\right\vert _{x^{i}=0}.
\end{equation}
Notice that a function being q-translated depends on two sets of coordinates.
In the above formulae the tensor factors corresponding to these coordinates
again have to be multiplied via the star product. The expressions for star
products given in the work of Ref. \cite{WW01} can easily be adapted to this
task. In the case of the quantum plane, for example, we have%
\begin{equation}
f(x^{i},x^{j})=q^{-\hat{n}_{x^{2}}\hat{n}_{y^{1}}}\left.  f(x^{i}%
,y^{i})\right\vert _{y^{i}\rightarrow x^{i}}.
\end{equation}
Finally, the property of the coproduct to be an algebra homomorphism [cf.
(\ref{Trans})] gives rise to%
\begin{align}
(f\circledast g)(x^{i}\oplus_{L}y^{j})  &  =f(x^{i}\oplus_{L}y^{j})\,\odot
_{L}\,g(x^{i}\oplus_{L}y^{j}),\nonumber\\
(f\circledast g)(x^{i}\oplus_{\bar{L}}y^{j})  &  =f(x^{i}\oplus_{\bar{L}}%
y^{j})\,\odot_{\bar{L}}\,g(x^{i}\oplus_{\bar{L}}y^{j}),\\[0.16in]
(f\circledast g)(x^{i}\oplus_{R}y^{j})  &  =f(x^{i}\oplus_{R}y^{j})\,\odot
_{R}\,g(x^{i}\oplus_{R}y^{j}),\nonumber\\
(f\circledast g)(x^{i}\oplus_{\bar{R}}y^{j})  &  =f(x^{i}\oplus_{\bar{R}}%
y^{j})\,\odot_{\bar{R}}\,g(x^{i}\oplus_{\bar{R}}y^{j}).
\end{align}
To understand the right-hand side of the above equalities recall that each
function being translated lives in a tensor product of two coordinate
algebras. So functions that undergo the same q-translation have to be
multiplied by the q-deformed tensor product of (\ref{BraTens1}) or
(\ref{BraTens2}).

As we know, classical translations form a group. Considering q-deformed
translations we can detect reminiscences of the group axioms satisfied by
classical translations. We wish to illustrate this observation by the
following calculation:%
\begin{align}
f((x^{i}\oplus_{L}y^{j})\oplus_{L}(\ominus_{L}y^{k}))  &  =f(x^{i}\oplus
_{L}(y^{j}\oplus_{L}(\ominus_{L}y^{k})))\nonumber\\
&  =f(x^{i}\oplus_{L}0)=f(x^{i}).
\end{align}
The first equality means associativity of q-deformed translations and follows
from coassociativity of comultiplication. The second equality results from the
identities in (\ref{qAddN}), which concern the existence of inverse elements.
The last equality, which is a consequence of the Hopf algebra axiom
\begin{equation}
(\text{id}\otimes\varepsilon)\circ\Delta=(\varepsilon\otimes\text{id}%
)\circ\Delta=\text{id},
\end{equation}
tells us that we have an identity element.

\subsection{Conjugation properties and crossing symmetries}

Now, we would like to discuss the conjugation properties of our quantum
spaces. First of all, let us recall that our quantum spaces are endowed with a
mapping $a\rightarrow\bar{a}$ that makes them $\ast$-algebras, i.e.
\begin{equation}
\overline{\alpha a+\beta b}=\bar{\alpha}\bar{a}+\bar{\beta}\bar{b}%
,\quad\overline{\bar{a}}=a,\quad\overline{ab}=\bar{b}\bar{a}, \label{AxiomCon}%
\end{equation}
with $\alpha,\beta\in\mathbb{C}$ and\ $a,b\in\mathcal{A}_{q}.$ This mapping
can be extended to tensor products of quantum spaces by%
\begin{equation}
\overline{(a\otimes b)}=\bar{b}\otimes\bar{a}. \label{ConTen}%
\end{equation}

A quantum space generator is said to be real if%
\begin{equation}
\overline{X^{i}}=X_{i}=g_{ij}X^{j}=\bar{X}_{i}, \label{RealCoor}%
\end{equation}
where $g_{ij}$ stands for the corresponding quantum metric. Notice that the
quantum metric allows us to raise and lower indices as usual (see also
Appendix \ref{AppQuan}). From Eq. (\ref{RealCoor}) we see that conjugating
coordinates changes their transformation properties, i.e. contravariant
coordinates become covariant and vice versa. However, there is one exception
from (\ref{RealCoor}) since the time element $X^{0}$ for q-deformed Minkowski
space is imaginary, i.e. it fulfills%
\begin{equation}
\overline{X^{0}}=-X_{0}=-g_{00}X^{0}=X^{0}=-\bar{X}_{0}.
\end{equation}
So i$X^{0}$ is a real coordinate. In order to simplify our considerations we
will assume that all quantum space coordinates are real. For q-deformed
Minkowski space this requires to deal with i$X^{0}$ instead of $X^{0}.$

Next, we wish to focus our attention on the conjugation properties of the
L-matrices. Applying the conjugation properties of the symmetry generators to
the explicit form of the L-matrices (see for example Ref. \cite{MSW04}) one
can check the identities%
\begin{align}
\overline{(\mathcal{L}_{x})_{j}^{i}}  &  =g_{ik}\,(\mathcal{\bar{L}}_{x}%
)_{l}^{k}\,g^{lj},\nonumber\\
\overline{(\mathcal{\bar{L}}_{x})_{j}^{i}}  &  =g_{ik}\,(\mathcal{L}_{x}%
)_{l}^{k}\,g^{lj}.
\end{align}
With these relations at hand we can proceed as follows:%
\begin{align}
\overline{X^{i}a}  &  =\overline{(1\otimes X^{i})(a\otimes1)}=\overline
{\big ((\mathcal{\bar{L}}_{x}\mathcal{)}_{j}^{i}\triangleright a\big )\otimes
X^{j}}\nonumber\\
&  =\bar{X}_{j}\otimes\big (\bar{a}\triangleleft\overline{(\mathcal{\bar{L}%
}_{x}\mathcal{)}_{j}^{i}}\big )=\bar{X}_{j}\otimes\big (\bar{a}\triangleleft
(\mathcal{L}_{x})_{l}^{k}\,g_{ik}g^{lj}\big )\nonumber\\
&  =g_{ik}\,\bar{X}^{l}\otimes\big (\bar{a}\triangleleft(\mathcal{L}_{x}%
)_{l}^{k}\big )=g_{ik}(1\otimes\bar{a})(\bar{X}^{k}\otimes1)\nonumber\\
&  =(1\otimes\bar{a})(\bar{X}_{i}\otimes1)=\bar{a}\,\bar{X}_{i}.
\label{KonXaN}%
\end{align}
For arbitrary quantum space elements we analogously get
\begin{align}
\overline{ab}  &  =\overline{(1\otimes a)(b\otimes1)}=\overline{(\mathcal{R}%
_{[2]}\triangleright b)\otimes(\mathcal{R}_{[1]}\triangleright a)}\nonumber\\
&  =(\bar{a}\triangleleft\mathcal{R}_{[2]})\otimes(\bar{b}\triangleleft
\mathcal{R}_{[1]})=(1\otimes\bar{b})(\bar{a}\otimes1)\nonumber\\
&  =\bar{b}\bar{a}, \label{Konab}%
\end{align}
where for the third equality we used that the universal R-matrix is real in
the sense
\begin{equation}
\overline{\mathcal{R}_{[1]}}\otimes\overline{\mathcal{R}_{[2]}}=\mathcal{R}%
_{[2]}\otimes\mathcal{R}_{[1]}.
\end{equation}
It should be mentioned that the result in (\ref{Konab}) remains unchanged if
we use the braiding induced by $\tau\circ\mathcal{R}^{-1}$. The product in
(\ref{KonXaN}) and (\ref{Konab}) should not be confused with that in
(\ref{AxiomCon}). In (\ref{AxiomCon}) $a$ and $b$ live in the same quantum
space, but in (\ref{KonXaN}) and (\ref{Konab}) this is not the case.
Introducing the braiding mappings%
\begin{align}
\Psi_{\mathcal{A}_{q},\mathcal{A}_{q}^{^{\prime}}}(a\otimes b)  &
\equiv(\mathcal{R}_{[2]}\triangleright b)\otimes(\mathcal{R}_{[1]}%
\triangleright a),\nonumber\\
\Psi_{\mathcal{A}_{q},\mathcal{A}_{q}^{^{\prime}}}^{-1}(a\otimes b)  &
\equiv(\mathcal{R}_{[1]}^{-1}\triangleright b)\otimes(\mathcal{R}_{[2]}%
^{-1}\triangleright a),
\end{align}
the results of the above considerations can be reformulated as%
\begin{align}
\overline{\Psi_{\mathcal{A}_{q},\mathcal{A}_{q}^{^{\prime}}}(a\otimes b)}  &
=\Psi_{\mathcal{A}_{q},\mathcal{A}_{q}^{^{\prime}}}(\bar{b}\otimes\bar
{a}),\nonumber\\
\overline{\Psi_{\mathcal{A}_{q},\mathcal{A}_{q}^{^{\prime}}}^{-1}(a\otimes
b)}  &  =\Psi_{\mathcal{A}_{q},\mathcal{A}_{q}^{^{\prime}}}^{-1}(\bar
{b}\otimes\bar{a}). \label{KonBraidN}%
\end{align}

In very much the same way as in (\ref{KonXaN}) we can compute for the
coproduct of single quantum space generators%
\begin{align}
\overline{\Delta_{L}(X^{i})}  &  =\overline{(X^{i}\otimes1+(\mathcal{L}%
_{x})_{j}^{i}\otimes X^{j})}\nonumber\\
&  =1\otimes\bar{X}_{i}+\bar{X}_{j}\otimes\overline{(\mathcal{L}_{x})_{j}^{i}%
}\nonumber\\
&  =1\otimes\bar{X}_{i}+\bar{X}_{j}\otimes g_{ik}\,(\mathcal{\bar{L}}_{x}%
)_{l}^{k}\,g^{lj}\nonumber\\
&  =1\otimes g_{ik}\bar{X}^{k}+\bar{X}^{l}\otimes g_{ik}\,(\mathcal{\bar{L}%
}_{x})_{l}^{k}\nonumber\\
&  =\Delta_{\bar{R}}(g_{ik}\bar{X}^{k})=\Delta_{\bar{R}}(\bar{X}_{i}).
\end{align}
This result confirms\ (cf. Ref. \cite{Maj95star}),
\begin{equation}
\tau\circ(\ast\otimes\ast)\circ\Delta_{L/\bar{L}}=\Delta_{\bar{R}/R}\circ\ast,
\label{KonCopr}%
\end{equation}
where $\ast$ symbolizes the operation for conjugation. Similar considerations
hold for the corresponding antipodes. As an example we give
\begin{align}
\overline{S_{L}(X^{i})}  &  =-\overline{S(\mathcal{L}_{x})_{j}^{i}\,X^{j}%
}=-\bar{X}_{j}\,\overline{S(\mathcal{L}_{x})_{j}^{i}}\nonumber\\
&  =-\bar{X}_{j}\,S^{-1}\overline{(\mathcal{L}_{x})_{j}^{i}}=-\bar{X}%
_{j}\,g_{ik}\,g^{lj}\,S^{-1}(\mathcal{\bar{L}}_{x})_{l}^{k}\nonumber\\
&  =-g_{ik}\,\bar{X}^{l}\,S^{-1}(\mathcal{\bar{L}}_{x})_{l}^{k}=S_{\bar{R}%
}(\bar{X}_{i}), \label{ConAntN}%
\end{align}
where the third equality makes use of the identity $\ast\circ S=S^{-1}%
\circ\ast$. The calculation in (\ref{ConAntN}) and its counterparts for
$S_{\bar{L}}$ and $S_{R}$ imply%
\begin{equation}
\ast\circ S_{L/\bar{L}}=S_{\bar{R}/R}\circ\ast. \label{KonAntipo}%
\end{equation}

Essentially for us is the fact that the conjugation on a quantum space
$\mathcal{A}_{q}$ induces a mapping on the corresponding commutative algebra
$\mathcal{A}$. This can be achieved through
\begin{equation}
\mathcal{W}(\,\overline{f(x^{i})}\,)=\overline{\mathcal{W}(f(x^{i}))}%
\qquad\text{or\qquad}\overline{f(x^{i})}=\mathcal{W}^{-1}(\,\overline
{\mathcal{W}(f(x^{i}))}\,), \label{ConAlgIso}%
\end{equation}
i.e. the algebra isomorphism $\mathcal{W}$ intertwines the conjugation on the
quantum space algebra with that on the corresponding commutative algebra.
Since we assume that a commutative function $f(x^{i})$ can be written as a
power series we especially have%
\begin{equation}
\overline{f(x^{i})}=\sum_{k_{1},\ldots,k_{n}}\bar{f}_{k_{1},\ldots,k_{n}%
}\,\overline{(x^{1})^{k_{1}}\ldots(x^{n})^{k_{n}}},
\end{equation}
where $\bar{f}_{k_{1},\ldots,k_{n}}$ denotes the complex conjugate of the
coefficient $f_{k_{1},\ldots,k_{n}}.$ Thus, it suffices to know how monomials
behave under conjugation:%
\begin{align}
\overline{(x^{1})^{k_{1}}\ldots(x^{n})^{k_{n}}}  &  =\mathcal{W}%
^{-1}\big (\,\overline{(X^{1})^{k_{1}}\ldots(X^{n})^{k_{n}}}%
\,\big )\nonumber\\
&  =\mathcal{W}^{-1}\big (\,\overline{(X^{n})^{k_{n}}}\ldots\overline
{(X^{1})^{k_{1}}}\,\big )\nonumber\\
&  =\mathcal{W}^{-1}\big (\,(X_{n})^{k_{n}}\ldots(X_{1})^{k_{1}}%
\,\big )\nonumber\\
&  =\mathcal{W}^{-1}\big ((g_{n\bar{n}}X^{\bar{n}})^{k_{n}}\ldots(g_{1\bar{1}%
}X^{\bar{1}})^{k_{1}}\big )\nonumber\\
&  =\mathcal{W}^{-1}\big ((g_{\bar{1}1}X^{1})^{k_{n}}\ldots(g_{\bar{n}n}%
X^{n})^{k_{1}}\big )\nonumber\\
&  =(g_{\bar{1}1}x^{1})^{k_{n}}\ldots(g_{\bar{n}n}x^{n})^{k_{1}}.
\label{ConMonN}%
\end{align}
Notice that in the third equality we have introduced for each index $i$ a
so-called conjugate index $\overline{i}(\equiv n+1-i)$. For the quantum spaces
we are considering for physical reasons the conjugate indices are given by

\begin{enumerate}
\item[(i)] (quantum plane) $\overline{i}\equiv3-i,\quad i=1,2,$

\item[(ii)] (three-dimensional Euclidean space) $\overline{(+,3,-)}=(-,3,+),$

\item[(iii)] (four-dimensional Euclidean space) $\overline{i}\equiv5-i,\quad
i=1,\ldots,4,$

\item[(iv)] (q-deformed Minkowski space) $\overline{(+,3,0,-)}=(-,3,0,+).$
\end{enumerate}

\noindent One should also realize that there is no summation over repeated
indices in (\ref{ConMonN}), since the non-vanishing entries of the quantum
metric are given by $g_{i\overline{i}}$ or $g_{\,\overline{i}i}$ for all
possible values of $i.$ Finally, we can summarize our considerations by%
\begin{equation}
\overline{f(x^{i})}=\bar{f}(\,\overline{x^{i}}\,)\equiv\left.  \bar{f}%
(x^{i})\right\vert _{x^{i}\rightarrow g_{i\overline{i}}\,x^{\overline{i}}},
\label{ConFunk}%
\end{equation}
where%
\begin{equation}
\bar{f}(x^{i})\sum_{k_{1},\ldots,k_{n}}\bar{f}_{k_{1},\ldots,k_{n}}%
\,(x^{1})^{k_{1}}\ldots(x^{n})^{k_{n}}.
\end{equation}

Now, we are in a position to describe the conjugation properties of star
products, braided products, and q-deformed translations.\ First of all, from
(\ref{AxiomCon}) and (\ref{ConAlgIso}) it follows directly that
\begin{align}
\overline{f(x^{i})\circledast g(x^{j})}  &  =\overline{\mathcal{W}%
^{-1}\big (\mathcal{W}\left(  f\right)  \mathcal{W}\left(  g\right)
\big )}=\mathcal{W}^{-1}\big (\,\overline{\mathcal{W}\left(  f\right)
\mathcal{W}\left(  g\right)  }\,\big )\nonumber\\
&  =\mathcal{W}^{-1}\big (\,\overline{\mathcal{W}\left(  g\right)  }%
\overline{\,\mathcal{W}\left(  f\right)  }\,\big )=\mathcal{W}^{-1}%
\big (\mathcal{W}\left(  \bar{g}\right)  \mathcal{W}(\bar{f}%
\,)\big )\nonumber\\
&  =\mathcal{W}^{-1}\big (\mathcal{W}\left(  \bar{g}\right)  \big )\mathcal{W}%
^{-1}\big (\mathcal{W}(\bar{f}\,)\big )\nonumber\\
&  =\overline{g(x^{i})}\circledast\overline{f(x^{j})}.
\end{align}
It should also be clear from what we have done so far that the relations in
(\ref{KonBraidN}) give rise to the following identities:%
\begin{align}
\overline{f(x^{i})\odot_{L}\,g(y^{j})}  &  =\overline{g(y^{j})}\odot_{\bar{R}%
}\,\overline{f(x^{i})}=\overline{g(y^{j})}\odot_{L}\,\overline{f(x^{i}%
)},\nonumber\\
\overline{f(x^{i})\odot_{\bar{L}}\,g(y^{j})}  &  =\overline{g(y^{j})}\odot
_{R}\,\overline{f(x^{i})}=\overline{g(y^{j})}\odot_{\bar{L}}\,\overline
{f(x^{i})}.
\end{align}
Furthermore, we can write down commutative counterparts of the relations in
(\ref{KonCopr}) and (\ref{KonAntipo}). Towards this end, we perform the
calculation%
\begin{align}
\overline{f(x^{i}\oplus_{L/\bar{L}}y^{j})}  &  \overset{(\ref{DefCoProN})}%
{=}\overline{\big (\mathcal{W}_{L}^{-1}\otimes\mathcal{W}_{L}^{-1}%
\big )\circ\Delta_{L/\bar{L}}(\mathcal{W}(f))}\nonumber\\
&  =\big (\mathcal{W}_{R}^{-1}\otimes\mathcal{W}_{R}^{-1}%
\big )\big (\,\overline{\Delta_{L/\bar{L}}(\mathcal{W}(f))}\,\big )\nonumber\\
&  \overset{(\ref{KonCopr})}{=}\big (\mathcal{W}_{R}^{-1}\otimes
\mathcal{W}_{R}^{-1}\big )\big (\Delta_{\bar{R}/R}(\,\overline{\mathcal{W}%
(f)}\,)\big )\nonumber\\
&  \overset{(\ref{ConAlgIso})}{=}\big (\mathcal{W}_{R}^{-1}\otimes
\mathcal{W}_{R}^{-1}\big )\big (\Delta_{\bar{R}/R}(\mathcal{W}(\bar
{f}\,))\big )\nonumber\\
&  =\bar{f}((\,\overline{y^{j}}\,)\oplus_{\bar{R}/R}\,(\,\overline{x^{i}}\,)),
\label{ConTransN}%
\end{align}
and similarly%
\begin{align}
\overline{f(\ominus_{L/\bar{L}}\,x^{i})}  &  \overset{(\ref{DefAntiN})}%
{=}\overline{\mathcal{W}_{R}^{-1}\big (S_{L/\bar{L}}(\mathcal{W}%
(f))\big )}=\mathcal{W}_{L}^{-1}\big (\,\overline{S_{L/\bar{L}}(\mathcal{W}%
(f))}\,\big )\nonumber\\
&  \overset{(\ref{KonAntipo})}{=}\mathcal{W}_{L}^{-1}\big (S_{\bar{R}%
/R}(\overline{\,\mathcal{W}(f)}\,)\big )\overset{(\ref{ConAlgIso})}%
{=}\mathcal{W}_{L}^{-1}\big (S_{\bar{R}/R}(\mathcal{W}(\bar{f}%
\,))\big )\nonumber\\
&  =\bar{f}(\ominus_{\bar{R}/R}\,(\,\overline{x^{i}}\,)). \label{ConInTranN}%
\end{align}
Notice that in (\ref{ConTransN}) and (\ref{ConInTranN}) we make use of
\begin{align}
\overline{\mathcal{W}_{L}^{-1}(ah)}  &  =\mathcal{W}_{R}^{-1}(\bar{h}\bar
{a}),\nonumber\\
\overline{\mathcal{W}_{R}^{-1}(ha)}  &  =\mathcal{W}_{L}^{-1}(\bar{a}\bar{h}),
\end{align}
where $a\in\mathcal{A}_{q}$ and $h\in\mathcal{H}$. To sum up, the above
considerations leave us with the rules%
\begin{align}
\overline{f(x^{i}\oplus_{L/\bar{L}}y^{j})}  &  =\bar{f}((\,\overline{y^{j}%
\,})\oplus_{\bar{R}/R}(\,\overline{x^{i}}\,)),\nonumber\\
\overline{f(x^{i}\oplus_{R/\bar{R}}y^{j})}  &  =\bar{f}((\,\overline{y^{j}%
}\,)\oplus_{\bar{L}/L}(\,\overline{x^{i}}\,)),\\[0.16in]
\overline{f(\ominus_{L/\bar{L}}\,x^{i})}  &  =\bar{f}(\ominus_{\bar{R}%
/R}\,(\,\overline{x^{i}}\,)),\nonumber\\
\overline{f(\ominus_{R/\bar{R}}\,x^{i})}  &  =\bar{f}(\ominus_{L/\bar{L}%
}\,(\,\overline{x^{i}}\,)).
\end{align}

Our last comment in this section concerns the existence of crossing
symmetries. They are based on the observation that the R-matrix and\ its
inverse are related to each other by%
\begin{equation}
\hat{R}_{kl}^{ij}\overset{{q\rightarrow1/q}}{\longleftrightarrow}(\hat{R}%
^{-1})_{\overline{k}\,\overline{l}}^{\overline{i}\,\overline{j}}.
\end{equation}
This means that substituting $1/q$ for $q$ we obtain from the entries of the
R-matrix those of its inverse, but now labeled by the corresponding conjugate
indices. The crossing symmetries are responsible for a number of remarkable
correspondences. In the following we would like to present some of these correspondences.

In Ref. \cite{WW01} we derived operator expressions for calculating star
products. These formulae refer to a given normal ordering, but due to the
crossing symmetries they can easily be transformed into expressions holding
for reversed normal ordering. In other words, if we use as algebra
homomorphism
\begin{align}
\widetilde{\mathcal{W}}  &  :\mathcal{A}\longrightarrow\mathcal{A}%
_{q},\nonumber\\
\widetilde{\mathcal{W}}((x^{1})^{i_{1}}\ldots(x^{n})^{i_{n}})  &
=(X^{n})^{i_{n}}\ldots(X^{1})^{i_{1}},
\end{align}
we get formulae for star products that can be obtained from those for the
algebra homomorphism $\mathcal{W}$ [cf. (\ref{AlgIsoN})] via the transition%
\begin{equation}
f\circledast g\overset{{%
\genfrac{}{}{0pt}{}{i}{q}%
}{%
\genfrac{}{}{0pt}{}{\leftrightarrow}{\leftrightarrow}%
}{%
\genfrac{}{}{0pt}{}{\overline{i}}{1/q}%
}}{\longleftrightarrow}f\,\widetilde{\circledast}\,g, \label{TransStarPro}%
\end{equation}
where the symbol $\overset{{%
\genfrac{}{}{0pt}{}{i}{q}%
}{%
\genfrac{}{}{0pt}{}{\leftrightarrow}{\leftrightarrow}%
}{%
\genfrac{}{}{0pt}{}{\overline{i}}{1/q}%
}}{\longleftrightarrow}$ now denotes the substitutions%
\begin{equation}
D_{q^{a}}^{i}\rightarrow D_{q^{-a}}^{\overline{i}},\quad\hat{n}_{x^{i}%
}\rightarrow\hat{n}_{x^{\overline{i}}},\quad q\rightarrow q^{-1}.
\label{TransConUn}%
\end{equation}
Notice that the tilde in Eq. (\ref{TransStarPro}) shall remind us of the fact
that the star product on the left-hand side refers to the algebra homomorphism
$\widetilde{\mathcal{W}}$. In the case of the Manin plane the relationship in
Eq. (\ref{TransStarPro}), for example, becomes%
\begin{align}
&  f(x^{i})\circledast g(x^{j})=\left[  q^{-\hat{n}_{x^{2}}\hat{n}_{y^{1}}%
}f(x^{i})g(y^{j})\right]  _{y\rightarrow x}\nonumber\\
\overset{{%
\genfrac{}{}{0pt}{}{i}{q}%
}{%
\genfrac{}{}{0pt}{}{\leftrightarrow}{\leftrightarrow}%
}{%
\genfrac{}{}{0pt}{}{\overline{i}}{1/q}%
}}{\longleftrightarrow}\hspace{0.15cm}  &  \left[  q^{\hat{n}_{x^{1}}\hat
{n}_{y^{2}}}f(x^{i})g(y^{j})\right]  _{y\rightarrow x}=f(x^{i})\,\widetilde
{\circledast}\,g(x^{j}).
\end{align}

In Refs. \cite{Wac04} and \cite{Wac05} we calculated explicit formulae for
q-translations and braided products. The crossing symmetries we could read off
from our results are sketched in Fig. \ref{ConBild1}. \begin{figure}[ptb]
\begin{center}
\setlength{\unitlength}{1.0cm} \begin{picture}(12,8)
\put(6,0.5){\vector(1,0){2.5}}
\put(6,0.5){\vector(-1,0){2.5}}
\put(6,6.5){\vector(1,0){2.5}}
\put(6,6.5){\vector(-1,0){2.5}}
\put(3,3.5){\vector(0,1){2.5}}
\put(3,3.5){\vector(0,-1){2.5}}
\put(9,3.5){\vector(0,1){2.5}}
\put(9,3.5){\vector(0,-1){2.5}}
\put(5.5,3){\vector(-1,-1){2}}
\put(5.5,4){\vector(-1,1){2}}
\put(6.5,3){\vector(1,-1){2}}
\put(6.5,4){\vector(1,1){2}}
\put(5.5,3){\makebox(1,0.5){
$ i \leftrightarrow \overline{i} $
}}
\put(5.5,3.5){\makebox(1,0.5){
$ x \leftrightarrow y $
}}
\put(0,4){\makebox(3,1)[b]{
$ q \leftrightarrow 1/q $
}}
\put(0,3.5){\makebox(3,1)[b]{
$ x \leftrightarrow y $
}}
\put(0.5,2.7){\parbox{2cm}{
\begin{center} {ordering \\ is reversed} \end{center}
}}
\put(9,4){\makebox(3,1)[b]{
$ q \leftrightarrow 1/q $
}}
\put(9,3.5){\makebox(3,1)[b]{
$ x \leftrightarrow y $
}}
\put(9.5,2.7){\parbox{2cm}{
\begin{center} {ordering \\ is reversed} \end{center}
}}
\put(4,0){\makebox(4,0.5)[b]{
ordering is reversed
}}
\put(4,6){\makebox(4,0.5)[b]{
ordering is reversed
}}
\put(4.5,0.7){\makebox(3,0.5)[b]{
$i \leftrightarrow \overline{i} $ \quad $ q \leftrightarrow 1/q $
}}
\put(4.5,6.7){\makebox(3,0.5)[b]{
$i \leftrightarrow \overline{i} $ \quad $ q \leftrightarrow 1/q $
}}
\put(1.3,0.5){\parbox{2cm}{
\begin{center} $\odot_R, \oplus_R,  \ominus_R$ \end{center}
}}
\put(1.3,6.5){\parbox{2cm}{
\begin{center} $\odot_L, \oplus_L,  \ominus_L$ \end{center}
}}
\put(8.7,0.5){\parbox{2cm}{
\begin{center} $\odot_{\bar{R}}, \oplus_{\bar{R}},  \ominus_{\bar{R}}$ \end{center}
}}
\put(8.7,6.5){\parbox{2cm}{
\begin{center} $\odot_{\bar{L}}, \oplus_{\bar{L}},  \ominus_{\bar{L}}$ \end{center}
}}
\end{picture}
\end{center}
\caption{Crossing-symmetries for braided products and q-translations}%
\label{ConBild1}%
\end{figure}
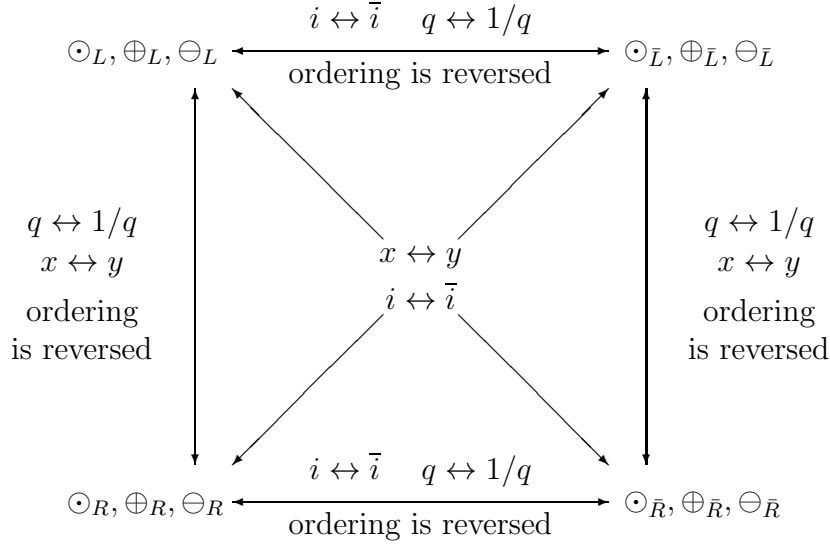Notice that expressions linked by an arrow can be transformed into
each other. The transition $i\leftrightarrow\overline{i}$ means that we have
to substitute for each coordinate index the conjugate one. In the same way,
$q\leftrightarrow1/q$ requires to interchange $q$ and its inverse. The symbol
$x\leftrightarrow y$ indicates that the variables corresponding to different
tensor factors have to be interchanged. Finally, the label 'ordering is
reversed' tells us that the corresponding expressions refer to different
algebra isomorphisms (given by $\mathcal{W}$ and $\widetilde{\mathcal{W}}$)$.$
We wish to illustrate these descriptions by the following example:%
\begin{align}
&  f(x^{i}\oplus_{L}y^{j})=\sum_{k_{1},k_{2}=0}^{\infty}\frac{(x^{1})^{k_{1}%
}(x^{2})^{k_{2}}}{[[k_{1}]]_{q^{-2}}![[k_{2}]]_{q^{-2}}!}\,\big ((D_{q^{-2}%
}^{1})^{k_{1}}(D_{q^{-2}}^{2})^{k_{2}}f\big )(q^{-k_{2}}y^{1})\nonumber\\
\overset{{%
\genfrac{}{}{0pt}{}{i}{q}%
}{%
\genfrac{}{}{0pt}{}{\leftrightarrow}{\leftrightarrow}%
}{%
\genfrac{}{}{0pt}{}{\overline{i}}{1/q}%
}}{\longleftrightarrow}\hspace{0.15cm}  &  f(x^{i}\oplus_{\bar{L}}y^{j}%
)=\sum_{k_{1},k_{2}=0}^{\infty}\frac{(x^{2})^{k_{2}}(x^{1})^{k_{1}}}%
{[[k_{1}]]_{q^{2}}![[k_{2}]]_{q^{2}}!}\,\big ((D_{q^{2}}^{1})^{k_{1}}%
(D_{q^{2}}^{2})^{k_{2}}f\big )(q^{k_{1}}y^{2})\nonumber\\
\overset{{%
\genfrac{}{}{0pt}{}{x}{q}%
}{%
\genfrac{}{}{0pt}{}{\leftrightarrow}{\leftrightarrow}%
}{%
\genfrac{}{}{0pt}{}{y}{1/q}%
}}{\longleftrightarrow}\hspace{0.15cm}  &  f(x^{i}\oplus_{\bar{R}}y^{j}%
)=\sum_{k_{1},k_{2}=0}^{\infty}\frac{(y^{1})^{k_{1}}(y^{2})^{k_{2}}}%
{[[k_{1}]]_{q^{-2}}![[k_{2}]]_{q^{-2}}!}\,\big ((D_{q^{-2}}^{1})^{k_{1}%
}(D_{q^{-2}}^{2})^{k_{2}}f\big )(q^{-k_{1}}y^{2}).
\end{align}

One should also keep in mind that the expressions for braided products and
q-deformed translations depend on the choice for the normal ordering.
Sometimes it is convenient to work with different normal orderings at the same
time. However, a physical theory should be formulated in terms of objects
referring to the same normal ordering. Thus, we need operators that enable us
to transform a commutative function to another one representing the same
quantum space element but now for reversed ordering. For the quantum spaces we
are interested in the explicit form of these operators can be found in the
work of Ref. \cite{BW01}. In the case of the Manin plane, for example,\ they
are given by%
\begin{align}
\tilde{f}(x^{2},x^{1})  &  =\hat{U}(f(x^{1},x^{2}))=q^{\hat{n}_{x^{1}}\hat
{n}_{x^{2}}}\,f(x^{1},x^{2}),\nonumber\\
f(x^{1},x^{2})  &  =\hat{U}^{-1}(\tilde{f}(x^{2},x^{1}))=q^{-\hat{n}_{x^{1}%
}\hat{n}_{x^{2}}}\,\tilde{f}(x^{2},x^{1}).
\end{align}
Notice that we take the convention that the underlying normal ordering is
indicated by the order in which the arguments of commutative functions are arranged.

\section{Partial derivatives on quantum spaces\label{SecPart}}

There are q-deformed analogs of partial derivatives, which act upon the
algebra of quantum space coordinates \cite{WZ91, CSW91, Song92}. In our
approach the role of q-deformed partial derivatives and coordinates is
completely symmetrical, i.e. partial derivatives show the same algebraic
properties as quantum space coordinates. In this manner, q-deformed partial
derivatives span quantum space algebras that are isomorphic to the quantum
spaces they act upon. Thus, we can conclude that the considerations of the
previous section also hold for the quantum space algebras of partial
derivatives. In the present section we focus attention on actions of partial derivatives.

\subsection{Leibniz rules and representations of partial derivatives}

In analogy to (\ref{VerRN}) and (\ref{VerRInN}) there are two possibilities
for commutation relations between partial derivatives and quantum space
coordinates. However, the action of partial derivatives on quantum space
coordinates requires to modify the commutation relations in (\ref{VerRN}) and
(\ref{VerRInN}) in such a way that they take the form%
\begin{align}
\partial^{i}X^{j}  &  =g^{ij}+k(\hat{R}^{-1})_{kl}^{ij}\,X^{k}\partial
^{l},\nonumber\\
\hat{\partial}^{i}X^{j}  &  =\bar{g}^{ij}+k^{-1}\hat{R}_{kl}^{ij}\,X^{k}%
\hat{\partial}^{l},\label{LeibRuleAnfN}\\[0.1in]
X^{i}\partial^{j}  &  =-g^{ij}+k(\hat{R}^{-1})_{kl}^{ij}\,\partial^{k}%
X^{l},\nonumber\\
X^{i}\hat{\partial}^{j}  &  =-\bar{g}^{ij}+k^{-1}\hat{R}_{kl}^{ij}%
\,\hat{\partial}^{k}X^{l}, \label{LeibRuleEndN}%
\end{align}
where we have introduced a conjugate quantum metric $\bar{g}^{ij}$. One should
notice that in the case of the quantum plane it holds $\bar{g}^{ij}=-g^{ij}.$
Otherwise we have $g^{ij}=\bar{g}^{ij}$. The constant $k$ is a power of $q$
and takes on the values

\begin{enumerate}
\item[(i)] (quantum plane) $k=q^{2},$

\item[(ii)] (three-dimensional Euclidean space) $k=1,$

\item[(iii)] (four-dimensional Euclidean space) $k=q,$

\item[(iv)] (q-deformed Minkowski space) $k=q^{-2}.$
\end{enumerate}

Let us say a few words about the relationship between the partial derivatives
$\partial^{i}$ and $\hat{\partial}^{i}.$ In the former literature \cite{OZ92,
LWW97} $\partial^{i}$ and $\hat{\partial}^{i}$ are seen to be conjugate to
each other. In our approach, however, the partial derivatives $\partial^{i}$
and $\hat{\partial}^{i}$ differ from each other by a normalization factor,
only. Concretely, we have
\begin{equation}
\hat{\partial}^{i}=(\alpha_{0})^{-1}k\partial^{i}, \label{norm}%
\end{equation}
where $\alpha_{0}$ denotes the eigenvalue to the one-dimensional subspace of
the vector representation of the universal R-matrix. For the quantum spaces we
are interested in, relation (\ref{norm}) becomes

\begin{enumerate}
\item[(i)] (quantum plane) $\hat{\partial}^{i}=q^{3}\partial^{i},$

\item[(ii)] (three-dimensional Euclidean space) $\hat{\partial}^{A}%
=q^{6}\partial^{A},$

\item[(iii)] (four-dimensional Euclidean space) $\hat{\partial}^{i}%
=q^{4}\partial^{i},$

\item[(iv)] (q-deformed Minkowski space) $\hat{\partial}^{\mu}=q^{-4}%
\partial^{\mu}.$
\end{enumerate}

From the q-deformed Leibniz rules in (\ref{LeibRuleAnfN}) and
(\ref{LeibRuleEndN}) we can derive right and left actions of partial
derivatives on quantum space elements. To this end, we repeatedly apply the
Leibniz rules to the product of a partial derivative with a normal ordered
monomial of coordinates, until we obtain an expression with all partial
derivatives standing to the right of all quantum space coordinates, i.e.%
\begin{equation}
\partial^{i}(X^{1})^{k_{1}}\ldots(X^{n})^{k_{n}}=\big (\partial_{(1)}%
^{i}\triangleright(X^{1})^{k_{1}}\ldots(X^{n})^{k_{n}}\big )\partial_{(2)}%
^{i}. \label{VerParX}%
\end{equation}
Taking the counit of all partial derivatives appearing on the right-hand side
finally yields the left action of $\partial^{i}$, since we have%
\begin{equation}
\big (\partial_{(1)}^{i}\triangleright(X^{1})^{k_{1}}\ldots(X^{n})^{k_{n}%
}\big )\varepsilon(\partial_{(2)}^{i})=\partial^{i}\triangleright
(X^{1})^{k_{1}}\ldots(X^{n})^{k_{n}}. \label{BerWirkPar}%
\end{equation}
Right actions of partial derivatives can be calculated in a similar way if we
start from a partial derivative standing to the right of a normal ordered
monomial and commute it to the left of all quantum space coordinates. Instead
of (\ref{VerParX}) and (\ref{BerWirkPar}), we now have
\begin{equation}
(X^{1})^{k_{1}}\ldots(X^{n})^{k_{n}}\partial^{i}=\partial_{(2)}^{i}%
\big ((X^{1})^{k_{1}}\ldots(X^{n})^{k_{n}}\triangleleft\partial_{(1)}%
^{i}\big )
\end{equation}
and%
\begin{equation}
\varepsilon(\partial_{(2)}^{i})\big ((X^{1})^{k_{1}}\ldots(X^{n})^{k_{n}%
}\triangleleft\partial_{(1)}^{i}\big )=(X^{1})^{k_{1}}\ldots(X^{n})^{k_{n}%
}\triangleleft\partial^{i}.
\end{equation}

The algebra isomorphism $\mathcal{W}$ allows us to introduce q-deformed
derivatives that act upon commutative functions. By means of the relations
\begin{align}
\mathcal{W}(\partial^{i}\triangleright f)  &  =\partial^{i}\triangleright
\mathcal{W}(f),\quad f\in\mathcal{A}\text{,}\nonumber\\
\mathcal{W}(f\triangleleft\partial^{i})  &  =\mathcal{W}(f)\triangleleft
\partial^{i},
\end{align}
or%
\begin{align}
\partial^{i}\triangleright f  &  \equiv\mathcal{W}^{-1}\left(  \partial
^{i}\triangleright\mathcal{W}(f)\right)  ,\nonumber\\
f\triangleleft\partial^{i}  &  \equiv\mathcal{W}^{-1}\left(  \mathcal{W}%
(f)\triangleleft\partial^{i}\right)  , \label{DefParAcN}%
\end{align}
the actions of partial derivatives on the quantum space algebra $\mathcal{A}%
_{q}$ carry over to the corresponding commutative algebra $\mathcal{A}$. It
should be obvious that each Leibniz rule in (\ref{LeibRuleAnfN}) and
(\ref{LeibRuleEndN}) leads to its own q-derivative:%
\begin{align}
\partial^{i}X^{j}  &  =g^{ij}+k(\hat{R}^{-1})_{kl}^{ij}\,X^{k}\partial^{l} &
&  \Rightarrow &  &  \partial^{i}\triangleright f,\nonumber\\
\hat{\partial}^{i}X^{j}  &  =\bar{g}^{ij}+k^{-1}\hat{R}_{kl}^{ij}\,X^{k}%
\hat{\partial}^{l} &  &  \Rightarrow &  &  \hat{\partial}^{i}\,\bar
{\triangleright}\,f,\label{FundWirkN}\\[0.16in]
X^{i}\partial^{j}  &  =-\bar{g}^{ij}+k(\hat{R}^{-1})_{kl}^{ij}\,\partial
^{k}X^{l} &  &  \Rightarrow &  &  f\,\bar{\triangleleft}\,\partial
^{i},\nonumber\\
X^{i}\hat{\partial}^{j}  &  =-g^{ij}+k^{-1}\hat{R}_{kl}^{ij}\,\hat{\partial
}^{k}X^{l} &  &  \Rightarrow &  &  f\triangleleft\hat{\partial}^{i}.
\label{FundWirk2N}%
\end{align}
In the work of Ref. \cite{BW01} we derived operator representations
for\ q-deformed partial derivatives by applying these ideas. Our results can
be viewed as multi-dimensional versions of the celebrated Jackson derivative
\cite{Jack08}. As an example we write down the left representations
for\ partial derivatives on the two-dimensional quantum plane:%
\begin{align}
\partial^{1}\triangleright f(x^{1},x^{2})  &  =-q^{-1/2}D_{q^{2}}^{2}%
f(qx^{1},x^{2}),\nonumber\\
\partial^{2}\triangleright f(x^{1},x^{2})  &  =q^{1/2}D_{q^{2}}^{1}%
f(x^{1},q^{2}x^{2}). \label{Par2dimNN}%
\end{align}

Next, we would like to discuss a point that is important for q-deformed
quantum mechanics. In our approach, as already mentioned, there is a symmetry
between coordinates and partial derivatives with the exception that under
conjugation coordinates are real objects, while partial derivatives have to be
imaginary (see the discussion in the next subsection). However, if we
introduce momentum variables by setting $P^{k}\equiv$ i$\partial^{k},$ we
obtain objects with the very same conjugation properties as quantum space
coordinates. With these new variables the Leibniz rules in (\ref{LeibRuleAnfN}%
) and (\ref{LeibRuleEndN}) now become%
\begin{align}
P^{k}(\text{i}X^{l})  &  =-g^{kl}+k(\hat{R}^{-1})_{mn}^{kl}\,(\text{i}%
X^{m})P^{n},\nonumber\\
P^{k}(\text{i}\hat{X}^{l})  &  =-\bar{g}^{kl}+k^{-1}\hat{R}_{mn}%
^{kl}\,(\text{i}\hat{X}^{m})P^{n},\label{LeiMom1N}\\[0.1in]
(\text{i}X^{k})P^{l}  &  =\bar{g}^{kl}+k(\hat{R}^{-1})_{mn}^{kl}%
\,P^{m}(\text{i}X^{n}),\nonumber\\
(\text{i}\hat{X}^{k})P^{l}  &  =g^{kl}+k^{-1}\hat{R}_{mn}^{kl}\,P^{m}%
(\text{i}\hat{X}^{n}). \label{LeibMom2N}%
\end{align}
This means that
\begin{equation}
\partial_{p}^{k}\equiv\text{i}X^{k}\text{\quad as well as\quad}\hat{\partial
}_{p}^{k}\equiv\text{i}\hat{X}^{k}%
\end{equation}
can play the role of partial derivatives on momentum space. In the case of the
Manin plane, however, we have to take into account an additional minus sign in
the definition of $\partial_{p}^{k}$ and\ $\hat{\partial}_{p}^{k}.$

Furthermore, we see by comparing (\ref{LeiMom1N}) with (\ref{LeibRuleAnfN})
and (\ref{LeibMom2N}) with (\ref{LeibRuleEndN}) that the Leibniz rules for
position and momentum space are linked via the substitutions%
\begin{equation}
X^{k}\longleftrightarrow P^{k},\quad\partial_{x}^{k}\longleftrightarrow
\partial_{p}^{k},\quad\hat{\partial}_{x}^{k}\longleftrightarrow\hat{\partial
}_{p}^{k}.
\end{equation}
Due to this fact, the actions of partial derivatives on momentum space take on
the same form as those for partial derivatives on coordinate space. To be more
specific we have the correspondences%
\begin{align}
\partial_{x}^{k}\triangleright f(x^{j})  &  \overset{x\leftrightarrow
p}{\longleftrightarrow}\partial_{p}^{k}\triangleright f(p^{j}),\nonumber\\
\hat{\partial}_{x}^{k}\,\bar{\triangleright}\,f(x^{j})  &  \overset
{x\leftrightarrow p}{\longleftrightarrow}\hat{\partial}_{p}^{k}\,\bar
{\triangleright}\,f(p^{j}),\\[0.16in]
f(x^{j})\triangleleft\hat{\partial}_{x}^{k}  &  \overset{x\leftrightarrow
p}{\longleftrightarrow}f(p^{j})\triangleleft\hat{\partial}_{p}^{k},\nonumber\\
f(x^{j})\,\bar{\triangleleft}\,\partial_{x}^{k}  &  \overset{x\leftrightarrow
p}{\longleftrightarrow}\,f(p^{j})\,\bar{\triangleleft}\,\partial_{p}^{k},
\end{align}
where $\overset{x\leftrightarrow p}{\longleftrightarrow}$ instructs us to
substitute the momentum variable $p^{i}$ for each space coordinate $x^{i}$ and
vice versa. These observations should tell us that all considerations for
position space pertain equally for momentum space.

\subsection{Conjugation properties and crossing symmetries}

As already mentioned, partial derivatives obey the same algebra relations as
the corresponding quantum space coordinates. Since we require for partial
derivatives to be imaginary, they have to satisfy%
\begin{equation}
\overline{\partial^{i}}=-\bar{\partial}_{i}=-\partial_{i}=-g_{ij}\partial^{j}.
\label{ConPart}%
\end{equation}
Again, there is one exception from this rule, since in the case of the
q-deformed Minkowski space we have%
\begin{equation}
\overline{\partial^{0}}=\bar{\partial}_{0}=\partial_{0}=g_{00}\partial
^{0}=-\partial^{0}.
\end{equation}

Next, we would like to say a few words about the conjugation properties of the
actions of partial derivatives. To this aim, we examine how the Leibniz rules
in (\ref{LeibRuleAnfN}) and (\ref{LeibRuleEndN}) change under conjugation. We
have%
\begin{align}
\overline{\partial^{i}X^{j}}  &  =\overline{g^{ij}+k(\hat{R}^{-1})_{kl}%
^{ij}\,X^{k}\partial^{l}}\nonumber\\
\Rightarrow-\bar{X}_{j}\bar{\partial}_{i}  &  =\bar{g}_{ij}-k(\hat{R}%
^{-1})_{ij}^{kl}\,\bar{\partial}_{l}\bar{X}_{k}\nonumber\\
\Rightarrow X_{j}\partial_{i}  &  =-\bar{g}_{ij}+k(\hat{R}^{-1})_{ij}%
^{kl}\,\partial_{l}X_{k}\nonumber\\
\Rightarrow X^{i}\partial^{j}  &  =-\bar{g}^{ij}+k(\hat{R}^{-1})_{kl}%
^{ij}\,\partial^{k}X^{l}, \label{ConLeib1N}%
\end{align}
and likewise%
\begin{align}
\overline{\hat{\partial}^{i}X^{j}}  &  =\overline{\bar{g}^{ij}+k^{-1}\hat
{R}_{kl}^{ij}\,X^{k}\hat{\partial}^{l}}\nonumber\\
\Rightarrow X^{i}\hat{\partial}^{j}  &  =-g^{ij}+k^{-1}\hat{R}_{kl}^{ij}%
\,\hat{\partial}^{k}X^{l}. \label{ConLeib2N}%
\end{align}
The derivation in (\ref{ConLeib1N}) makes use of (\ref{AxiomCon}),
(\ref{RealCoor}), and (\ref{ConPart}). In addition to this, we need the
identities%
\begin{gather}
\overline{g^{ij}}=g^{ij}=\bar{g}_{ij},\quad\overline{\bar{g}^{ij}}=\bar
{g}^{ij}=g_{ij},\nonumber\\
\overline{(\hat{R}^{\pm1})_{ij}^{kl}}=(\hat{R}^{\pm1})_{ij}^{kl}=(\hat{R}%
^{\pm1})_{kl}^{ij},\nonumber\\
g^{ml}g^{nk}(\hat{R}^{\pm1})_{kl}^{ij}=(\hat{R}^{\pm1})_{kl}^{mn}g^{li}g^{kj}.
\end{gather}
The above considerations tell us that under conjugation the Leibniz rules
leading to left actions transform into those for right actions and vice versa.
However, this statement is only valid, if we deal with coordinates and partial
derivatives being respectively real and imaginary under conjugation. For
q-deformed Minkowski space this means that we have to work with i$X^{0}$ and
i$\partial^{0}$ instead of $X^{0}$ and $\partial^{0}$. In doing so, the
quantum metric and the R-matrices for q-deformed Minkowski space must be
modified as follows:%
\begin{gather}
g^{\mu\nu}\rightarrow\text{i}^{\delta_{\mu0}+\delta_{\nu0}}g^{\mu\nu
},\nonumber\\
(\hat{R}^{\pm1})_{\rho\sigma}^{\mu\nu}\rightarrow\text{i}^{\delta_{\mu
0}+\delta_{\nu0}}(\hat{R}^{\pm1})_{\rho\sigma}^{\mu\nu}\,\text{i}%
^{\delta_{\rho0}+\delta_{\sigma0}}. \label{WickN}%
\end{gather}
The substitutions in (\ref{WickN}) can be viewed as a q-analog of classical
Wick rotation. By virtue of the identifications in (\ref{FundWirkN}) and
(\ref{FundWirk2N}), we finally conclude that the results of (\ref{ConLeib1N})
and (\ref{ConLeib2N}) imply the identities%
\begin{align}
\overline{\partial^{i}\triangleright f(x^{j})}  &  =-\overline{f(x^{j})}%
\,\bar{\triangleleft}\,\partial_{i}, & \overline{f(x^{j})\,\bar{\triangleleft
}\,\partial^{i}}  &  =-\partial_{i}\triangleright\overline{f(x^{j}%
)},\nonumber\\
\overline{\hat{\partial}^{i}\,\bar{\triangleright}\,f(x^{j})}  &
=-\overline{f(x^{j})}\triangleleft\hat{\partial}_{i}, & \overline
{f(x^{j})\triangleleft\hat{\partial}^{i}}  &  =-\hat{\partial}_{i}%
\,\bar{\triangleright}\,\overline{f(x^{j})}. \label{RegConAblN}%
\end{align}

Next, we turn to the crossing symmetries for partial derivatives, since they
give a simple method to transform the different representations\ of partial
derivatives into each other. In the case of partial derivatives crossing
symmetries represent the fact that we can make a transition between a Leibniz
rule and its conjugate version by applying the substitutions%
\begin{equation}
\partial^{i}\leftrightarrow\hat{\partial}^{\overline{i}},\quad X^{i}%
\leftrightarrow X^{\overline{i}},\quad q\leftrightarrow q^{-1}.
\label{SubCross}%
\end{equation}
For this to become more clear we write down the explicit form of the Leibniz
rules in the case of the two-dimensional quantum plane:%
\begin{align}
\partial^{1}X^{1}  &  =qX^{1}\partial^{1},\nonumber\\
\partial^{1}X^{2}  &  =-q^{-1/2}+q^{2}X^{2}\partial^{1},\\[0.16in]
\partial^{2}X^{1}  &  =q^{1/2}+q^{2}X^{1}\partial^{2}-q^{2}\lambda
X^{2}\partial^{1},\nonumber\\
\partial^{2}X^{2}  &  =qX^{2}\partial^{2}.
\end{align}
The substitutions in (\ref{SubCross}) indeed give the conjugate Leibniz rules,
i.e.%
\begin{align}
\hat{\partial}^{1}X^{1}  &  =q^{-1}X^{1}\hat{\partial}^{1},\nonumber\\
\hat{\partial}^{1}X^{2}  &  =q^{-1/2}+q^{-2}X^{2}\hat{\partial}^{1}%
+q^{-2}\lambda X^{1}\hat{\partial}^{2},\\[0.16in]
\hat{\partial}^{2}X^{1}  &  =-q^{1/2}+q^{-2}X^{1}\hat{\partial}^{2}%
,\nonumber\\
\hat{\partial}^{2}X^{2}  &  =q^{-1}X^{2}\hat{\partial}^{2}.
\end{align}
This correspondence is the deeper reason for the transition rules:%
\begin{align}
\partial^{1}\triangleright f(x^{1},x^{2})  &  =-q^{-1/2}D_{q^{2}}^{2}%
f(qx^{1},x^{2})\nonumber\\
\overset{{%
\genfrac{}{}{0pt}{}{i}{q}%
}{%
\genfrac{}{}{0pt}{}{\rightarrow}{\rightarrow}%
}{%
\genfrac{}{}{0pt}{}{\overline{i}}{1/q}%
}}{\longleftrightarrow}\quad\hat{\partial}^{2}\,\bar{\triangleright}%
\,f(x^{2},x^{1})  &  =-q^{1/2}D_{q^{-2}}^{1}f(q^{-1}x^{2},x^{1}%
),\label{Cros2dim1N}\\[0.16in]
\partial^{2}\triangleright f(x^{1},x^{2})  &  =q^{1/2}D_{q^{2}}^{1}%
f(x^{1},q^{2}x^{2})\nonumber\\
\overset{{%
\genfrac{}{}{0pt}{}{i}{q}%
}{%
\genfrac{}{}{0pt}{}{\rightarrow}{\rightarrow}%
}{%
\genfrac{}{}{0pt}{}{\overline{i}}{1/q}%
}}{\longleftrightarrow}\quad\hat{\partial}^{1}\,\bar{\triangleright}%
\,f(x^{2},x^{1})  &  =q^{-1/2}D_{q^{-2}}^{2}f(x^{2},q^{-2}x^{1}),
\label{Cros2dim2N}%
\end{align}
where the transition symbol has the same meaning as in (\ref{TransConUn}). It
should be noted that the above transition rules require to reverse the
underlying normal ordering.

It is also possible to establish a relationship between right and left actions
of partial derivatives. To this end, let us recall that the identities in
(\ref{RegConAblN}) enable us to derive right representations from left ones
and vice versa. Exploiting this idea one can show that in the case of the
Manin plane we have%
\begin{align}
\partial^{1}\triangleright f(x^{1},x^{2})  &  =-q^{-1/2}D_{q^{2}}^{2}%
f(qx^{1},x^{2})\nonumber\\
\overset{i\leftrightarrow\overline{i}}{\longleftrightarrow}\quad-f(x^{1}%
,x^{2})\,\bar{\triangleleft}\,\partial^{2}  &  =q^{-1/2}D_{q^{2}}^{1}%
f(x^{1},qx^{2}),\\[0.16in]
\partial^{2}\triangleright f(x^{1},x^{2})  &  =q^{1/2}D_{q^{2}}^{1}%
f(x^{1},q^{2}x^{2})\nonumber\\
\overset{i\leftrightarrow\overline{i}}{\longleftrightarrow}\quad-f(x^{1}%
,x^{2})\,\bar{\triangleleft}\,\partial^{1}  &  =-q^{1/2}D_{q^{2}}^{2}%
f(q^{2}x^{1},x^{2}),
\end{align}
where the symbol $\overset{i\leftrightarrow\overline{i}}{\longleftrightarrow}$
stands for the substitutions%
\begin{equation}
D_{q^{a}}^{i}\rightarrow D_{q^{a}}^{\overline{i}},\quad\hat{n}_{x^{i}%
}\rightarrow\hat{n}_{x^{\overline{i}}},\quad x^{i}\rightarrow x^{\overline{i}%
}.
\end{equation}

These reasonings carry over to the other quantum spaces we are considering
without any problem. In general, the correspondences between the different
representations diagrammatically appear as in Fig. \ref{ConBild2}.
\begin{figure}[ptb]
\begin{center}
\setlength{\unitlength}{1.0cm} \begin{picture}(12,8)
\put(6,0.5){\vector(1,0){2.5}}
\put(6,0.5){\vector(-1,0){2.5}}
\put(6,6.5){\vector(1,0){2.5}}
\put(6,6.5){\vector(-1,0){2.5}}
\put(3,3.5){\vector(0,1){2.5}}
\put(3,3.5){\vector(0,-1){2.5}}
\put(9,3.5){\vector(0,1){2.5}}
\put(9,3.5){\vector(0,-1){2.5}}
\put(5.5,3){\vector(-1,-1){2}}
\put(5.5,4){\vector(-1,1){2}}
\put(6.5,3){\vector(1,-1){2}}
\put(6.5,4){\vector(1,1){2}}
\put(5.5,3.25){\makebox(1,0.5){
$ i \leftrightarrow \overline{i} $
}}
\put(0,3.25){\makebox(3,1)[b]{
$ q \leftrightarrow 1/q $
}}
\put(9,3.25){\makebox(3,1)[b]{
$ q \leftrightarrow 1/q $
}}
\put(4,0){\makebox(4,0.5)[b]{
ordering is reversed
}}
\put(4,6){\makebox(4,0.5)[b]{
ordering is reversed
}}
\put(4.5,0.7){\makebox(3,0.5)[b]{
$i \leftrightarrow \overline{i} $ \quad $ q \leftrightarrow 1/q $
}}
\put(4.5,6.7){\makebox(3,0.5)[b]{
$i \leftrightarrow \overline{i} $ \quad $ q \leftrightarrow 1/q $
}}
\put(1.7,0.5){\parbox{2cm}{
\begin{center} $\lhd(-\hat{\partial^i})$ \end{center}
}}
\put(2.0,6.3){\parbox{2cm}{
\begin{center} $\partial^i \rhd$ \end{center}
}}
\put(8.3,0.5){\parbox{2cm}{
\begin{center} $\bar{\lhd}(-\partial^{\overline{i}})$ \end{center}
}}
\put(8.1,6.3){\parbox{2cm}{
\begin{center} $\hat{\partial}^{\overline{i}} \;\bar{\rhd}$ \end{center}
}}
\end{picture}
\end{center}
\caption{Crossing-symmetries for q-deformed partial derivatives}%
\label{ConBild2}%
\end{figure}
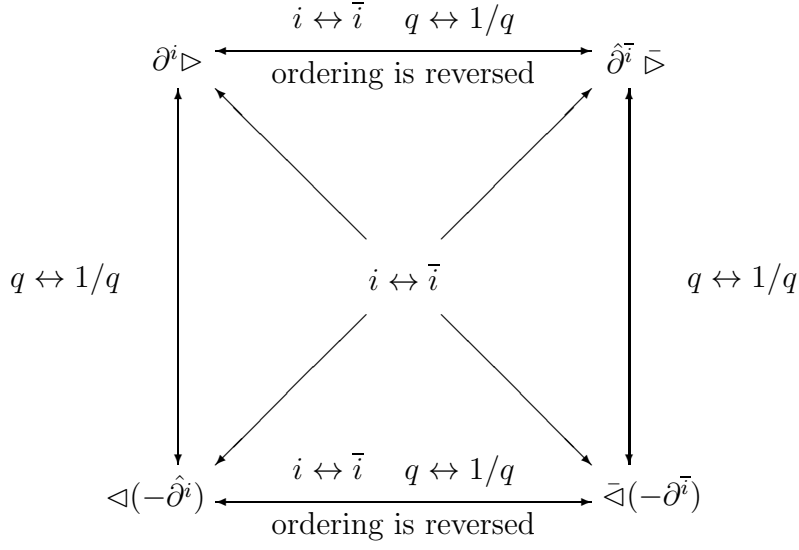For the sake of completeness we want to list the normal orderings
that lead to the simplest expressions for unconjugate left actions of partial derivatives:

\begin{enumerate}
\item[(i)] (quantum plane) $(X^{1})^{n_{1}}(X^{2})^{n_{2}},$

\item[(ii)] (three-dimensional Euclidean space) $(X^{+})^{n_{+}}(X^{3}%
)^{n_{3}}(X^{-})^{n_{-}},$

\item[(iii)] (four-dimensional Euclidean space) $(X^{4})^{n_{4}}(X^{3}%
)^{n_{3}}(X^{2})^{n_{2}}(X^{1})^{n_{1}},$

\item[(iv)] (q-deformed Minkowski space) $(X^{-})^{n_{-}}(X^{3/0})^{n_{3/0}%
}(X^{3})^{n_{3}}(X^{+})^{n_{+}}.$
\end{enumerate}

\subsection{Further correspondences between actions of partial derivatives}

Next, we would like to mention that the operations introduced in
(\ref{DefAntiN}) intertwine the actions of partial derivatives with their
conjugate counterparts, as was pointed out in Ref. \cite{Maj95star}. Using the
explicit form for the actions of partial derivatives and that for the
operations in (\ref{DefAntiN}) one directly verifies that%
\begin{align}
\hat{\partial}^{i}\,\bar{\triangleright}\,\big (f(\ominus_{R}\,x^{j})\big )
&  =-(\partial^{i}\triangleright f)(\ominus_{R}\,x^{j}),\nonumber\\
\partial^{i}\triangleright\big (f(\ominus_{\bar{R}}\,x^{j})\big )  &
=-(\hat{\partial}^{i}\,\bar{\triangleright}\,f)(\ominus_{\bar{R}}%
\,x^{j}),\label{IntAntAnfN}\\[0.16in]
\big (f(\ominus_{\bar{L}}\,x^{j})\big )\triangleleft\hat{\partial}^{i}  &
=-(f\,\bar{\triangleleft}\,\partial^{i})(\ominus_{\bar{L}}\,x^{j}),\nonumber\\
\big (f(\ominus_{L}\,x^{j})\big )\,\bar{\triangleleft}\,\partial^{i}  &
=-(f\triangleleft\hat{\partial}^{i})(\ominus_{L}\,x^{j}). \label{IntAntEndN}%
\end{align}
It should be noticed that in the case of the Manin plane the minus sign on the
right-hand side must be dropped.

From (\ref{IntAntAnfN}) and (\ref{IntAntEndN}) we can derive further relations
which also intertwine the actions of partial derivatives with their conjugate
versions. To clarify this assertion we start from the first identity in
(\ref{IntAntAnfN}) and proceed as follows:%
\begin{align}
&  \hat{\partial}^{i}\,\bar{\triangleright}\,\big (f(\ominus_{R}%
\,x^{j})\big )=-(\partial^{i}\triangleright f)(\ominus_{R}\,x^{j}),\nonumber\\
\Rightarrow\quad &  \hat{\partial}^{i}\,\bar{\triangleright}\,S^{-1}%
(\mathcal{W}(f))=-S^{-1}(\partial^{i}\triangleright\mathcal{W}(f)),\nonumber\\
\Rightarrow\quad &  S\big (\hat{\partial}^{i}\,\bar{\triangleright}%
\,S^{-1}(\mathcal{W}(f))\big )=-\partial^{i}\triangleright\mathcal{W}%
(f),\nonumber\\
\Rightarrow\quad &  S\big (\hat{\partial}^{i}\,\bar{\triangleright
}\,\mathcal{W}(f)\big )=-\partial^{i}\triangleright S(\mathcal{W}%
(f)),\nonumber\\
\Rightarrow\quad &  (\hat{\partial}^{i}\,\bar{\triangleright}\,f)(\ominus
_{L}\,x^{j})=-\partial^{i}\triangleright\big (f(\ominus_{L}\,x^{j})\big ).
\end{align}
The first step uses (\ref{DefAntiN}) and (\ref{DefParAcN}). For the second
step we hit both sides of the second relation by the antipode. For the third
step we substitute $S(\mathcal{W}(f))$ for $\mathcal{W}(f)$ and the final
result is obtained by applying the definitions in (\ref{DefParAcN}). Repeating
the same steps for the other identities in (\ref{IntAntAnfN}) and
(\ref{IntAntEndN}) gives us%
\begin{align}
(\partial^{i}\triangleright f)(\ominus_{\bar{L}}\,x^{j})  &  =-\hat{\partial
}^{i}\,\bar{\triangleright}\,\big (f(\ominus_{\bar{L}}\,x^{j}%
)\big ),\\[0.16in]
(f\triangleleft\hat{\partial}^{i})(\ominus_{\bar{R}}\,x^{j})  &
=-\big (f(\ominus_{\bar{R}}\,x^{j})\big )\,\bar{\triangleleft}\,\partial
^{i},\nonumber\\
(f\,\bar{\triangleleft}\,\partial^{i})(\ominus_{R}\,x^{j})  &
=-\big (f(\ominus_{R}\,x^{j})\big )\triangleleft\hat{\partial}^{i}.
\end{align}

Our last comment in this section concerns the fact that we can switch between
right and left representations of partial derivatives by means of the antipode
and its inverse. Concretely, we have the identities%
\begin{align}
\partial^{i}\triangleright f(x^{k})  &  =f(x^{k})\triangleleft S_{L}%
(\partial^{i})=-\big (f(x^{k})\triangleleft S(\mathcal{L}_{\partial})_{j}%
^{i}\big )\triangleleft\partial^{j}\nonumber\\
&  =-\big ((\mathcal{L}_{\partial})_{j}^{i}\triangleright f(x^{k}%
)\big )\triangleleft\partial^{j},\nonumber\\
\partial^{i}\,\bar{\triangleright}\,f(x^{k})  &  =f(x^{k})\,\bar
{\triangleleft}\,S_{\bar{L}}(\partial^{i})=-\big (f(x^{k})\triangleleft
S(\mathcal{\bar{L}}_{\partial})_{j}^{i}\big )\,\bar{\triangleleft}%
\,\partial^{j}\nonumber\\
&  =-\big ((\mathcal{\bar{L}}_{\partial})_{j}^{i}\triangleright f(x^{k}%
)\big )\,\bar{\triangleleft}\,\partial^{j},\label{LinkRechtDarN}\\[0.16in]
f(x^{k})\triangleleft\partial^{i}  &  =S_{R}(\partial^{i})\triangleright
f(x^{k})=-\partial^{j}\triangleright\big (S^{-1}(\mathcal{L}_{\partial}%
)_{j}^{i}\triangleright f(x^{k})\big )\nonumber\\
&  =-\partial^{j}\triangleright\big (f(x^{k})\triangleleft(\mathcal{L}%
_{\partial})_{j}^{i}\big ),\nonumber\\
f(x^{k})\,\bar{\triangleleft}\,\partial^{i}  &  =S_{\bar{R}}(\partial
^{i})\,\bar{\triangleright}\,f(x^{k})=-\partial^{j}\,\bar{\triangleright
}\,\big (S^{-1}(\mathcal{\bar{L}}_{\partial})_{j}^{i}\triangleright
f(x^{k})\big )\nonumber\\
&  =-\partial^{j}\,\bar{\triangleright}\,\big (f(x^{k})\triangleleft
(\mathcal{\bar{L}}_{\partial})_{j}^{i}\big ), \label{RechtsLinksDarN}%
\end{align}
where we used (\ref{SExplN}) and (\ref{S-1ExplN}).

\section{Dual pairings and q-exponentials\label{SecExp}}

In this section we deal with dual pairings and exponentials for q-deformed
quantum spaces. Let us recall that the comodules of a given quantum group form
the objects of a braided tensor category [cf. Section \ref{Sec1}]. In what
follows we suppose that for each quantum space algebra $\mathcal{A}_{q}$ in
the braided tensor category we have a dual object $\mathcal{A}_{q}^{\ast}$
living in the same category. Our assumption means that in our tensor category
we have dual pairings%
\begin{align}
\left\langle .\,,.\right\rangle  &  :\mathcal{A}_{q}\otimes\mathcal{A}%
_{q}^{\ast}\rightarrow\mathbb{C}\quad\text{with\quad}\big \langle e_{a}%
,f^{b}\big \rangle=\delta_{a}^{b},\nonumber\\
\left\langle .\,,.\right\rangle ^{\prime}  &  :\mathcal{A}_{q}^{\ast}%
\otimes\mathcal{A}_{q}\rightarrow\mathbb{C}\quad\text{with\quad}%
\big \langle f^{b},e_{a}\big \rangle^{\prime}=\delta_{a}^{b},
\end{align}
where $\{e_{a}\}$ is a basis in $\mathcal{A}_{q}$ and $\{f^{b}\}$ a dual basis
in $\mathcal{A}_{q}^{\ast}.$

\subsection{Definition and basic properties of dual pairings}

In Ref. \cite{Maj93-5} it was shown that the algebra of quantum space
coordinates and that of the corresponding partial derivatives are dual to each
other. The dual pairing is given by
\begin{equation}
\left\langle .\,,.\right\rangle :\mathcal{A}_{q}^{\ast}\otimes\mathcal{A}%
_{q}\rightarrow\mathbb{C}\quad\text{with\quad}\big \langle f(\partial
^{i}),g(X^{j})\big \rangle\equiv\varepsilon(f(\partial^{i})\triangleright
g(X^{j})), \label{DefParAb1}%
\end{equation}
or%
\begin{equation}
\left\langle .\,,.\right\rangle ^{\prime}:\mathcal{A}_{q}\otimes
\mathcal{A}_{q}^{\ast}\rightarrow\mathbb{C}\quad\text{with\quad}%
\big \langle f(X^{i}),g(\partial^{j})\big \rangle^{\prime}\equiv
\varepsilon(f(X^{i})\triangleleft g(\partial^{j})). \label{DefParAb2}%
\end{equation}
As usual the above pairings carry over to commutative algebras by means of the
algebra isomorphism $\mathcal{W},$ i.e.%
\begin{equation}
\left\langle .\,,.\right\rangle :\mathcal{A}^{\ast}\otimes\mathcal{A}%
\rightarrow\mathbb{C}\quad\text{with\quad}\big \langle f(\partial^{i}%
),g(x^{j})\big \rangle\equiv\big \langle\mathcal{W}(f(\partial^{i}%
)),\mathcal{W}(g(x^{j}))\big \rangle,
\end{equation}
and likewise%
\begin{equation}
\left\langle .\,,.\right\rangle ^{\prime}:\mathcal{A}\otimes\mathcal{A}^{\ast
}\rightarrow\mathbb{C}\quad\text{with}\quad\big \langle f(x^{i}),g(\partial
^{j})\big \rangle^{\prime}\equiv\big \langle\mathcal{W}(f(x^{i})),\mathcal{W}%
(g(\partial^{j}))\big \rangle^{\prime}.
\end{equation}
Since the partial derivatives can act on coordinates in four different ways,
there are four possibilities for defining a pairing between coordinates and
derivatives. More concretely, we have the pairings%
\begin{align}
\big \langle f(\partial^{i}),g(x^{j})\big \rangle_{L,\bar{R}}  &
\equiv(f(\partial^{i})\triangleright g(x^{j}))|_{x^{j}=0}=(f(\partial
^{i})\,\bar{\triangleleft}\,g(x^{j}))|_{\partial^{i}=0},\nonumber\\
\big \langle f(\hat{\partial}^{i}),g(x^{j})\big \rangle_{\bar{L},R}  &
\equiv(f(\hat{\partial}^{i})\,\bar{\triangleright}\,g(x^{j}))|_{x^{j}%
=0}=(f(\hat{\partial}^{i})\triangleleft g(x^{j}))|_{\partial^{i}%
=0},\label{ExpDualAnfN}\\[0.16in]
\big \langle f(x^{i}),g(\partial^{j})\big \rangle_{L,\bar{R}}  &
\equiv(f(x^{i})\,\bar{\triangleleft}\,g(\partial^{j}))|_{x^{i}=0}%
=(f(x^{i})\triangleright g(\partial^{j}))|_{\partial^{j}=0},\nonumber\\
\big \langle f(x^{i}),g(\hat{\partial}^{j})\big \rangle_{\bar{L},R}  &
\equiv(f(x^{i})\triangleleft g(\hat{\partial}^{j}))|_{x^{i}=0}=(f(x^{i}%
)\,\bar{\triangleright}\,g(\hat{\partial}^{j}))|_{\partial^{j}=0}.
\label{ExplDualEndN}%
\end{align}

Notice that in (\ref{ExpDualAnfN}) and (\ref{ExplDualEndN}) a label is
introduced to characterize the different pairings. Its meaning can be
understood in the following way. The left and right indices refer to the left
and right arguments of the dual parings, respectively. If an argument of a
given dual paring acts on the other argument via the conjugate representation
the index for\ the acting argument is overlined. One should also notice that
each pairing can be calculated in two different ways. This observation shows
us the symmetry between derivatives and coordinates once again. As an example
we give the explicit form of a pairing for the Manin plane. In this case the
first pairing in (\ref{ExpDualAnfN}) reads on normal ordered monomials as
follows:%
\begin{equation}
\big\langle(\partial_{1})^{n_{1}}(\partial_{2})^{n_{2}},(x^{2})^{m_{2}}%
(x^{1})^{m_{1}}\big\rangle_{L,\bar{R}}=\delta_{m_{1},n_{1}}\delta_{m_{2}%
,n_{2}}\,[[n_{1}]]_{q^{2}}!\,[[n_{2}]]_{q^{2}}!. \label{ExplDualn}%
\end{equation}

From their definitions in (\ref{ExpDualAnfN}) and (\ref{ExplDualEndN}) we see
that the different pairings are determined by the actions of derivatives on
coordinates. On the other hand the dual pairings together with the coaddition
on quantum spaces allow us to write down the so-called coregular actions,
which are identical with the actions in (\ref{FundWirkN}) and
(\ref{FundWirk2N}). Concretely, we have%
\begin{align}
\partial^{i}\triangleright f(x^{j})  &  =\big\langle\partial^{i},f_{(\bar
{L},1)}\big\rangle_{L,\bar{R}}\,f_{(\bar{L},2)}=\big (\partial^{i}\overset
{y}{\triangleright}f(y^{k}\oplus_{\bar{L}}x^{j})\big )\big |_{y^{k}%
=0},\nonumber\\
\hat{\partial}^{i}\,\bar{\triangleright}\,f(x^{j})  &  =\big\langle\hat
{\partial}^{i},f_{(L,1)}\big\rangle_{\bar{L},R}\,f_{(L,2)}=\big (\hat
{\partial}^{i}\,\overset{y}{\bar{\triangleright}}\,f(y^{k}\oplus_{L}%
x^{j})\big )\big |_{y^{k}=0},\label{CorWirkAnfN}\\[0.16in]
f(x^{j})\,\bar{\triangleleft}\,\partial^{i}  &  =f_{(R,1)}%
\,\big\langle f_{(R,2)},\partial^{i}\big\rangle_{L,\bar{R}}=\big (f(x^{j}%
\oplus_{R}y^{k})\overset{y}{\,\bar{\triangleleft}}\,\partial^{i}%
\big )\big |_{y^{k}=0},\nonumber\\
f(x^{j})\triangleleft\hat{\partial}^{i}  &  =f_{(\bar{R},1)}%
\,\big\langle f_{(\bar{R},2)},\hat{\partial}^{i}\big\rangle_{\bar{L}%
,R}=\big (f(x^{j}\oplus_{\bar{R}}y^{k})\overset{y}{\triangleleft}\hat
{\partial}^{i}\big )\big |_{y^{k}=0}, \label{CorWirkEndN}%
\end{align}
where
\begin{align}
f_{(L/\bar{L},1)}\otimes f_{(L/\bar{L},2)}  &  \equiv\left(  \mathcal{W}%
_{L}^{-1}\otimes\mathcal{W}_{L}^{-1}\right)  \circ\Delta_{L/\bar{L}%
}(\mathcal{W}(f)),\nonumber\\
f_{(R/\bar{R},1)}\otimes f_{(R/\bar{R},2)}  &  \equiv\left(  \mathcal{W}%
_{R}^{-1}\otimes\mathcal{W}_{R}^{-1}\right)  \circ\Delta_{R/\bar{R}%
}(\mathcal{W}(f)).
\end{align}
Notice that in (\ref{CorWirkAnfN}) and (\ref{CorWirkEndN}) the coordinate on
top of the symbol for the action indicates the tensor factor the partial
derivative acts upon.

There are several properties of the pairings in (\ref{ExpDualAnfN}) and
(\ref{ExplDualEndN}) worth recording here. As can be seen from their
definition the values of the pairings are numbers. Due to this fact they
should behave like scalars, i.e. the pairings obey%
\begin{align}
h\triangleright\big\langle f(\partial^{i}),g(x^{j})\big\rangle_{L,\bar{R}}  &
=\varepsilon(h)\big\langle f(\partial^{i}),g(x^{j})\big\rangle_{L,\bar{R}%
},\nonumber\\
h\triangleright\big\langle f(\hat{\partial}^{i}),g(x^{j})\big\rangle_{\bar
{L},R}  &  =\varepsilon(h)\big\langle f(\hat{\partial}^{i}),g(x^{j}%
)\big\rangle_{\bar{L},R},\\[0.16in]
h\triangleright\big\langle f(x^{i}),g(\partial^{j})\big\rangle_{L,\bar{R}}  &
=\varepsilon(h)\big\langle f(x^{i}),g(\partial^{j})\big\rangle_{L,\bar{R}%
},\nonumber\\
h\triangleright\big\langle f(x^{i}),g(\hat{\partial}^{j})\big\rangle_{\bar
{L},R}  &  =\varepsilon(h)\big\langle f(x^{i}),g(\hat{\partial}^{j}%
)\big\rangle_{\bar{L},R}.
\end{align}
Furthermore, from the relations in (\ref{CorWirkAnfN}) and (\ref{CorWirkEndN})
together with the definition of the dual pairings\ in (\ref{ExpDualAnfN}) and
(\ref{ExplDualEndN}) we find%
\begin{align}
\big\langle\partial^{i}\partial^{j},f(x^{k})\big\rangle_{L,\bar{R}}  &
=\big (\partial^{i}\partial^{j}\triangleright f(x^{k})\big )\big |_{x^{k}%
=0}\nonumber\\
&  =\big (\partial^{i}\triangleright(\partial^{j}\triangleright f(x^{k}%
))\big )\big |_{x^{k}=0}\nonumber\\
&  =\big\langle\partial^{j},f_{(\bar{L},1)}\big\rangle_{L,\bar{R}%
}\,\big\langle\partial^{i},f_{(\bar{L},2)}\big\rangle_{L,\bar{R}}\nonumber\\
&  =\big (\partial^{j}\overset{x}{\triangleright}[\partial^{i}\overset
{y}{\triangleright}f(x^{k}\oplus_{\bar{L}}y^{l})]\big )\big |_{x^{k}=y^{l}%
=0},\nonumber\\
\big\langle\hat{\partial}^{i}\hat{\partial}^{j},f(x^{k})\big\rangle_{\bar
{L},R}  &  =\big\langle\hat{\partial}^{j},f_{(L,1)}\big\rangle_{\bar{L}%
,R}\,\big\langle\hat{\partial}^{i},f_{(L,2)}\big\rangle_{\bar{L},R}\nonumber\\
&  =\big (\hat{\partial}^{j}\,\overset{x}{\bar{\triangleright}}\,[\hat
{\partial}^{i}\,\overset{y}{\bar{\triangleright}}\,f(x^{k}\oplus_{L}%
y^{l})]\big )\big |_{x^{k}=y^{l}=0},\label{AxDual1N}\\[0.16in]
\big\langle f(x^{k}),\partial^{j}\partial^{i}\big\rangle_{L,\bar{R}}  &
=\big\langle f_{(R,1)},\partial^{i}\big\rangle_{L,\bar{R}}%
\,\big\langle f_{(R,2)},\partial^{j}\big\rangle_{L,\bar{R}}\nonumber\\
&  =\big ([f(y^{l}\oplus_{R}x^{k})\,\overset{y}{\bar{\triangleleft}}%
\,\partial^{j}]\,\overset{x}{\bar{\triangleleft}}\,\partial^{i}%
\big )\big |_{x^{k}=y^{l}=0},\nonumber\\
\big\langle f(x^{k}),\hat{\partial}^{j}\hat{\partial}^{i}\big\rangle_{\bar
{L},R}  &  =\big\langle f_{(\bar{R},1)},\hat{\partial}^{i}\big\rangle_{\bar
{L},R}\,\big\langle f_{(\bar{R},2)},\hat{\partial}^{j}\big\rangle_{\bar{L}%
,R}\nonumber\\
&  =\big ([f(y^{l}\oplus_{\bar{R}}x^{k})\overset{y}{\triangleleft}%
\hat{\partial}^{j}]\overset{x}{\triangleleft}\hat{\partial}^{i}%
\big )\big |_{x^{k}=y^{l}=0}. \label{AxDual2N}%
\end{align}
In addition to this, we have
\begin{align}
\left\langle 1,f(x^{i})\right\rangle _{L,\bar{R}}  &  =f(x^{i})|_{x^{i}=0}, &
\left\langle 1,f(x^{i})\right\rangle _{\bar{L},R}  &  =f(x^{i})|_{x^{i}%
=0},\nonumber\\
\left\langle f(x^{i}),1\right\rangle _{L,\bar{R}}  &  =f(x^{i})|_{x^{i}=0}, &
\left\langle f(x^{i}),1\right\rangle _{\bar{L},R}  &  =f(x^{i})|_{x^{i}=0},
\label{NorPaarN}%
\end{align}
which again is a direct consequence of the definition of the dual pairings in
(\ref{ExpDualAnfN}) and (\ref{ExplDualEndN}).

There is another axiom the dual pairing has to fulfill. For this purpose we
perform the following calculation:%
\begin{align}
&  \big\langle f(\ominus_{R}\,\partial^{i}),g(x^{j})\big\rangle_{L,\bar{R}%
}=\big (f(\ominus_{R}\,\partial^{i})\triangleright g(x^{j})\big )\big |_{x=0}%
\nonumber\\
=\, &  \big ([\mathcal{R}_{[1]}^{-1}\triangleright g(x^{j})]\triangleleft
\lbrack(\mathcal{R}_{[2]}^{-1}\triangleright f(\partial^{i}%
)]\big )\big |_{x^{j}=0}\nonumber\\
=\, &  \big ([\mathcal{R}_{[1]}^{-1}\triangleright g(x^{j})]\,\bar
{\triangleright}\,[\mathcal{R}_{[2]}^{-1}\triangleright f(\partial
^{i})]\big )\big |_{\partial^{i}=0}\nonumber\\
=\, &  \big (f(\partial^{i})\,\bar{\triangleleft}\,g(\ominus_{\bar{L}}%
\,x^{j})\big )\big |_{\partial^{i}=0}=\big\langle f(\partial^{i}%
),g(\ominus_{\bar{L}}\,x^{j})\big\rangle_{L,\bar{R}},\label{AntiPaarAnfN}%
\end{align}
where the second and fourth identity hold due to (\ref{LinkRechtDarN}) and
(\ref{RechtsLinksDarN}), respectively. With the same reasonings we can also
show that%
\begin{align}
\big\langle f(\ominus_{\bar{R}}\,\hat{\partial}^{i}),g(x^{j})\big\rangle_{\bar
{L},R} &  =\big\langle f(\hat{\partial}^{i}),g(\ominus_{L}\,x^{j}%
)\big\rangle_{\bar{L},R},\nonumber\\
\big\langle f(\partial^{i}),g(\ominus_{\bar{R}}\,x^{j})\big\rangle_{L,\bar{R}}
&  =\big\langle f(\ominus_{L}\,\partial^{i}),g(x^{j})\big\rangle_{L,\bar{R}%
},\nonumber\\
\big\langle f(\hat{\partial}^{i}),g(\ominus_{R}\,x^{j})\big\rangle_{\bar{L},R}
&  =\big\langle f(\ominus_{\bar{L}}\,\hat{\partial}^{i}),g(x^{j}%
)\big\rangle_{\bar{L},R}.
\end{align}
Interchanging the roles of coordinates and derivatives we additionally obtain
\begin{align}
\big\langle f(\ominus_{R}\,x^{i}),g(\partial^{j})\big\rangle_{L,\bar{R}} &
=\big\langle f(x^{i}),g(\ominus_{\bar{L}}\,\partial^{j})\big\rangle_{L,\bar
{R}},\nonumber\\
\big\langle f(\ominus_{\bar{R}}\,x^{i}),g(\hat{\partial}^{j})\big\rangle_{\bar
{L},R} &  =\big\langle f(x^{i}),g(\ominus_{L}\,\hat{\partial}^{j}%
)\big\rangle_{\bar{L},R},\\[0.16in]
\big\langle f(x^{i}),g(\ominus_{\bar{R}}\,\partial^{j})\big\rangle_{L,\bar{R}}
&  =\big\langle f(\ominus_{L}\,x^{i}),g(\partial^{j})\big\rangle_{L,\bar{R}%
},\nonumber\\
\big\langle f(x^{i}),g(\ominus_{R}\,\hat{\partial}^{j})\big\rangle_{\bar{L},R}
&  =\big\langle f(\ominus_{\bar{L}}\,x^{i}),g(\hat{\partial}^{j}%
)\big\rangle_{\bar{L},R}.\label{AntiPaarEndN}%
\end{align}

\subsection{Definition and basic properties of q-deformed exponentials}

Let us make contact with another very important ingredient of braided tensor
categories, i.e. an exponential. From an abstract point of view an exponential
is nothing other than an object whose dualization is the pairing in
(\ref{DefParAb1}) or (\ref{DefParAb2}). Thus, the exponential is given by the
mapping%
\begin{equation}
\exp:\mathbb{C}\longrightarrow\mathcal{A}_{q}\otimes\mathcal{A}_{q}^{\ast
},\quad\text{with\quad}\exp(\alpha)=\alpha\sum_{a}e_{a}\otimes f^{a},
\label{AbsExp1}%
\end{equation}
or%
\begin{equation}
\exp^{\prime}:\mathbb{C}\longrightarrow\mathcal{A}_{q}^{\ast}\otimes
\mathcal{A}_{q},\quad\text{with\quad}\exp^{\prime}(\alpha)=\alpha\sum_{a}%
f^{a}\otimes e_{a}. \label{AbsExp2}%
\end{equation}
Again, the algebra isomorphism $\mathcal{W}$ enables us to introduce
q-deformed exponentials that live on a tensor product of commutative algebras.
More concretely, this is achieved by the expressions%
\begin{equation}
\exp(x^{i}|\partial^{j})\equiv\sum_{a}\mathcal{W}(e_{a})\otimes\mathcal{W}%
(f^{a}), \label{ExpAll}%
\end{equation}
and%
\begin{equation}
\exp(\partial^{i}|x^{j})\equiv\sum_{a}\mathcal{W}(f^{a})\otimes\mathcal{W}%
(e_{a}). \label{ExpAl2}%
\end{equation}

If we want to derive explicit formulae for q-deformed exponentials it is our
task to determine a basis of the coordinate algebra $\mathcal{A}_{q}$ being
dual to a given one of the derivative algebra $\mathcal{A}_{q}^{\ast}$.
Inserting the elements of the two bases into the formulae of (\ref{ExpAll})
and (\ref{ExpAl2}) will then provide us with explicit expressions for
q-deformed exponentials. It should be obvious that the two bases being dually
paired depend on the choice of the pairing. Thus, each pairing in
(\ref{ExpDualAnfN}) and (\ref{ExplDualEndN}) leads to its own q-exponential:%
\begin{align}
&  \big\langle f(\partial^{i}),g(x^{j})\big\rangle_{L,\bar{R}} &  &
\Rightarrow &  &  \exp(x^{i}|\partial^{j})_{\bar{R},L},\nonumber\\
&  \big\langle f(\hat{\partial}^{i}),g(x^{j})\big\rangle_{\bar{L},R} &  &
\Rightarrow &  &  \exp(x^{i}|\hat{\partial}^{j})_{R,\bar{L}}%
,\label{DefExpAnfN}\\[0.16in]
&  \big\langle f(x^{i}),g(\partial^{j})\big\rangle_{L,\bar{R}} &  &
\Rightarrow &  &  \exp(\partial^{i}|x^{j})_{\bar{R},L},\nonumber\\
&  \big\langle f(x^{i}),g(\hat{\partial}^{j})\big\rangle_{\bar{L},R} &  &
\Rightarrow &  &  \exp(\hat{\partial}^{i}|x^{j})_{R,\bar{L}}.
\end{align}
In Ref. \cite{Wac03} we presented explicit formulae for q-deformed
exponentials. Once again, the results for the two-dimensional quantum plane
shall serve as an example. For the second exponential in (\ref{DefExpAnfN}) we
found the expression%
\begin{equation}
\exp(x^{i}|\hat{\partial}^{j})_{R,\bar{L}}=\sum_{n_{1},n_{2}=0}^{\infty}%
\frac{(x^{1})^{n_{1}}(x^{2})^{n_{2}}\otimes(\hat{\partial}_{2})^{n_{2}}%
(\hat{\partial}_{1})^{n_{1}}}{[[n_{1}]]_{q^{-2}}!\,[[n_{2}]]_{q^{-2}}!}.
\label{ExplExp}%
\end{equation}

Next, we wish to explain, in more detail, what it means that pairings and
exponentials are dual to each other. To this aim, we should realize that the
existence of two bases being dual to each other gives rise to completeness
relations of the form
\begin{equation}
u=\sum_{a}\left\langle u,e_{a}\right\rangle f^{a}\in\mathcal{A}_{q}^{\ast
},\quad y=\sum_{a}e_{a}\left\langle f^{a},y\right\rangle \in\mathcal{A}_{q},
\label{Voll1}%
\end{equation}
and%
\begin{equation}
u=\sum_{a}\left\langle e_{a},u\right\rangle ^{\prime}f^{a}\in\mathcal{A}%
_{q}^{\ast},\quad y=\sum_{a}e_{a}\left\langle y,f^{a}\right\rangle ^{\prime
}\in\mathcal{A}_{q}. \label{Voll2}%
\end{equation}
From our definitions so far, it should be clear that the relations in
(\ref{Voll1}) and (\ref{Voll2}) imply%
\begin{align}
\text{id}  &  =\big (\left\langle .\,,.\right\rangle _{L,\bar{R}}%
\otimes\text{id}\big )\circ\big (\text{id}\otimes\exp(x^{i}|\partial
^{j})_{\bar{R},L}\big ),\nonumber\\
\text{id}  &  =\big (\left\langle .\,,.\right\rangle _{\bar{L},R}%
\otimes\text{id}\big )\circ\big (\text{id}\otimes\exp(x^{i}|\hat{\partial}%
^{j})_{R,\bar{L}}\big ),\label{VollExp1N}\\[0.16in]
\text{id}  &  =\big (\text{id}\otimes\left\langle .\,,.\right\rangle
_{L,\bar{R}}\big )\circ\big (\exp(x^{i}|\partial^{j})_{\bar{R},L}%
\otimes\text{id}\big ),\nonumber\\
\text{id}  &  =\big (\text{id}\otimes\left\langle .\,,.\right\rangle _{\bar
{L},R}\big )\circ\big (\exp(x^{i}|\hat{\partial}^{j})_{R,\bar{L}}%
\otimes\text{id}\big ). \label{VollExp2N}%
\end{align}
Let us note that here and in what follows we limit ourselves to the
exponentials of (\ref{DefExpAnfN}), since the relations for the other types of
q-deformed exponentials are easily obtained by interchanging the roles for
coordinates and derivatives.

Now, we would like to use the properties in (\ref{VollExp1N}) and
(\ref{VollExp2N}) to derive the addition laws for q-exponentials. To this end,
we show that the mapping in (\ref{ExpAll}) has to be subject to%
\begin{align}
(\Delta\otimes\text{id})\exp &  =\exp_{23}\exp_{13}=\sum_{a}e_{a}%
\otimes\Big(\sum_{b}e_{b}\otimes f^{b}\Big)f^{a},\nonumber\\
(\text{id}\otimes\Delta)\exp &  =\exp_{13}\exp_{12}=\sum_{a}e_{a}\Big(\sum
_{b}e_{b}\otimes f^{b}\Big)\otimes f^{a}, \label{CoExpN}%
\end{align}
where
\begin{align}
\exp_{12}(\alpha)  &  \equiv\alpha\sum_{a}e_{a}\otimes f^{a}\otimes
1,\nonumber\\
\exp_{23}(\alpha)  &  \equiv\alpha\sum_{a}1\otimes e_{a}\otimes f^{a}%
,\nonumber\\
\exp_{13}(\alpha)  &  \equiv\alpha\sum_{a}e_{a}\otimes1\otimes f^{a}.
\end{align}
It should be mentioned that the relations in (\ref{CoExpN}) can be viewed as
dualized versions of the identities [cf. (\ref{AxDual1N}) and (\ref{AxDual2N}%
)]%
\begin{align}
\big\langle uv,x\big\rangle  &  =\big\langle u,x_{(2)}%
\big\rangle\big\langle v,x_{(1)}\big\rangle,\nonumber\\
\big\langle u,xy\big\rangle  &  =\big\langle u_{(2)}%
,x\big\rangle\big\langle u_{(1)},y\big\rangle. \label{EigPaarN}%
\end{align}
For this reason the proof of the relations in (\ref{CoExpN}) is mainly based
on the identities in (\ref{EigPaarN}), as can be seen by the following
calculations:%
\begin{align}
&  \exp_{23}\exp_{13}=\sum_{a,b}e_{a}\otimes e_{b}\otimes f^{b}f^{a}%
\nonumber\\
\overset{(\ref{Voll1})}{=}  &  \sum_{a,b,c}e_{a}\otimes e_{b}\otimes
\big\langle f^{b}f^{a},e_{c}\big\rangle f^{c}\nonumber\\
\overset{(\ref{EigPaarN})}{=}  &  \sum_{a,b,c}e_{a}\otimes e_{b}%
\otimes\big\langle f^{b},(e_{c})_{(2)}\big\rangle\big\langle f^{a}%
,(e_{c})_{(1)}\big\rangle f^{c}\nonumber\\
\overset{(\ref{Voll1})}{=}  &  \sum_{c}(e_{c})_{(1)}\otimes(e_{c}%
)_{(2)}\otimes f^{c}=(\Delta\otimes\text{id})\exp,
\end{align}
and%
\begin{align}
&  \exp_{13}\exp_{12}=\sum_{a,b}e_{a}e_{b}\otimes f^{b}\otimes f^{a}%
\nonumber\\
\overset{(\ref{Voll1})}{=}  &  \sum_{a,b,c}e_{c}\big\langle e_{a}e_{b}%
,f^{c}\big\rangle^{\prime}\otimes f^{b}\otimes f^{a}\nonumber\\
\overset{(\ref{EigPaarN})}{=}  &  \sum_{a,b,c}e_{c}\big\langle e_{a}%
,f_{(2)}^{c}\big\rangle^{\prime}\big\langle e_{b},f_{(1)}^{c}%
\big\rangle^{\prime}\otimes f^{b}\otimes f^{a}\nonumber\\
\overset{(\ref{Voll1})}{=}  &  \sum_{c}e_{c}\otimes f_{(1)}^{c}\otimes
f_{(2)}^{c}=(\text{id}\otimes\Delta)\exp.
\end{align}
It should be obvious that the relations in (\ref{CoExpN}) carry over to
q-deformed exponentials on commutative algebras. This way, they can be
reformulated as%
\begin{align}
\exp(x^{i}\oplus_{L}y^{j}|\hat{\partial}^{k})_{R,\bar{L}}  &  =\exp(x^{i}%
|\exp(y^{j}|\hat{\partial}^{k})_{R,\bar{L}}\overset{\partial}{\circledast}%
\hat{\partial}^{l})_{R,\bar{L}},\nonumber\\
\exp(x^{i}\oplus_{\bar{L}}y^{j}|\partial^{k})_{\bar{R},L}  &  =\exp(x^{i}%
|\exp(y^{j}|\partial^{k})_{\bar{R},L}\overset{\partial}{\circledast}%
\partial^{l})_{\bar{R},L},\label{AddExp1N}\\[0.16in]
\exp(x^{k}|\hat{\partial}^{j}\oplus_{\bar{R}}\hat{\partial}^{\prime
i})_{R,\bar{L}}  &  =\exp(x^{l}\overset{x}{\circledast}\exp(x^{k}%
|\hat{\partial}^{j})_{R,\bar{L}}|\hat{\partial}^{\prime i})_{R,\bar{L}%
},\nonumber\\
\exp(x^{k}|\partial^{j}\oplus_{R}\partial^{\prime i})_{\bar{R},L}  &
=\exp(x^{l}\overset{x}{\circledast}\exp(x^{k}|\partial^{j})_{\bar{R}%
,L}|\partial^{\prime i})_{\bar{R},L}. \label{AddExp2N}%
\end{align}

To understand the physical meaning of q-exponentials it is recorded that they
can be regarded as q-analogs of classical plane waves. In other words,
q-exponentials are in some sense eigenfunctions of partial derivatives, since
they obey%
\begin{align}
\partial^{i}\overset{x}{\triangleright}\exp(x^{k}|\partial^{l})_{\bar{R},L}
&  =\exp(x^{k}|\partial^{l})_{\bar{R},L}\overset{\partial}{\circledast
}\partial^{i},\nonumber\\
\hat{\partial}^{i}\,\overset{x}{\bar{\triangleright}}\,\exp(x^{k}%
|\hat{\partial}^{l})_{R,\bar{L}}  &  =\exp(x^{k}|\hat{\partial}^{l}%
)_{R,\bar{L}}\overset{\partial}{\circledast}\hat{\partial}^{i},
\label{EigGl1N}\\[0.16in]
\exp(\partial^{l}|x^{k})_{\bar{R},L}\,\overset{x}{\bar{\triangleleft}%
}\,\partial^{i}  &  =\partial^{i}\overset{\partial}{\circledast}\exp
(\partial^{l}|x^{k})_{\bar{R},L},\nonumber\\
\exp(\hat{\partial}^{l}|x^{k})_{R,\bar{L}}\overset{x}{\triangleleft}%
\hat{\partial}^{i}  &  =\hat{\partial}^{i}\overset{\partial}{\circledast}%
\exp(\hat{\partial}^{l}|x^{k})_{R,\bar{L}},
\end{align}
and%
\begin{align}
\exp(x^{k}|\partial^{l})_{\bar{R},L}\,\overset{\partial}{\bar{\triangleleft}%
}\,x^{i}  &  =x^{i}\overset{x}{\circledast}\exp(x^{k}|\partial^{l})_{\bar
{R},L},\nonumber\\
\exp(x^{k}|\hat{\partial}^{l})_{R,\bar{L}}\overset{\partial}{\triangleleft
}x^{i}  &  =x^{i}\overset{x}{\circledast}\exp(x^{k}|\hat{\partial}%
^{l})_{R,\bar{L}},\label{EigGl2N}\\[0.16in]
x^{i}\overset{\partial}{\triangleright}\exp(\partial^{l}|x^{k})_{\bar{R},L}
&  =\exp(\partial^{l}|x^{k})_{\bar{R},L}\overset{x}{\circledast}%
x^{i},\nonumber\\
x^{i}\,\overset{\partial}{\bar{\triangleright}}\,\exp(\hat{\partial}^{l}%
|x^{k})_{R,\bar{L}}  &  =\exp(\hat{\partial}^{l}|x^{k})_{R,\bar{L}}\overset
{x}{\circledast}x^{i}. \label{EigGl2EndN}%
\end{align}
For a proof of the above identities we need the coregular actions in
(\ref{CorWirkAnfN}) and (\ref{CorWirkEndN}), which more abstractly take the
form%
\begin{align}
\partial^{i}\triangleright f  &  =\left\langle \partial^{i},f_{(1)}%
\right\rangle f_{(2)},\nonumber\\
f\triangleleft\partial^{i}  &  =f_{(2)}\left\langle f_{(1)},\partial
^{i}\right\rangle . \label{CorActN}%
\end{align}
With these identities at hand a straightforward computation yields%
\begin{align}
&  \partial^{i}\triangleright\exp(x^{k}|\partial^{l})=\sum_{a}\big (\partial
^{i}\triangleright\mathcal{W}(e_{a})\big )\otimes\mathcal{W}(f^{a})\nonumber\\
\overset{(\ref{CorActN})}{=}  &  \sum_{a}\big\langle\partial^{i}%
,\mathcal{W}(e_{a})_{(1)}\big\rangle\mathcal{W}(e_{a})_{(2)}\otimes
\mathcal{W}(f^{a})\nonumber\\
\overset{(\ref{CoExpN})}{=}  &  \sum_{a,b}\big\langle\partial^{i}%
,\mathcal{W}(e_{a})\big\rangle\mathcal{W}(e_{b})\otimes\mathcal{W}%
(f^{b})\mathcal{W}(f^{a})\nonumber\\
\overset{(\ref{Voll1})}{=}  &  \sum_{a}\mathcal{W}(e_{a})\otimes
\mathcal{W}(f^{a})\partial^{i}=\exp(x^{k}|\partial^{l})\overset{\partial
}{\circledast}\partial^{i}.
\end{align}
Through a slight modification to these arguments one can prove the
corresponding identities concerning right actions of $x^{i}.$

For the sake of completeness let us note that our q-exponentials are
normalized in such a way that%
\begin{align}
\exp(x^{i}|\partial^{j})_{\bar{R},L}|_{x^{i}=0}  &  =\exp(x^{i}|\partial
^{j})_{\bar{R},L}|_{\partial^{j}=0}=1,\nonumber\\
\exp(x^{i}|\hat{\partial}^{j})_{R,\bar{L}}|_{x^{i}=0}  &  =\exp(x^{i}%
|\hat{\partial}^{j})_{R,\bar{L}}|_{\partial^{j}=0}=1. \label{NorExpN}%
\end{align}
These equalities result from
\begin{align}
\exp(x^{i}|\partial^{j})|_{x^{i}=0}\overset{(\ref{NorPaarN})}{=}  &  \sum
_{a}\big\langle1,\mathcal{W}(e_{a})\big\rangle\otimes\mathcal{W}%
(f^{a})\nonumber\\
\overset{(\ref{Voll1})}{=}  &  \,1\otimes1=1,
\end{align}
and%
\begin{align}
\exp(x^{i}|\partial^{j})|_{x^{i}=0}\overset{(\ref{NorPaarN})}{=}  &  \sum
_{a}\mathcal{W}(e_{a})\otimes\big\langle\mathcal{W}(f^{a}%
),1\big\rangle\nonumber\\
\overset{(\ref{Voll1})}{=}  &  \,1\otimes1=1.
\end{align}

It is well-known that the generators of the quantum algebras describing the
underlying symmetry can be realized within the Weyl-algebras $\mathcal{A}%
_{q}^{\ast}\otimes\mathcal{A}_{q}$ and $\mathcal{A}_{q}\otimes\mathcal{A}%
_{q}^{\ast}$, i.e. they can be represented in terms of coordinates and partial
derivatives \cite{LWW97, Oca96}. Due to this fact, it immediately follows from
(\ref{EigGl1N}) and (\ref{EigGl2N}) that q-deformed exponentials satisfy%
\begin{align}
h\overset{x}{\triangleright}\exp(x^{i}|\partial^{j})_{\bar{R},L}  &
=\exp(x^{i}|\partial^{j})_{\bar{R},L}\overset{\partial}{\triangleleft
}h,\nonumber\\
h\overset{x}{\triangleright}\exp(x^{i}|\hat{\partial}^{j})_{R,\bar{L}}  &
=\exp(x^{i}|\hat{\partial}^{j})_{R,\bar{L}}\overset{\partial}{\triangleleft}h,
\label{TexpN}%
\end{align}
where $h$ stands for an element of the symmetry algebra. With these relations
at hand we can now prove that q-deformed exponentials behave like scalars:%
\begin{align}
&  h\triangleright\big(\exp(x^{i}|\partial^{j})\big)\nonumber\\
=  &  \sum_{a}\big(h_{(1)}\triangleright\mathcal{W}(e_{a})\big)\otimes
\big(h_{(2)}\triangleright\mathcal{W}(f^{a})\big)\nonumber\\
\overset{(\ref{TexpN})}{=}  &  \sum_{a}\mathcal{W}(e_{a})\otimes
\big(h_{(2)}\triangleright(\mathcal{W}(f^{a})\triangleleft h_{(1)}%
)\big)\nonumber\\
=  &  \sum_{a}\mathcal{W}(e_{a})\otimes\big(h_{(2)}S^{-1}(h_{(1)}%
)\triangleright\mathcal{W}(f^{a})\big)\nonumber\\
=  &  \,\varepsilon(h)\sum_{a}\mathcal{W}(e_{a})\otimes\mathcal{W}%
(f^{a})=\varepsilon(h)\exp(x^{i}|\partial^{j}),\label{ScalPropN}\\[0.16in]
&  \big(\exp(x^{i}|\partial^{j})\big)\triangleleft h\nonumber\\
=  &  \sum_{a}\big(\mathcal{W}(e_{a})\triangleleft h_{(2)}\big)\otimes
\big(\mathcal{W}(f^{a})\triangleleft h_{(1)}\big)\nonumber\\
\overset{(\ref{TexpN})}{=}  &  \sum_{a}\big((h_{(1)}\triangleright
\mathcal{W}(e_{a}))\triangleleft h_{(2)}\big)\otimes\mathcal{W}(f^{a}%
)\nonumber\\
=  &  \sum_{a}\big(\mathcal{W}(e_{a})\triangleleft S(h_{(1)})h_{(2)}%
\big)\otimes\mathcal{W}(f^{a})\nonumber\\
=  &  \,\varepsilon(h)\sum_{a}\mathcal{W}(e_{a})\otimes\mathcal{W}%
(f^{a})=\varepsilon(h)\exp(x^{i}|\partial^{j}). \label{ScalProp2N}%
\end{align}

Next, we come to the dualized versions of the relations in (\ref{AntiPaarAnfN}%
)-(\ref{AntiPaarEndN}). They are given by%
\begin{align}
\exp(\ominus_{\bar{L}}\,x^{i}|\partial^{j})_{\bar{R},L}  &  =\exp
(x^{i}|\ominus_{R}\partial^{j})_{\bar{R},L},\nonumber\\
\exp(\ominus_{L}\,x^{i}|\partial^{j})_{R,\bar{L}}  &  =\exp(x^{i}%
|\ominus_{\bar{R}}\partial^{j})_{R,\bar{L}},\label{ExpAnti1N}\\[0.16in]
\exp(\ominus_{\bar{R}}\,x^{i}|\partial^{j})_{\bar{R},L}  &  =\exp
(x^{i}|\ominus_{L}\partial^{j})_{\bar{R},L},\nonumber\\
\exp(\ominus_{R}\,x^{i}|\partial^{j})_{R,\bar{L}}  &  =\exp(x^{i}%
|\ominus_{\bar{L}}\partial^{j})_{R,\bar{L}}. \label{ExpAnti2N}%
\end{align}
To prove them we first perform calculations like the following one:%
\begin{align}
&  \sum_{a}\big\langle g(\partial^{i}),\bar{S}(e_{a})\big\rangle f^{a}%
\nonumber\\
&  =\sum_{a}\big\langle S^{-1}(g(\partial^{i})),e_{a}\big\rangle f^{a}%
\nonumber\\
&  =S^{-1}(g(\partial^{i}))=\sum_{a}\big\langle g(\partial^{i}),e_{a}%
\big\rangle S^{-1}(f^{a}), \label{HerInvNN}%
\end{align}
where $\bar{S}$ and $S^{-1}$ respectively denote conjugate and inverse of the
antipode of the braided Hopf algebra $\mathcal{A}$ [cf. (\ref{UebReg1}) and
(\ref{UebReg2})]. For the first identity in (\ref{HerInvNN}) we applied the
result of (\ref{AntiPaarAnfN}), while for the second and third equality we
made use of (\ref{Voll1}). Realizing that the pairing is non-degenerate we can
finally read off from (\ref{HerInvNN}) that
\begin{equation}
\sum_{a}\bar{S}(e_{a})\otimes f^{a}=\sum_{a}e_{a}\otimes S^{-1}(f^{a}),
\end{equation}
which is equivalent to the first relation in (\ref{ExpAnti1N}). Similar
considerations lead to the other relations in (\ref{ExpAnti1N}) and
(\ref{ExpAnti2N}).

It should also be mentioned that in some sense the exponentials in
(\ref{ExpAnti1N}) are inverse to those introduced in (\ref{DefExpAnfN}), since
they obey the identities
\begin{align}
&  \exp(x^{i}\overset{x}{\circledast}\exp(\ominus_{\bar{L}}\,x^{j}%
|\partial^{k})_{\bar{R},L}\overset{\partial}{\circledast}\partial^{l}%
)_{\bar{R},L}\nonumber\\
\overset{(\ref{AddExp1N})}{=}  &  \exp(x^{i}\oplus_{\bar{L}}(\ominus_{\bar{L}%
}\,x^{j})|\partial^{k})_{\bar{R},L}\nonumber\\
\overset{(\ref{qAddN})}{=}  &  \exp(x^{i}|\partial^{k})_{\bar{R},L}|_{x^{i}%
=0}=1\nonumber\\
\overset{(\ref{qAddN})}{=}  &  \exp((\ominus_{\bar{L}}\,x^{j})\oplus_{\bar{L}%
}x^{i}|\partial^{k})_{\bar{R},L}\nonumber\\
\overset{(\ref{AddExp1N})}{=}  &  \exp((\ominus_{\bar{L}}\,x^{j})\overset
{x}{\circledast}\exp(x^{i}|\partial^{k})_{\bar{R},L}\overset{\partial
}{\circledast}\partial^{l})_{\bar{R},L},
\end{align}
and%
\begin{align}
&  \exp(x^{i}\overset{x}{\circledast}\exp(x^{j}|\ominus_{R}\partial^{k}%
)_{\bar{R},L}\overset{\partial}{\circledast}\partial^{l})_{\bar{R}%
,L}\nonumber\\
\overset{(\ref{AddExp2N})}{=}  &  \exp(x^{i}|(\ominus_{R}\,\partial^{k}%
)\oplus_{R}\partial^{l})_{\bar{R},L}\nonumber\\
\overset{(\ref{qAdd2N})}{=}  &  \exp(x^{i}|\partial^{k})_{\bar{R}%
,L}|_{\partial^{k}=0}=1\nonumber\\
\overset{(\ref{qAdd2N})}{=}  &  \exp(x^{i}|\partial^{k}\oplus_{R}(\ominus
_{R}\,\partial^{l}))_{\bar{R},L}\nonumber\\
\overset{(\ref{AddExp2N})}{=}  &  \exp(x^{i}\overset{x}{\circledast}\exp
(x^{j}|\partial^{k})_{\bar{R},L}\overset{\partial}{\circledast}(\ominus
_{R}\,\partial^{l}))_{\bar{R},L}.
\end{align}

Before proceeding further we would like to examine wether the inverse
exponentials also show the property of being eigenfunctions of partial
derivatives. This is indeed the case, but compared to (\ref{EigGl1N}) in a
somewhat different way, as can be shown by the following calculation:%
\begin{align}
&  \sum_{a}(\partial^{i}\triangleright e_{a})\otimes S^{-1}(f^{a})\nonumber\\
&  =\sum_{a}e_{a}\otimes S^{-1}(f^{a}\partial^{i})\nonumber\\
&  =-\sum_{a}e_{a}\otimes\partial^{j}S^{-1}(f^{a}\triangleleft\mathcal{L}%
_{j}^{i})\nonumber\\
&  =-\sum_{a}(1\otimes\partial^{k})\big ((e_{a}\triangleleft\mathcal{L}%
_{k}^{j})\otimes S^{-1}(f^{a}\triangleleft\mathcal{L}_{j}^{i}%
)\big )\nonumber\\
&  =-(1\otimes\partial^{i})\sum_{a}e_{a}\otimes S^{-1}(f^{a})\nonumber\\
&  =-\sum_{a}(\mathcal{L}_{j}^{i}\triangleright e_{a})\otimes\partial
^{j}S^{-1}(f^{a}). \label{HerEigInN}%
\end{align}
The first equality is nothing else than (\ref{EigGl1N}) and the second
equality uses (\ref{Hopf1EndN}) together with $S^{-1}(\partial^{i}%
)=-\partial^{i}$. For the third equality we commute the partial derivative to
the left of the exponential. The fourth equality results from the fact that
exponentials behave like scalars [cf. (\ref{ScalPropN}) and (\ref{ScalProp2N}%
)]. Thus, exponentials are invariant under the action of $\mathcal{L}_{j}^{i}$
(notice that for the entries of the L-matrix we have $\Delta(\mathcal{L}%
_{j}^{i})=\mathcal{L}_{k}^{i}\otimes\mathcal{L}_{j}^{k}$ and $\varepsilon
(\mathcal{L}_{j}^{i})=\delta_{j}^{i})$. It should be clear that on commutative
algebras the result in (\ref{HerEigInN}) reads as%
\begin{equation}
\partial^{i}\overset{x}{\triangleright}\exp(x^{k}|\ominus_{R}\partial
^{l})_{\bar{R},L}=-\partial^{i}\overset{\partial|x}{\odot}_{L}\exp
(x^{k}|\ominus_{R}\partial^{l})_{\bar{R},L}, \label{InEigGlAnf}%
\end{equation}
where the factors of the braided product are indicated by the variables on top
of the corresponding symbol. Likewise, we have
\begin{equation}
\hat{\partial}^{i}\,\overset{x}{\bar{\triangleright}}\,\exp(x^{k}%
|\ominus_{\bar{R}}\hat{\partial}^{l})_{R,\bar{L}}=-\hat{\partial}^{i}%
\overset{\partial|x}{\odot}_{\bar{L}}\exp(x^{k}|\ominus_{\bar{R}}\hat
{\partial}^{l})_{R,\bar{L}},
\end{equation}
and%
\begin{align}
\exp(x^{k}|\ominus_{R}\partial^{l})_{\bar{R},L}\,\overset{\partial}%
{\bar{\triangleleft}}\,x^{i}  &  =-\exp(x^{k}|\ominus_{R}\partial^{l}%
)_{\bar{R},L}\overset{\partial|x}{\odot}_{\bar{R}}x^{i},\nonumber\\
\exp(x^{k}|\ominus_{\bar{R}}\hat{\partial}^{l})_{R,\bar{L}}\overset{\partial
}{\triangleleft}x^{i}  &  =-\exp(x^{k}|\ominus_{\bar{R}}\hat{\partial}%
^{l})_{R,\bar{L}}\overset{\partial|x}{\odot}_{R}x^{i}. \label{InvEigGlEndN}%
\end{align}

So far it seems that the inverse exponentials give another set of plane-waves,
since the identities in (\ref{InEigGlAnf})-(\ref{InvEigGlEndN}) differ from
those in (\ref{EigGl1N}) and (\ref{EigGl2N}). Remarkably, the inverse
exponentials can be identified with the exponentials introduced in
(\ref{DefExpAnfN}). For this to become clear, we show that the inverse
exponentials satisfy the same relations as the original ones. With the help of
the identities in (\ref{IntAntAnfN}), we get%
\begin{align}
&  \sum_{a}(\partial^{i}\triangleright\bar{S}^{-1}(e_{a}))\otimes
f^{a}\nonumber\\
&  =-\sum_{a}S^{-1}(\hat{\partial}^{i}\,\bar{\triangleright}\,e_{a})\otimes
f^{a}\nonumber\\
&  =-\sum_{a}\bar{S}^{-1}(e_{a})\otimes(f^{a}\hat{\partial}^{i})
\label{IdInvN}%
\end{align}
being tantamount to
\begin{equation}
\partial^{i}\overset{x}{\triangleright}\exp(\ominus_{\bar{R}}\,x^{k}%
|\hat{\partial}^{l})_{R,\bar{L}}=\exp(\ominus_{\bar{R}}\,x^{k}|\hat{\partial
}^{l})_{R,\bar{L}}\overset{\partial}{\circledast}(-\hat{\partial}^{i}).
\label{IdInv0}%
\end{equation}
The point now is that the exponentials in (\ref{DefExpAnfN}) are completely
characterized by their relations in (\ref{EigGl1N}). Thus, comparing the
result of (\ref{IdInvN}) to the relations in (\ref{EigGl1N}) yields
\begin{equation}
\exp(\ominus_{\bar{R}}\,x^{i}|\hat{\partial}^{j})_{R,\bar{L}}=\exp(x^{i}%
|-\hat{\partial}^{j})_{\bar{R},L}\equiv\exp(x^{i}|\partial^{j})_{\bar{R}%
,L}|_{\partial^{j}\rightarrow-\hat{\partial}^{j}}, \label{IdInv1}%
\end{equation}
which by virtue of $f(\ominus_{\bar{L}}(\ominus_{\bar{R}}\,x^{i}))=f(x^{i})$
leads to the equality%
\begin{equation}
\exp(\ominus_{\bar{L}}x^{i}|\partial^{j})_{\bar{R},L}=\exp(x^{i}|-\partial
^{j})_{R,\bar{L}}\equiv\exp(x^{i}|\hat{\partial}^{j})_{R,\bar{L}}%
|_{\hat{\partial}^{j}\rightarrow-\partial^{j}}.
\end{equation}
In the same way we can verify that%
\begin{equation}
\exp(\ominus_{R}x^{i}|\partial^{j})_{\bar{R},L}=\exp(x^{i}|-\partial
^{j})_{R,\bar{L}}\equiv\exp(x^{i}|\hat{\partial}^{j})_{R,\bar{L}}%
|_{\hat{\partial}^{j}\rightarrow-\partial^{j}},
\end{equation}
and%
\begin{equation}
\exp(\ominus_{L}x^{i}|\hat{\partial}^{j})_{R,\bar{L}}=\exp(x^{i}%
|-\hat{\partial}^{j})_{\bar{R},L}\equiv\exp(x^{i}|\partial^{j})_{\bar{R}%
,L}|_{\partial^{j}\rightarrow-\hat{\partial}^{j}}. \label{IdInv2}%
\end{equation}
Notice that for the Manin plane the minus sign in the expressions
(\ref{IdInv0})-(\ref{IdInv2}) must be dropped.

\subsection{q-Deformed analogs of Taylor rules}

Now, we come to a very useful application of our q-deformed exponentials, i.e.
q-deformed Taylor rules. First of all, we should note that our exponentials
generate q-deformed translations in the following sense:%
\begin{align}
\exp(x^{i}|\partial^{j})_{\bar{R},L}\overset{\partial|y}{\triangleright
}g(y^{k})  &  =g(x^{i}\oplus_{\bar{L}}y^{k}),\nonumber\\
\exp(x^{i}|\hat{\partial}^{j})_{R,\bar{L}}\,\overset{\partial|y}%
{\bar{\triangleright}}\,g(y^{k})  &  =g(x^{i}\oplus_{L}y^{k}), \label{q-TayN}%
\end{align}
and%
\begin{align}
g(y^{k})\,\overset{y|\partial}{\bar{\triangleleft}}\,\exp(\partial^{j}%
|x^{i})_{\bar{R},L}  &  =g(y^{k}\oplus_{R}x^{i}),\nonumber\\
g(y^{k})\overset{y|\partial}{\triangleleft}\exp(\hat{\partial}^{j}%
|x^{i})_{R,\bar{L}}  &  =g(y^{k}\oplus_{\bar{R}}x^{i}), \label{q-TayRecN}%
\end{align}
where we have extended our notation in order to indicate the acting object.
The above identities can be proved by straightforward calculations, i.e.%
\begin{align}
\exp(x^{i}|\partial^{j})_{\bar{R},L}\overset{\partial|y}{\triangleright
}g(y^{k})  &  =\sum_{a}\mathcal{W}(e_{a})\otimes\big\langle\mathcal{W}%
(f^{a}),g(y^{k})_{(\bar{L},1)}\big\rangle_{L,\bar{R}}\,g(y^{k})_{(\bar{L}%
,2)}\nonumber\\
&  =\sum_{a}\mathcal{W}(e_{a})\otimes\varepsilon\big (\mathcal{W}(f^{a}%
)\,\bar{\triangleleft}\,g(y^{k})_{(\bar{L},1)}\big )g(y^{k})_{(\bar{L}%
,2)}\nonumber\\
&  =\sum_{a}\big (g(x^{i})_{(\bar{L},1)}\overset{x}{\circledast}%
\mathcal{W}(e_{a})\big )\otimes\varepsilon\left(  \mathcal{W}(f^{a})\right)
g(y^{k})_{(\bar{L},2)}\nonumber\\
&  =g(x^{i})_{(\bar{L},1)}\otimes g(y^{k})_{(\bar{L},2)}=g(x^{i}\oplus
_{\bar{L}}y^{k}),
\end{align}
and%
\begin{align}
g(y^{k})\,\overset{y|\partial}{\bar{\triangleleft}}\,\exp(\partial^{j}%
|x^{i})_{\bar{R},L}  &  =\sum_{a}g(y^{k})_{(R,1)}\big\langle g(y^{k}%
)_{(R,2)},\mathcal{W}(f^{a})\big\rangle_{L,\bar{R}}\otimes\mathcal{W}%
(e_{a})\nonumber\\
&  =\sum_{a}g(y^{k})_{(R,1)}\,\varepsilon\big (g(y^{k})_{(R,2)}\triangleright
\mathcal{W}(f^{a})\big )\otimes\mathcal{W}(e_{a})\nonumber\\
&  =\sum_{a}g(y^{k})_{(R,1)}\,\varepsilon\left(  \mathcal{W}(f^{a})\right)
\otimes\big (\mathcal{W}(e_{a})\circledast g(x^{i})_{(R,2)}\big )\nonumber\\
&  =g(y^{k})_{(R,1)}\otimes g(x^{i})_{(R,2)}=g(y^{k}\oplus_{R}x^{i}).
\end{align}
For the first step we have rewritten the action of the partial derivatives by
(\ref{CorWirkAnfN}) and (\ref{CorWirkEndN}). The second equality is the
definition of the pairing. The third and fourth equality respectively use the
properties in (\ref{EigGl1N}) and the normalization conditions in
(\ref{NorExpN}).

It is very instructive to expand the q-exponentials in (\ref{q-TayN}) and
(\ref{q-TayRecN}) up to terms linear in $x^{i}$. In doing so, we obtain%
\begin{align}
f(x^{i}\oplus_{L}y^{j})  &  =1\otimes f(y^{j})+x^{k}\otimes\hat{\partial}%
_{k}\,\bar{\triangleright}\,f(y^{j})+O(x^{2}),\nonumber\\
f(x^{i}\oplus_{\bar{L}}y^{j})  &  =1\otimes f(y^{j})+x^{k}\otimes\partial
_{k}\triangleright f(y^{j})+O(x^{2}),\\[0.16in]
f(y^{i}\oplus_{\bar{R}}x^{j})  &  =f(y^{i})\otimes1-f(y^{i})\triangleleft
\hat{\partial}^{k}\otimes x_{k}+O(x^{2}),\nonumber\\
f(y^{i}\oplus_{R}x^{j})  &  =f(y^{i})\otimes1-f(y^{i})\,\bar{\triangleleft
}\,\partial^{k}\otimes x_{k}+O(x^{2}).
\end{align}
As one expects we see that the partial derivatives indeed generate
infinitesimal translations. Furthermore, the above relations give a hint to
another possibility for defining the action of partial derivatives. In analogy
to the undeformed situation, we have%
\begin{align}
\partial_{i}\triangleright f(x^{j})  &  =\lim_{y^{k}\rightarrow0}\frac
{f(y^{k}\oplus_{\bar{L}}x^{j})-f(x^{j})}{y^{i}},\nonumber\\
\hat{\partial}_{i}\,\bar{\triangleright}\,f(x^{j})  &  =\lim_{y^{k}%
\rightarrow0}\frac{f(y^{k}\oplus_{L}x^{j})-f(x^{j})}{y^{i}},\\[0.16in]
f(x^{j})\,\bar{\triangleleft}\,\partial^{i}  &  =\lim_{y^{k}\rightarrow0}%
\frac{f(x^{j})-f(x^{j}\oplus_{R}y^{k})}{y_{i}},\nonumber\\
f(x^{j})\triangleleft\hat{\partial}^{i}  &  =\lim_{y^{k}\rightarrow0}%
\frac{f(x^{j})-f(x^{j}\oplus_{\bar{R}}y^{k})}{y_{i}}.
\end{align}

Now, we are in a position to write down q-deformed Taylor rules, which allow
us to determine the value of a function at a certain point from its values at
a different point. From what we have done so far, it is not very difficult to
convince oneself that we have%
\begin{align}
f(y^{i})  &  =f((y^{i}\oplus_{\bar{L}}(\ominus_{\bar{L}}\,x^{j}))\oplus
_{\bar{L}}x^{k})\nonumber\\
&  =\exp(y^{i}\oplus_{\bar{L}}(\ominus_{\bar{L}}\,x^{j})|\partial^{l}%
)_{\bar{R},L}\overset{\partial|x}{\triangleright}f(x^{k}),\nonumber\\[0.16in]
f(y^{i})  &  =f((y^{i}\oplus_{L}(\ominus_{L}\,x^{j}))\oplus_{L}x^{k}%
)\nonumber\\
&  =\exp(y^{i}\oplus_{L}(\ominus_{L}\,x^{j})|\hat{\partial}^{l})_{R,\bar{L}%
}\,\overset{\partial|x}{\bar{\triangleright}}\,f(x^{k}),
\end{align}
and%
\begin{align}
f(y^{i})  &  =f(x^{k}\oplus_{R}((\ominus_{R}\,x^{j})\oplus_{R}y^{i}%
))\nonumber\\
&  =f(x^{k})\overset{x|\partial}{\triangleleft}\exp(\partial^{l}|(\ominus
_{R}\,x^{j})\oplus_{R}y^{i})_{\bar{R},L},\nonumber\\[0.16in]
f(y^{i})  &  =f(x^{k}\oplus_{\bar{R}}((\ominus_{\bar{R}}\,x^{j})\oplus
_{\bar{R}}y^{i}))\nonumber\\
&  =f(x^{k})\,\overset{x|\partial}{\bar{\triangleleft}}\,\exp(\hat{\partial
}^{l}|(\ominus_{\bar{R}}\,x^{j})\oplus_{\bar{R}}y^{i})_{R,\bar{L}}.
\end{align}

\subsection{Conjugation properties and crossing symmetries}

Next, we turn to the conjugation properties of dual pairings and q-deformed
exponentials. We start our considerations with the following calculation:%
\begin{align}
\overline{\big\langle f(\partial^{i}),g(x^{j})\big\rangle}_{L,\bar{R}}  &
=\overline{\varepsilon(f(\partial^{i})\triangleright g(x^{j}))}=\varepsilon
(\overline{f(\partial^{i})\triangleright g(x^{j})})\nonumber\\
&  =\varepsilon(\overline{g(x^{i})}\,\bar{\triangleleft}\overline
{\,f(\partial^{j})})=\big\langle\overline{g(x^{i})},\overline{f(\partial^{j}%
)}\big\rangle_{L,\bar{R}}.
\end{align}
In the same way, we can verify%
\begin{equation}
\overline{\big\langle f(\hat{\partial}^{i}),g(x^{j})\big\rangle}_{\bar{L}%
,R}=\big\langle\overline{g(x^{j})},\overline{f(\hat{\partial}^{i}%
)}\big\rangle_{\bar{L},R},
\end{equation}
and
\begin{align}
\overline{\big\langle f(x^{i}),g(\partial^{j})\big\rangle}_{L,\bar{R}}  &
=\big\langle\overline{g(\partial^{j})},\overline{f(x^{i})}\big\rangle_{L,\bar
{R}},\nonumber\\
\overline{\big\langle f(x^{i}),g(\hat{\partial}^{j})\big\rangle}_{\bar{L},R}
&  =\big\langle\overline{g(\hat{\partial}^{j})},\overline{f(x^{i}%
)}\big\rangle_{\bar{L},R}.
\end{align}

In order to derive the conjugation properties for q-deformed exponentials, let
us recall that q-deformed exponentials are completely characterized by the
equations in (\ref{EigGl1N}) and (\ref{EigGl2N}). Thus, the conjugation
properties of q-deformed exponentials can easily be read off if we compare
these equations with their conjugate versions. For this to become more clear
we proceed as follows:
\begin{align}
&  \overline{\partial^{k}\overset{x}{\triangleright}\exp(x^{i}|\partial
^{j})_{\bar{R},L}}=\overline{\exp(x^{i}|\partial^{j})_{\bar{R},L}%
\overset{\partial}{\circledast}\partial^{k}},\nonumber\\
\Rightarrow\hspace{0.15in}  &  \overline{\exp(x^{i}|\partial^{j})}_{\bar{R}%
,L}\overset{x}{\,\bar{\triangleleft}}\,\partial^{k}=\partial^{k}%
\overset{\partial}{\circledast}\overline{\exp(x^{i}|\partial^{j})}_{\bar{R}%
,L}.
\end{align}
Comparing the last result with
\begin{equation}
\exp(\partial^{j}|x^{i})_{\bar{R},L}\,\overset{x}{\bar{\triangleleft}%
}\,\partial^{k}=\partial^{k}\overset{\partial}{\circledast}\exp(\partial
^{j}|x^{i})_{\bar{R},L}%
\end{equation}
yields the identification%
\begin{equation}
\overline{\exp(x^{i}|\partial^{j})}_{\bar{R},L}=\exp(\partial^{j}|x^{i}%
)_{\bar{R},L}.
\end{equation}
Similar considerations additionally give
\begin{equation}
\overline{\exp(x^{i}|\hat{\partial}^{j})}_{R,\bar{L}}=\exp(\hat{\partial}%
^{j}|x^{i})_{R,\bar{L}},
\end{equation}
and%
\begin{align}
\overline{\exp(\partial^{i}|x^{j})}_{\bar{R},L}  &  =\exp(x^{j}|\partial
^{i})_{\bar{R},L},\nonumber\\
\overline{\exp(\hat{\partial}^{i}|x^{j})}_{R,\bar{L}}  &  =\exp(x^{j}%
|\hat{\partial}^{i})_{R,\bar{L}}.
\end{align}

Now we come to the crossing symmetries for pairings and exponentials. As we
know there are different versions of dual pairings and q-deformed
exponentials. Again, the crossing symmetries are responsible for the fact that
the expressions for the different versions of dual pairings and q-deformed
exponentials can be transformed into each other by applying some simple
substitutions. Fig. \ref{ConBild3} shows the concrete relationship between the
different types of pairings and exponentials. \begin{figure}[ptb]
\begin{center}
\setlength{\unitlength}{1.0cm} \begin{picture}(12,8)
\put(6,0.5){\vector(1,0){2.0}}
\put(6,0.5){\vector(-1,0){2.0}}
\put(6,6.5){\vector(1,0){2.0}}
\put(6,6.5){\vector(-1,0){2.0}}
\put(3,3.5){\vector(0,1){2.0}}
\put(3,3.5){\vector(0,-1){2.0}}
\put(9,3.5){\vector(0,1){2.0}}
\put(9,3.5){\vector(0,-1){2.0}}
\put(0,4.2){\makebox(3,1)[b]{
$ q \leftrightarrow 1/q $
}}
\put(0,3.5){\makebox(3,1)[b]{
$ {\hat{\partial}}_i \leftrightarrow {\partial}_{\overline{i}} $
}}
\put(0.5,2.95){\parbox{2cm}{
\begin{center} {$x^i \leftrightarrow x^{\overline{i}}$} \end{center}
}}
\put(3.5,3.5){\parbox{2cm}{
\begin{center} {ordering \\ is reversed} \end{center}
}}
\put(9,4.2){\makebox(3,1)[b]{
$ q \leftrightarrow 1/q $
}}
\put(9,3.5){\makebox(3,1)[b]{
$ {\hat{\partial}}^i \leftrightarrow {\partial}^{\overline{i}} $
}}
\put(9.5,2.7){\parbox{2cm}{
\begin{center} {$x_i \leftrightarrow x_{\overline{i}}$} \end{center}
}}
\put(6.5,3.5){\parbox{2cm}{
\begin{center} {ordering \\ is reversed} \end{center}
}}
\put(4.5,0.7){\makebox(3,0.5)[b]{
$x^i \leftrightarrow -\partial^i $ \quad $ \partial_i \leftrightarrow x_i $
}}
\put(4.5,6.7){\makebox(3,0.5)[b]{
$x^i \leftrightarrow -{\hat{\partial}}^i $ \quad $ {\hat{\partial}}_i
\leftrightarrow  x_i$
}}
\put(0.8,0.5){\parbox{2cm}{
\begin{center} $\big \langle f(\partial^i),g(x^j) \big \rangle_{L,\bar{R}}$ \\
$\exp (x^i|\partial^j)_{\bar{R},L}$
\end{center}
}}
\put(0.8,6.5){\parbox{2cm}{
\begin{center} $\big \langle f(\hat{\partial}^i),g(x^j) \big \rangle_{\bar{L},R}$ \\
$\exp (x^i|\hat{\partial}^j)_{R,\bar{L}}$
\end{center}
}}
\put(8.2,0.5){\parbox{2cm}{
\begin{center} $\big \langle f(x^i),g(-\partial^j) \big \rangle_{L,\bar{R}}$ \\
$\exp ({\partial}^i|x^j)_{\bar{R},L}$
\end{center}
}}
\put(8.2,6.5){\parbox{2cm}{
\begin{center} $\big \langle f(x^i),g(-\hat{\partial}^j) \big \rangle_{\bar{L},R}$ \\
$\exp (\hat{{\partial}}^i|x^j)_{R,\bar{L}}$
\end{center}
}}
\end{picture}
\end{center}
\caption{Crossing-symmetries for braided products and q-translations}%
\label{ConBild3}%
\end{figure}Again, we illustrate the transition rules by an example referring
to the Manin plane \cite{Wac03}:%
\begin{align}
&  \big\langle(\hat{\partial}_{2})^{n_{2}}(\hat{\partial}_{1})^{n_{1}}%
,(x^{1})^{m_{1}}(x^{2})^{m_{2}}\big\rangle_{\bar{L},R}\nonumber\\
&  =\delta_{m_{1},n_{1}}\delta_{m_{2},n_{2}}\,[[n_{1}]]_{q^{-2}}%
!\,[[n_{2}]]_{q^{-2}}!\nonumber\\
\longleftrightarrow\hspace{0.15in}\,  &  \big\langle(\partial_{1})^{n_{1}%
}(\partial_{2})^{n_{2}},(x^{2})^{m_{2}}(x^{1})^{m_{1}}\big\rangle_{L,\bar{R}%
}\nonumber\\
&  =\delta_{m_{1},n_{1}}\delta_{m_{2},n_{2}}\,[[n_{1}]]_{q^{2}}!\,[[n_{2}%
]]_{q^{2}}!\nonumber\\
\longleftrightarrow\hspace{0.15in}\,  &  \big\langle(x_{1})^{n_{1}}%
(x_{2})^{n_{2}},(-\partial^{2})^{m_{2}}(-\partial^{1})^{m_{1}}%
\big\rangle_{L,\bar{R}}\nonumber\\
&  =\delta_{m_{1},n_{1}}\delta_{m_{2},n_{2}}\,[[n_{1}]]_{q^{2}}!\,[[n_{2}%
]]_{q^{2}}!,
\end{align}
and%
\begin{align}
&  \exp(x^{i}|\hat{\partial}^{j})_{R,\bar{L}}=\sum_{n_{1},n_{2}=0}^{\infty
}\frac{(x^{1})^{n_{1}}(x^{2})^{n_{2}}\otimes(\hat{\partial}_{2})^{n_{2}}%
(\hat{\partial}_{1})^{n_{1}}}{[[n_{1}]]_{q^{-2}}![[n_{2}]]_{q^{-2}}%
!}\nonumber\\
\longleftrightarrow\hspace{0.15in}\,  &  \exp(x^{i}|\partial^{j})_{\bar{R}%
,L}=\sum_{n_{1},n_{2}=0}^{\infty}\frac{(x^{2})^{n_{2}}(x^{1})^{n_{1}}%
\otimes(\partial_{1})^{n_{1}}(\partial_{2})^{n_{2}}}{[[n_{1}]]_{q^{2}}%
![[n_{2}]]_{q^{2}}!}\nonumber\\
\longleftrightarrow\hspace{0.15in}\,  &  \exp(\partial^{i}|x^{j})_{\bar{R}%
,L}=\sum_{n_{1},n_{2}=0}^{\infty}\frac{(-\partial^{2})^{n_{2}}(-\partial
^{1})^{n_{1}}\otimes(x_{1})^{n_{1}}(x_{2})^{n_{2}}}{[[n_{1}]]_{q^{2}}%
![[n_{2}]]_{q^{2}}!}.
\end{align}

We would like to close this section with a comment on q-deformed exponentials
in terms of momentum variables. It should be clear that there is no reason to
prevent us from substituting i$^{-1}p^{k}$ for $\partial^{k}$ in the
expressions for q-deformed exponentials.

\section{Integration on quantum spaces\label{SecInt}}

In this section we introduce a powerful concept of integration on q-deformed
quantum spaces. We start with presenting the basic ideas and illustrate them
by the results for the two-dimensional quantum plane. We shall then explore
the main features of q-deformed integrals. For other concepts of integration
on quantum spaces we refer the reader to Refs. \cite{CSW93, KM94, Chry96,
Sta96}.

\subsection{Inverse partial derivatives and their representations}

Let us recall that integrals can be seen as operations being inverse to
partial derivatives. Thus, we first try to extend the algebra of partial
derivatives by introducing inverse elements. In the case of the
two-dimensional quantum plane the algebra of partial derivatives is now
characterized by the following relations:%
\begin{align}
\partial^{2}\partial^{1}  &  =q^{-1}\partial^{1}\partial^{2}%
,\label{VerExdParAl}\\[0.1in]
\partial^{i}(\partial^{i})^{-1}  &  =(\partial^{i})^{-1}\partial^{i}=1,\quad
i=1,2,\nonumber\\
\partial^{2}(\partial^{1})^{-1}  &  =q(\partial^{1})^{-1}\partial
^{2},\nonumber\\
\partial^{1}(\partial^{2})^{-1}  &  =q(\partial^{2})^{-1}\partial
^{1},\label{ZwVEPAN}\\[0.1in]
(\partial^{2})^{-1}(\partial^{1})^{-1}  &  =q^{-1}(\partial^{1})^{-1}%
(\partial^{2})^{-1}. \label{VerExdEnd}%
\end{align}
It should be mentioned that the commutation relations involving inverse
partial derivatives can be derived most easily by means of the operator
expressions for star products. To understand this more properly we give the
following calculation holding for the Manin plane:%
\begin{align}
(X^{i})^{\pm1}(X^{j})^{\pm1}  &  =\mathcal{W}\big (\mathcal{W}^{-1}%
((X^{i})^{\pm1})\circledast\mathcal{W}^{-1}((X^{j})^{\pm1})\big )\nonumber\\
&  =\mathcal{W}\big ((x^{i})^{\pm1}\circledast(x^{j})^{\pm1}\big )\nonumber\\
&  =\mathcal{W}\big (\big [q^{-\hat{n}_{x^{2}}\hat{n}_{y^{1}}}(x^{i})^{\pm
1}(y^{j})^{\pm1}\big ]_{y^{j}\rightarrow x^{i}}\big )\nonumber\\
&  =q^{-(\pm\delta_{i2})(\pm\delta_{j1})}\mathcal{W}\big ((x^{i})^{\pm1}%
(x^{j})^{\pm1}\big ).
\end{align}

For the sake of completeness we would like to mention that inverse partial
derivatives show well-defined transformation properties. In the case of the
two-dimensional quantum plane, for example, the relations (\ref{VerExdParAl}%
)-(\ref{VerExdEnd}) together with the commutation relations between
$U_{q}(su_{2})$-generators and partial derivatives imply%
\begin{align}
T^{+}(\partial^{1})^{-1}  &  =q^{-1}(\partial^{1})^{-1}T^{+}-q^{-1}%
(\partial^{1})^{-2}\partial^{2},\nonumber\\
T^{+}(\partial^{2})^{-1}  &  =q(\partial^{2})^{-1}T^{+},\\[0.16in]
T^{-}(\partial^{1})^{-1}  &  =q^{-1}(\partial^{1})^{-1}T^{-},\nonumber\\
T^{-}(\partial^{2})^{-1}  &  =q(\partial^{2})^{-1}T^{-}-q^{2}\partial
^{1}(\partial^{2})^{-2},\\[0.16in]
\tau^{1/2}(\partial^{1})^{-1}  &  =q^{-1}(\partial^{1})^{-1}\tau
^{1/2},\nonumber\\
\tau^{1/2}(\partial^{2})^{-1}  &  =q(\partial^{2})^{-1}\tau^{1/2},\\[0.16in]
\Lambda(\partial^{i})^{-1}  &  =q^{-2}(\partial^{i})^{-1}\Lambda,\qquad i=1,2,
\end{align}
where $\Lambda$ denotes a scaling operator.

As a next step we would like to find representations for the inverse partial
derivatives. This can be achieved in the following way. In Ref. \cite{BW01} it
was shown that according to
\begin{equation}
\partial^{i}\triangleright F=\left(  \partial_{\text{cl}}^{i}+\partial
_{\text{cor}}^{i}\right)  F \label{SplitAbl}%
\end{equation}
the representations of our partial derivatives can be divided up into a
classical part and corrections vanishing in the undeformed limit
$q\rightarrow1$. Thus, seeking a solution to the equation
\begin{equation}
\partial^{i}\triangleright F=f \label{DiffEuq}%
\end{equation}
for given $f$ it is reasonable to consider the following expression:
\begin{align}
F  &  =(\partial^{i})^{-1}\triangleright f=\frac{1}{\partial_{\text{cl}}%
^{i}+\partial_{\text{cor}}^{i}}f\nonumber\\
&  =\frac{1}{\partial_{\text{cl}}^{i}\left(  1+(\partial_{\text{cl}}^{i}%
)^{-1}\partial_{\text{cor}}^{i}\right)  }f\nonumber\\
&  =\frac{1}{1+(\partial_{\text{cl}}^{i})^{-1}\partial_{\text{cor}}^{i}}%
\cdot\frac{1}{\partial_{\text{cl}}^{i}}f\nonumber\\
&  =\sum_{k=0}^{\infty}\left(  -1\right)  ^{k}\left[  (\partial_{\text{cl}%
}^{i})^{-1}\partial_{\text{cor}}^{i}\right]  ^{k}(\partial_{\text{cl}}%
^{i})^{-1}f. \label{IntegralE3N}%
\end{align}
To apply this formula, we need to identify the contributions $\partial
_{\text{cl}}^{i}$ and $\partial_{\text{cor}}^{i}$ the representations of our
partial derivatives consist of. In the two-dimensional case, for example, we
can read off from the expressions in (\ref{Par2dimNN}) that%
\begin{gather}
(\partial_{\text{cl}}^{1})f=-q^{-1/2}D_{q^{2}}^{2}f(qx^{1}),\quad
(\partial_{\text{cor}}^{1})f=0,\nonumber\\
(\partial_{\text{cl}}^{2})f=q^{1/2}D_{q^{2}}^{1}f(q^{2}x^{2}),\quad
(\partial_{\text{cor}}^{2})f=0.
\end{gather}
Plugging this into formula (\ref{IntegralE3N}) leaves us with
\begin{align}
(\partial^{1})^{-1}\triangleright f  &  =(\partial_{\text{cl}}^{1}%
)^{-1}\triangleright f=-q^{1/2}(D_{q^{2}}^{2})^{-1}f(q^{-1}x^{1}),\nonumber\\
(\partial^{2})^{-1}\triangleright f  &  =(\partial_{\text{cl}}^{2}%
)^{-1}\triangleright f=q^{-1/2}(D_{q^{2}}^{1})^{-1}f(q^{-2}x^{2}),
\label{Int2dimN}%
\end{align}
where $(D_{q^{a}}^{i})^{-1}$ denotes the \textit{Jackson integral} operator
\cite{Jack27}. For $a>0,$ $q>1,$ and $x^{i}>0,$ the definition of the Jackson
integral is given by%
\begin{align}
(D_{q^{a}}^{i})^{-1}\big |_{0}^{x^{i}}f  &  =-(1-q^{a})\sum_{k=1}^{\infty
}(q^{-ak}x^{i})f(q^{-ak}x^{i}),\nonumber\\
(D_{q^{a}}^{i})^{-1}\big |_{x^{i}}^{\infty}f  &  =-(1-q^{a})\sum_{k=0}%
^{\infty}(q^{ak}x^{i})f(q^{ak}x^{i}),\nonumber\\
(D_{q^{-a}}^{i})^{-1}\big |_{0}^{x^{i}}f  &  =(1-q^{-a})\sum_{k=0}^{\infty
}(q^{-ak}x^{i})f(q^{-ak}x^{i}),\nonumber\\
(D_{q^{-a}}^{i})^{-1}\big |_{x^{i}}^{\infty}f  &  =(1-q^{-a})\sum
_{k=1}^{\infty}(q^{ak}x^{i})f(q^{ak}x^{i}), \label{Jackson1N}%
\end{align}
and likewise for $a>0,$ $q>1,$ and $x^{i}<0,$
\begin{align}
(D_{q^{a}}^{i})^{-1}\big |_{x^{i}}^{0}f  &  =(1-q^{a})\sum_{k=1}^{\infty
}(q^{-ak}x^{i})f(q^{-ak}x^{i}),\nonumber\\
(D_{q^{a}}^{i})^{-1}\big |_{-\infty}^{x^{i}}f  &  =(1-q^{a})\sum_{k=0}%
^{\infty}(q^{ak}x^{i})f(q^{ak}x^{i}),\nonumber\\
(D_{q^{-a}}^{i})^{-1}\big |_{x^{i}}^{0}f  &  =-(1-q^{-a})\sum_{k=0}^{\infty
}(q^{-ak}x^{i})f(q^{-ak}x^{i}),\nonumber\\
(D_{q^{-a}}^{i})^{-1}\big |_{-\infty}^{x^{i}}f  &  =-(1-q^{-a})\sum
_{k=1}^{\infty}(q^{ak}x^{i})f(q^{ak}x^{i}). \label{Jackson2N}%
\end{align}

From the discussion so far it is apparent that on an algebra of commutative
functions the representations of inverse partial derivatives are given by
expressions that depend on Jackson integrals and Jackson derivatives. But this
is not the whole story, since the explicit form of a Jackson integral is
determined by the choice of its limits. Realizing that the expression for
$(\partial_{\text{cor}}^{i})^{-1}$ is mainly given by a Jackson integral we
can introduce definite integrals\ by%
\begin{equation}
(\partial^{i})^{-1}\big |_{b}^{x^{\overline{i}}}\triangleright f=\sum
_{k=0}^{\infty}\left(  -1\right)  ^{k}\big [(\partial_{\text{cl}}^{i}%
)^{-1}\big |_{b}^{x^{\overline{i}}}\partial_{\text{cor}}^{i}\big ]^{k}%
(\partial_{\text{cl}}^{i})^{-1}\big |_{b}^{x^{\overline{i}}}f,
\label{RepInPar}%
\end{equation}
i.e. the lower and upper limits of all Jackson integrals appearing on the left
side of Eq. (\ref{RepInPar}) are respectively given by a constant $b$ and the
variable $x^{\overline{i}}$ (notice that the partial derivative $\partial^{i}$
does not refer to the coordinate $x^{i}$ but $x^{\overline{i}}$).

Now we would like to explore in which sense the operator in (\ref{RepInPar})
is inverse to that in (\ref{SplitAbl}). To this end we perform the following
calculations:%
\begin{align}
&  \partial^{i}\triangleright\big ((\partial^{i})^{-1}\big |_{b}%
^{x^{\overline{i}}}\triangleright f\big )\nonumber\\
=\,  &  \sum_{k=0}^{\infty}\left(  -1\right)  ^{k}(\partial_{\text{cl}}%
^{i})\big [(\partial_{\text{cl}}^{i})^{-1}\big |_{b}^{x^{\overline{i}}%
}\partial_{\text{cor}}^{i}\big ]^{k}\hspace{-0.02in}\left.  (\partial
_{\text{cl}}^{i})^{-1}\right\vert _{b}^{x^{\overline{i}}}f\nonumber\\
\,  &  +\sum_{k=0}^{\infty}\left(  -1\right)  ^{k}(\partial_{\text{cor}}%
^{i})\big [\hspace{-0.02in}\left.  (\partial_{\text{cl}}^{i})^{-1}\right\vert
_{b}^{x^{\overline{i}}}\partial_{\text{cor}}^{i}\big ]^{k}(\partial
_{\text{cl}}^{i})^{-1}\big |_{b}^{x^{\overline{i}}}f\nonumber\\
=\,  &  \sum_{k=0}^{\infty}\left(  -1\right)  ^{k}\big [\partial_{\text{cor}%
}^{i}(\partial_{\text{cl}}^{i})^{-1}\big |_{b}^{x^{\overline{i}}}%
\big ]^{k}f\nonumber\\
\,  &  -\sum_{k=1}^{\infty}\left(  -1\right)  ^{k}\big [\partial_{\text{cor}%
}^{i}(\partial_{\text{cl}}^{i})^{-1}\big |_{b}^{x^{\overline{i}}}%
\big ]^{k}f\nonumber\\
=\,  &  f,
\end{align}
and
\begin{align}
&  \left.  (\partial^{i})^{-1}\right\vert _{b}^{x^{\overline{i}}%
}\triangleright\big (\partial^{i}\triangleright f\big )\nonumber\\
=\,  &  \sum_{k=0}^{\infty}\left(  -1\right)  ^{k}\big [(\partial_{\text{cl}%
}^{i})^{-1}\big |_{b}^{x^{\overline{i}}}\partial_{\text{cor}}^{i}%
\big ]^{k}(\partial_{\text{cl}}^{i})^{-1}\big |_{b}^{x^{\overline{i}}}%
\partial_{\text{cl}}^{i}f\nonumber\\
\,  &  +\sum_{k=0}^{\infty}\left(  -1\right)  ^{k}\big [(\partial_{\text{cl}%
}^{i})^{-1}\big |_{b}^{x^{\overline{i}}}\partial_{\text{cor}}^{i}%
\big ]^{k}(\partial_{\text{cl}}^{i})^{-1}\big |_{b}^{x^{\overline{i}}}%
\partial_{\text{cor}}^{i}f\nonumber\\
=\,  &  \sum_{k=0}^{\infty}\left(  -1\right)  ^{k}\big [(\partial_{\text{cl}%
}^{i})^{-1}\big |_{b}^{x^{\overline{i}}}\partial_{\text{cor}}^{i}%
\big ]^{k}\big (f\big |_{b}^{x^{\overline{i}}}\big )\nonumber\\
\,  &  -\sum_{k=1}^{\infty}\left(  -1\right)  ^{k}\big [(\partial_{\text{cl}%
}^{i})^{-1}\big |_{b}^{x^{\overline{i}}}\partial_{\text{cor}}^{i}%
\big ]^{k}f\nonumber\\
=\,  &  \sum_{k=0}^{\infty}\left(  -1\right)  ^{k}\big [(\partial_{\text{cl}%
}^{i})^{-1}\big |_{b}^{x^{\overline{i}}}\partial_{\text{cor}}^{i}%
\big ]^{k}\big (f-f|_{x^{\overline{i}}=b}\big )\nonumber\\
\,  &  -\sum_{k=1}^{\infty}\left(  -1\right)  ^{k}\big [(\partial_{\text{cl}%
}^{i})^{-1}\big |_{b}^{x^{\overline{i}}}\partial_{\text{cor}}^{i}%
\big ]^{k}f\nonumber\\
=\,  &  f-\sum_{k=0}^{\infty}\left(  -1\right)  ^{k}\big [(\partial
_{\text{cl}}^{i})^{-1}\big |_{b}^{x^{\overline{i}}}\partial_{\text{cor}}%
^{i}\big ]^{k}\big (f|_{x^{\overline{i}}=b}\big ).
\end{align}
In the above derivations we make use of the identities%
\begin{equation}
\left.  \partial_{\text{cl}}^{i}(\partial_{\text{cl}}^{i})^{-1}\right\vert
_{b}^{x^{\overline{i}}}f=f,\quad\left.  (\partial_{\text{cl}}^{i}%
)^{-1}\right\vert _{b}^{x^{\overline{i}}}\partial_{\text{cl}}^{i}%
f=f\big|_{b}^{x^{\overline{i}}},
\end{equation}
which are a direct consequence of the well-known relations%
\begin{align}
\left.  D_{q^{a}}^{i}(D_{q^{a}}^{i})^{-1}\right\vert _{b}^{x^{i}}f  &
=f,\nonumber\\
\left.  (D_{q^{a}}^{i})^{-1}\right\vert _{b}^{x^{i}}D_{q^{a}}^{i}f  &
=f\big|_{b}^{x^{i}}.
\end{align}
We can now summarize that in contradiction to $\partial^{i}(\partial^{i}%
)^{-1}=(\partial^{i})^{-1}\partial^{i}=1$ we\ instead have%
\begin{align}
\partial^{i}\triangleright\big ((\partial^{i})^{-1}\big |_{b}^{x^{\overline
{i}}}\triangleright f\big )  &  =f,\\
(\partial^{i})^{-1}\big |_{b}^{x^{\overline{i}}}\triangleright\big (\partial
^{i}\triangleright f\big )  &  =f-\hat{B}^{i}\big (f|_{x^{\overline{i}}%
=b}\big ), \label{HauDifIntN}%
\end{align}
where, for brevity, we have introduced%
\begin{equation}
\hat{B}^{i}\equiv\sum_{k=0}^{\infty}\left(  -1\right)  ^{k}\big [(\partial
_{\text{cl}}^{i})^{-1}\big |_{b}^{x^{\overline{i}}}\partial_{\text{cor}}%
^{i}\big ]^{k}.
\end{equation}

However, if the lower limit$\ b$ tends to $-\infty$ and if we restrict
attention to functions $f\in\widetilde{\mathcal{A}}$ with
\begin{equation}
\widetilde{\mathcal{A}}\equiv\Big \{f\in\mathcal{A},\text{ }\lim
_{x^{j}\rightarrow-\infty}(x^{j})^{\alpha}f=0,\text{\quad for all }\alpha
\in\mathbb{R}\text{ and }j\Big \}, \label{RandBed}%
\end{equation}
the representations (\ref{RepInPar}) now respect all algebra relations for
inverse partial derivatives [cf. (\ref{ZwVEPAN}) and (\ref{VerExdEnd})]. This
observation motivates us to define the q-deformed integral over the interval
$]-\infty,x^{\overline{i}}\,]$ as the mapping%
\begin{equation}
\int_{-\infty}^{y^{\overline{i}}}d_{L}x^{\overline{i}}:\widetilde{\mathcal{A}%
}\longrightarrow\widetilde{\mathcal{A}},\quad\int_{-\infty}^{y^{\overline{i}}%
}d_{L}x^{\overline{i}}\,f(x^{j})\equiv\lim_{b\rightarrow-\infty}%
\big ((\partial^{i})^{-1}\big |_{b}^{x^{\overline{i}}}\triangleright
f(x^{j})\big ). \label{UnInt}%
\end{equation}
If we want to deal with definite integrals over arbitrary intervals it is
reasonable to generalize the definition of (\ref{UnInt}) by
\begin{gather}
\int_{z^{\overline{i}}}^{y^{\overline{i}}}d_{L}x^{\overline{i}}:\widetilde
{\mathcal{A}}\longrightarrow\widetilde{\mathcal{A}}\otimes\widetilde
{\mathcal{A}},\nonumber\\
\int_{z^{\overline{i}}}^{y^{\overline{i}}}d_{L}x^{\overline{i}}\,f(x^{j}%
)\equiv\Big (\int_{-\infty}^{y^{\overline{i}}}d_{L}x^{\overline{i}}%
\,f(x^{j})\Big )\otimes1-1\otimes\Big (\int_{-\infty}^{z^{\overline{i}}}%
d_{L}x^{\overline{i}}\,f(x^{j})\Big ). \label{GenqIntN}%
\end{gather}

We started our considerations for introducing q-deformed integrals from one
specific representation of partial derivatives. But what we have done so far
applies to the other representations for partial derivatives as well. In this
manner, q-deformed integrals can be placed into four categories:%
\begin{align}
&  \partial^{i}\triangleright f &  &  \Rightarrow &  &  (\partial^{i}%
)^{-1}\big |_{b}^{x^{\overline{i}}}\triangleright f, &  &  \int_{-\infty
}^{y^{\overline{i}}}d_{L}x^{\overline{i}}\,f(x^{j})\equiv\ \lim_{b\rightarrow
-\infty}\big ((\partial^{i})^{-1}\big |_{b}^{x^{\overline{i}}}\triangleright
f\big ),\nonumber\\
&  \hat{\partial}^{i}\,\bar{\triangleright}\,f &  &  \Rightarrow &  &
(\hat{\partial}^{i})^{-1}\big |_{b}^{x^{\overline{i}}}\,\bar{\triangleright
}\,f, &  &  \int_{-\infty}^{y^{\overline{i}}}d_{\bar{L}}x^{\overline{i}%
}\,f(x^{j})\equiv\ \lim_{b\rightarrow-\infty}\big ((\partial^{i}%
)^{-1}\big |_{b}^{x^{\overline{i}}}\,\bar{\triangleright}\,f\big ),\\[0.16in]
&  f\triangleleft\hat{\partial}^{i} &  &  \Rightarrow &  &  f\triangleleft
(\hat{\partial}^{i})^{-1}\big |_{b}^{x^{\overline{i}}}, &  &  \int_{-\infty
}^{y^{\overline{i}}}d_{R}x^{\overline{i}}\,f(x^{j})\equiv\lim_{b\rightarrow
-\infty}\big (f\triangleleft(\hat{\partial}^{i})^{-1}\big |_{b}^{x^{\overline
{i}}}\big ),\nonumber\\
&  f\,\bar{\triangleleft}\,\partial^{i} &  &  \Rightarrow &  &  f\,\bar
{\triangleleft}\,(\partial^{i})^{-1}\big |_{b}^{x^{\overline{i}}}, &  &
\int_{-\infty}^{y^{\overline{i}}}d_{\bar{R}}x^{\overline{i}}\,f(x^{j}%
)\equiv\lim_{b\rightarrow-\infty}\big (f\,\bar{\triangleleft}\,(\partial
^{i})^{-1}\big |_{b}^{x^{\overline{i}}}\big ).
\end{align}
In complete analogy to (\ref{GenqIntN}) we have
\begin{equation}
\int_{z^{\overline{i}}}^{y^{\overline{i}}}d_{\bar{L}}x^{\overline{i}}%
\,f(x^{j})\equiv\Big (\int_{-\infty}^{y^{\overline{i}}}d_{\bar{L}}%
x^{\overline{i}}\,f(x^{j})\Big )\otimes1-1\otimes\Big (\int_{-\infty
}^{z^{\overline{i}}}d_{\bar{L}}x^{\overline{i}}\,f(x^{j})\Big ),
\end{equation}
but in contrast to this we now require for right integrals to hold%
\begin{align}
\int_{z^{\overline{i}}}^{y^{\overline{i}}}d_{R}x^{\overline{i}}\,f(x^{j})  &
\equiv1\otimes\Big (\int_{-\infty}^{y^{\overline{i}}}d_{R}x^{\overline{i}%
}\,f(x^{j})\Big )-\Big (\int_{-\infty}^{z^{\overline{i}}}d_{R}x^{\overline{i}%
}\,f(x^{j})\Big )\otimes1,\nonumber\\
\int_{z^{\overline{i}}}^{y^{\overline{i}}}d_{\bar{R}}x^{\overline{i}}%
\,f(x^{j})  &  \equiv1\otimes\Big (\int_{-\infty}^{y^{\overline{i}}}d_{\bar
{R}}x^{\overline{i}}\,f(x^{j})\Big )-\Big (\int_{-\infty}^{z^{\overline{i}}%
}d_{\bar{R}}x^{\overline{i}}\,f(x^{j})\Big )\otimes1.
\end{align}

It should be clear that the crossing-symmetries for representations of partial
derivatives carry over to the corresponding integrals. For this reason, the
relationship between the expressions for the various types of q-deformed
integrals can be graphed as in Fig. \ref{ConBild4}. \begin{figure}[ptb]
\begin{center}
\setlength{\unitlength}{1.0cm} \begin{picture}(12,8)
\put(6,0.5){\vector(1,0){2.0}}
\put(6,0.5){\vector(-1,0){2.0}}
\put(6,6.5){\vector(1,0){2.0}}
\put(6,6.5){\vector(-1,0){2.0}}
\put(3,3.5){\vector(0,1){2.2}}
\put(3,3.5){\vector(0,-1){2.2}}
\put(9,3.5){\vector(0,1){2.2}}
\put(9,3.5){\vector(0,-1){2.2}}
\put(5.5,3){\vector(-1,-1){1.8}}
\put(5.5,4){\vector(-1,1){1.8}}
\put(6.5,3){\vector(1,-1){1.8}}
\put(6.5,4){\vector(1,1){1.8}}
\put(5.5,3.25){\makebox(1,0.5){
$ i \leftrightarrow \overline{i} $
}}
\put(0,3.25){\makebox(3,1)[b]{
$ q \leftrightarrow 1/q $
}}
\put(9,3.25){\makebox(3,1)[b]{
$ q \leftrightarrow 1/q $
}}
\put(4,0){\makebox(4,0.5)[b]{
ordering is reversed
}}
\put(4,6){\makebox(4,0.5)[b]{
ordering is reversed
}}
\put(4.5,0.7){\makebox(3,0.5)[b]{
$i \leftrightarrow \overline{i} $ \quad $ q \leftrightarrow 1/q $
}}
\put(4.5,6.7){\makebox(3,0.5)[b]{
$i \leftrightarrow \overline{i} $ \quad $ q \leftrightarrow 1/q $
}}
\put(1.7,0.5){\parbox{2cm}{
\begin{center} $\lhd(-\hat{\partial^i})^{-1} \big |^{x^{\overline{i}}}_b$ \end{center}
}}
\put(1.8,6.5){\parbox{2cm}{
\begin{center} $(\partial^i)^{-1} \big |^{x^{\overline{i}}}_b \rhd$ \end{center}
}}
\put(8.3,0.5){\parbox{2cm}{
\begin{center} $\bar{\lhd}(-\partial^{\overline{i}})^{-1} \big |^{x^{i}}_b$ \end{center}
}}
\put(8.3,6.5){\parbox{2cm}{
\begin{center} $(\hat{\partial}^{\overline{i}})^{-1} \big |^{x^{i}}_b \;\bar{\rhd}$ \end{center}
}}
\end{picture}
\end{center}
\caption{Crossing-symmetries for q-deformed integrals}%
\label{ConBild4}%
\end{figure}
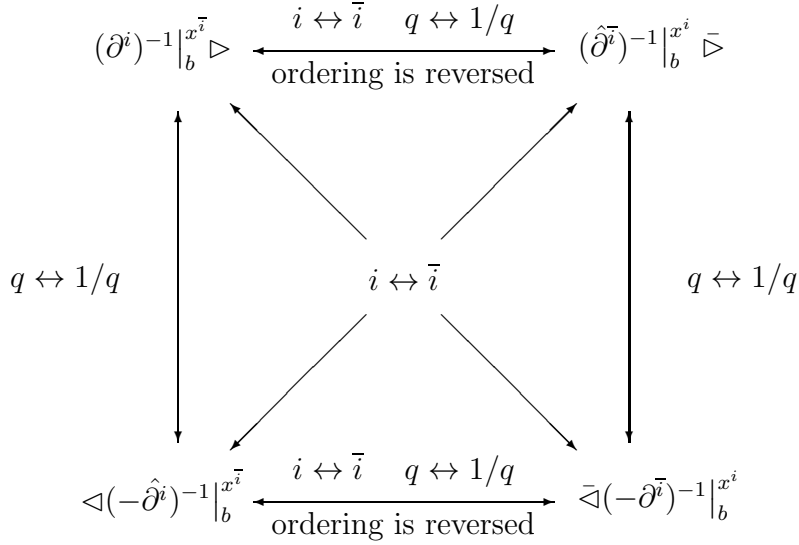Applying these transition rules to the first relation in
(\ref{Int2dimN}), we obtain as example%
\begin{align}
&  (\partial^{1})^{-1}\big |_{b}^{x^{2}}\triangleright f(x^{1},x^{2}%
)=-q^{1/2}(D_{q^{2}}^{2})^{-1}\big |_{b}^{x^{2}}f(q^{-1}x^{1},x^{2}%
)\nonumber\\
\longleftrightarrow\hspace{0.15in}  &  (\hat{\partial}^{2})^{-1}%
\big |_{b}^{x^{1}}\,\bar{\triangleright}\,f(x^{2},x^{1})=-q^{-1/2}(D_{q^{-2}%
}^{1})^{-1}\big |_{b}^{x^{1}}f(qx^{2},x^{1}),\nonumber\\
\longleftrightarrow\hspace{0.15in}  &  f(x^{1},x^{2})\,\bar{\triangleleft
}\,(\partial^{2})^{-1}\big |_{b}^{x^{1}}=-q^{1/2}(D_{q^{2}}^{1})^{-1}%
\big |_{b}^{x^{1}}f(x^{1},q^{-1}x^{2}). \label{ExpTransIntN}%
\end{align}

\subsection{Basic properties of q-deformed integrals}

Next we would like to discuss some important properties of q-deformed
integrals. First of all, the fact that the antipodes for the quantum space
algebra intertwine the actions of partial derivatives [cf. (\ref{IntAntAnfN})
and (\ref{IntAntEndN})] also holds for the representations of inverse partial
derivatives. This can be seen as follows:%
\begin{align}
&  \hat{\partial}^{i}\,\bar{\triangleright}\,\big (f(\ominus_{R}%
\,x^{j})\big )=-\big (\partial^{i}\triangleright f\big )(\ominus_{R}%
\,x^{j})\nonumber\\
\Rightarrow\  &  \hat{\partial}^{i}\,\bar{\triangleright}%
\,\big [\big ((\partial^{i})^{-1}\triangleright f\big )(\ominus_{R}%
\,x^{j})\big ]=-\big (\partial^{i}\triangleright\big ((\partial^{i}%
)^{-1}\triangleright f\big )\big )(\ominus_{R}\,x^{j})\nonumber\\
\Rightarrow\  &  \hat{\partial}^{i}\,\bar{\triangleright}\,\big ((\partial
^{i})^{-1}\triangleright f\big )(\ominus_{R}\,x^{j})=-f(\ominus_{R}%
\,x^{j})\nonumber\\
\Rightarrow\  &  (\hat{\partial}^{i})^{-1}\,\bar{\triangleright}%
\,\big [\hat{\partial}^{i}\,\bar{\triangleright}\,\big [\big ((\partial
^{i})^{-1}\triangleright f\big )(\ominus_{R}\,x^{j})\big ]\big ]=-(\hat
{\partial}^{i})^{-1}\,\bar{\triangleright}\,\big (f(\ominus_{R}\,x^{j}%
)\big )\nonumber\\
\Rightarrow\  &  \big ((\partial^{i})^{-1}\triangleright f\big )(\ominus
_{R}\,x^{j})=-(\hat{\partial}^{i})^{-1}\,\bar{\triangleright}\,\big (f(\ominus
_{R}\,x^{j})\big ),
\end{align}
where again the minus sign has to be dropped in the case of the Manin plane.
In very much the same way, we obtain%
\begin{equation}
\big ((\hat{\partial}^{i})^{-1}\,\bar{\triangleright}\,f\big )(\ominus
_{\bar{R}}\,x^{j})=-(\partial^{i})^{-1}\triangleright\big (f(\ominus_{\bar{R}%
}\,x^{j})\big )
\end{equation}
and%
\begin{align}
\big (f\,\bar{\triangleleft}\,(\partial^{i})^{-1}\big )(\ominus_{\bar{L}%
}\,x^{j})  &  =-\big (f(\ominus_{\bar{L}}\,x^{j})\big )\triangleleft
(\hat{\partial}^{i})^{-1},\nonumber\\
\big (f\triangleleft(\hat{\partial}^{i})^{-1}\big )(\ominus_{L}\,x^{j})  &
=-\big (f(\ominus_{L}\,x^{j})\big )\,\bar{\triangleleft}\,(\partial^{i})^{-1}.
\end{align}

What we are doing next is to consider rules for integration by parts. To this
end, let us first recall that the coproduct for partial derivatives leads to a
kind of q-deformed product rule, since we have%
\begin{align}
\partial^{i}\triangleright(fg)  &  =(\partial^{i}\triangleright
f)g+((\mathcal{L}_{\partial}\mathcal{)}_{j}^{i}\triangleright f)(\partial
^{j}\triangleright g),\nonumber\\
\hat{\partial}^{i}\,\bar{\triangleright}\,(fg)  &  =(\hat{\partial}^{i}%
\,\bar{\triangleright}\,f)g+((\mathcal{\bar{L}}_{\partial}\mathcal{)}_{j}%
^{i}\triangleright f)(\hat{\partial}^{j}\,\bar{\triangleright}\,g),\\[0.14in]
(fg)\triangleleft\hat{\partial}^{i}  &  =f(g\triangleleft\hat{\partial}%
^{i})+(f\triangleleft\hat{\partial}^{j})(g\triangleleft(\mathcal{L}_{\partial
}\mathcal{)}_{j}^{i}),\nonumber\\
(fg)\,\bar{\triangleleft}\,\partial^{i}  &  =f(g\,\bar{\triangleleft
}\,\partial^{i})+(f\,\bar{\triangleleft}\,\partial^{j})(g\triangleleft
(\mathcal{\bar{L}}_{\partial}\mathcal{)}_{j}^{i}).
\end{align}
Hitting the above equations with the corresponding integral operator and
rearranging terms, we get%
\begin{align}
\int_{-\infty}^{y^{\overline{i}}}d_{L}x^{\overline{i}}\,(\partial
^{i}\triangleright f)g  &  =\left.  fg\right\vert _{-\infty}^{y^{\overline{i}%
}}-\int_{-\infty}^{y^{\overline{i}}}d_{L}x^{\overline{i}}\,((\mathcal{L}%
_{\partial}\mathcal{)}_{j}^{i}\triangleright f)(\partial^{j}\triangleright
g),\nonumber\\
\int_{-\infty}^{y^{\overline{i}}}d_{\bar{L}}x^{\overline{i}}\,(\hat{\partial
}^{i}\,\bar{\triangleright}\,f)g  &  =\left.  fg\right\vert _{-\infty
}^{y^{\overline{i}}}-\int_{-\infty}^{y^{\overline{i}}}d_{\bar{L}}%
x^{\overline{i}}\,((\mathcal{\bar{L}}_{\partial}\mathcal{)}_{j}^{i}%
\triangleright f)(\hat{\partial}^{j}\,\bar{\triangleright}%
\,g),\label{ParInt1N}\\[0.16in]
\int_{-\infty}^{y^{\overline{i}}}d_{R}x^{\overline{i}}\,f(g\triangleleft
\hat{\partial}^{i})  &  =\left.  fg\right\vert _{-\infty}^{y^{\overline{i}}%
}-\int_{-\infty}^{y^{\overline{i}}}d_{R}x^{\overline{i}}\,(f\triangleleft
\hat{\partial}^{j})(g\triangleleft(\mathcal{L}_{\partial}\mathcal{)}_{j}%
^{i}),\nonumber\\
\int_{-\infty}^{y^{\overline{i}}}d_{\bar{R}}x^{\overline{i}}\,f(g\,\bar
{\triangleleft}\,\partial^{i})  &  =\left.  fg\right\vert _{-\infty
}^{y^{\overline{i}}}-\int_{-\infty}^{y^{\overline{i}}}d_{\bar{R}}%
x^{\overline{i}}\,(f\,\bar{\triangleleft}\,\partial^{j})(g\triangleleft
(\mathcal{\bar{L}}_{\partial}\mathcal{)}_{j}^{i}). \label{PartInt2N}%
\end{align}
It is not very difficult to convince oneself that the above identities also
hold for integrals over arbitrary intervals. In the case of the
two-dimensional quantum plane the rules in (\ref{ParInt1N}) and
(\ref{PartInt2N}) explicitly read as%
\begin{align}
&  \int_{-\infty}^{y^{2}}d_{L}x^{2}\,(\partial^{1}\triangleright f)\circledast
g=\nonumber\\
&  \qquad\left.  f\circledast g\right\vert _{-\infty}^{y^{2}}-\int_{-\infty
}^{y^{2}}d_{L}x^{2}\,(\Lambda^{-3/4}\tau^{-1/4}\triangleright f)\circledast
(\partial^{1}\triangleright g),\\[0.16in]
&  \int_{-\infty}^{y^{1}}d_{L}x^{1}\,(\partial^{2}\triangleright f)\circledast
g=\nonumber\\
&  \qquad\left.  f\circledast g\right\vert _{-\infty}^{y^{1}}-\int_{-\infty
}^{y^{1}}d_{L}x^{1}\,(\Lambda^{-3/4}\tau^{1/4}\triangleright f)\circledast
(\partial^{2}\triangleright g)\nonumber\\
&  \qquad+q\lambda\int_{-\infty}^{y^{1}}d_{L}x^{1}\,(\Lambda^{-3/4}\tau
^{-1/4}T^{+}\triangleright f)\circledast(\partial^{1}\triangleright
g),\\[0.16in]
&  \int_{-\infty}^{y^{2}}d_{\bar{L}}x^{2}\,(\hat{\partial}^{1}\,\bar
{\triangleright}\,f)\circledast g=\nonumber\\
&  \qquad\left.  f\circledast g\right\vert _{-\infty}^{y^{2}}-\int_{-\infty
}^{y^{2}}d_{\bar{L}}x^{2}\,(\Lambda^{3/4}\tau^{1/4}\triangleright
f)\circledast(\hat{\partial}^{1}\,\bar{\triangleright}\,g)\nonumber\\
&  \qquad-q^{-1}\lambda\int_{-\infty}^{y^{2}}d_{\bar{L}}x^{2}\,(\Lambda
^{3/4}\tau^{-1/4}T^{-}\triangleright f)\circledast(\hat{\partial}^{2}%
\,\bar{\triangleright}\,g),\\[0.16in]
&  \int_{-\infty}^{y^{1}}d_{\bar{L}}x^{1}\,(\hat{\partial}^{2}\,\bar
{\triangleright}\,f)g=\nonumber\\
&  \qquad\left.  f\circledast g\right\vert _{-\infty}^{y^{1}}-\int_{-\infty
}^{y^{1}}d_{\bar{L}}x^{1}\,(\Lambda^{3/4}\tau^{-1/4}\triangleright
f)\circledast(\hat{\partial}^{2}\,\bar{\triangleright}\,g),
\end{align}
and likewise%
\begin{align}
&  \int_{-\infty}^{y^{2}}d_{R}x^{2}\,f\circledast(g\triangleleft\hat{\partial
}^{1})=\nonumber\\
&  \qquad\left.  f\circledast g\right\vert _{-\infty}^{y^{2}}-\int_{-\infty
}^{y^{2}}d_{R}x^{2}\,(f\triangleleft\hat{\partial}^{1})\circledast
(g\triangleleft\Lambda^{-3/4}\tau^{-1/4}),\\[0.16in]
&  \int_{-\infty}^{y^{1}}d_{R}x^{1}\,f\circledast(g\triangleleft\hat{\partial
}^{2})=\nonumber\\
&  \qquad\left.  f\circledast g\right\vert _{-\infty}^{y^{1}}-\int_{-\infty
}^{y^{1}}d_{R}x^{1}\,(f\triangleleft\hat{\partial}^{2})\circledast
(g\triangleleft\Lambda^{-3/4}\tau^{1/4})\nonumber\\
&  \qquad+q\lambda\int_{-\infty}^{y^{1}}d_{R}x^{1}\,(f\triangleleft
\hat{\partial}^{1})\circledast(g\triangleleft\Lambda^{-3/4}\tau^{-1/4}%
T^{+}),\\[0.16in]
&  \int_{-\infty}^{y^{2}}d_{\bar{R}}x^{2}\,f\circledast(g\,\bar{\triangleleft
}\,\partial^{1})=\nonumber\\
&  \qquad\left.  f\circledast g\right\vert _{-\infty}^{y^{2}}-\int_{-\infty
}^{y^{2}}d_{\bar{R}}x^{2}\,(f\,\bar{\triangleleft}\,\partial^{1}%
)\circledast(g\triangleleft\Lambda^{3/4}\tau^{1/4})\nonumber\\
&  \qquad-q^{-1}\lambda\int_{-\infty}^{y^{2}}d_{\bar{R}}x^{2}\,(f\,\bar
{\triangleleft}\,\partial^{2})\circledast(g\triangleleft\Lambda^{3/4}%
\tau^{-1/4}T^{-}),\\[0.16in]
&  \int_{-\infty}^{y^{1}}d_{\bar{R}}x^{1}\,f\circledast(g\,\bar{\triangleleft
}\,\partial^{2})=\nonumber\\
&  \qquad\left.  f\circledast g\right\vert _{-\infty}^{y^{1}}-\int_{-\infty
}^{y^{1}}d_{\bar{R}}x^{1}\,(f\,\bar{\triangleleft}\,\partial^{2}%
)\circledast(g\triangleleft\Lambda^{3/4}\tau^{-1/4}).
\end{align}
One readily checks the above identities by inserting the expressions of
(\ref{LMa2dimN}) and (\ref{LMa2dimKonNN}) into (\ref{ParInt1N}) and
(\ref{PartInt2N}).

Now, we come to another important point concerning translations of integrals.
To this end, we take up the relations in (\ref{EigGl1N})-(\ref{EigGl2EndN})
again and proceed as follows:%
\begin{align}
&  \partial^{i}\overset{x}{\triangleright}\exp(x^{k}|\partial^{l})_{\bar{R}%
,L}=\exp(x^{k}|\partial^{l})_{\bar{R},L}\overset{\partial}{\circledast
}\partial^{i}\nonumber\\[0.09in]
\Rightarrow\hspace{0.15in}  &  \partial^{i}\overset{x}{\triangleright}%
\exp(x^{k}|\partial^{l})_{\bar{R},L}\overset{\partial|y}{\triangleright}%
\int\limits_{-\infty}^{y^{\overline{i}}}d_{L}y^{\overline{i}}\,f(y^{j}%
)=\nonumber\\
&  \qquad\big (\exp(x^{k}|\partial^{l})_{\bar{R},L}\overset{\partial
}{\circledast}\partial^{i}\big )\overset{\partial|y}{\triangleright}%
\int\limits_{-\infty}^{y^{\overline{i}}}d_{L}\tilde{y}^{\overline{i}%
}\,f(\tilde{y}^{j})\nonumber\\[0.09in]
\Rightarrow\hspace{0.15in}  &  \int\limits_{-\infty}^{x^{\overline{i}}}%
d_{L}\tilde{x}^{\overline{i}}\,\tilde{\partial}^{i}\overset{\tilde{\partial
}|\tilde{x}}{\triangleright}\exp(\tilde{x}^{k}|\partial^{l})_{\bar{R}%
,L}\overset{\partial|y}{\triangleright}\int\limits_{-\infty}^{y^{\overline{i}%
}}d_{L}\tilde{y}^{\overline{i}}\,f(\tilde{y}^{j})=\nonumber\\
&  \qquad\int\limits_{-\infty}^{x^{\overline{i}}}d_{L}\tilde{x}^{\overline{i}%
}\,\exp(\tilde{x}^{k}|\partial^{l})_{\bar{R},L}\overset{\partial
|y}{\triangleright}f(y^{j})\nonumber\\[0.09in]
\Rightarrow\hspace{0.15in}  &  \exp(x^{k}|\partial^{l})_{\bar{R},L}%
\overset{\partial|y}{\triangleright}\int\limits_{-\infty}^{y^{\overline{i}}%
}d_{L}\tilde{y}^{\overline{i}}\,f(\tilde{y}^{j})=\int\limits_{-\infty
}^{x^{\overline{i}}}d_{L}\tilde{x}^{\overline{i}}\,\exp(\tilde{x}^{k}%
|\partial^{l})_{\bar{R},L}\overset{\partial|y}{\triangleright}f(y^{j}%
)\nonumber\\[0.09in]
\Rightarrow\hspace{0.15in}  &  \int\limits_{-\infty}^{(x^{k}\oplus_{\bar{L}%
}y^{j})^{\overline{i}}}d_{L}\tilde{y}^{\overline{i}}\,f(\tilde{y}^{j}%
)=\int\limits_{-\infty}^{x^{\overline{i}}}d_{L}\tilde{x}^{\overline{i}%
}\,f(\tilde{x}^{k}\oplus_{\bar{L}}y^{j}). \label{HerAddIntN}%
\end{align}
In the above derivation we suppose $f\in\widetilde{\mathcal{A}}$ and
apply\ (\ref{HauDifIntN}). For the last step we make use of (\ref{q-TayN}).

If we consider integrals over finite intervals, we have to work a little bit
harder. First of all, it is necessary to realize that%
\begin{align}
\partial^{i}\overset{y,z}{\triangleright}\int\limits_{z^{\overline{i}}%
}^{y^{\overline{i}}}d_{L}\tilde{y}^{\overline{i}}\,f(\tilde{y}^{k})  &
=\partial^{i}\overset{y}{\triangleright}\int\limits_{-\infty}^{y^{\overline
{i}}}d_{L}\tilde{y}^{\overline{i}}\,f(\tilde{y}^{k})\otimes1-1\otimes
\partial^{i}\overset{z}{\triangleright}\int\limits_{-\infty}^{z^{\overline{i}%
}}d_{L}\tilde{z}^{\overline{i}}\,f(\tilde{z}^{l})\nonumber\\
&  =f(y^{k})\otimes1-1\otimes f(z^{l}),
\end{align}
and analogously%
\begin{align}
&  \exp(x^{k}|\partial^{l})_{\bar{R},L}\overset{\partial|y,z}{\triangleright
}\int\limits_{z^{\overline{i}}}^{y^{\overline{i}}}d_{L}\tilde{y}^{\overline
{i}}\,f(\tilde{y}^{j})\nonumber\\
=\,  &  \exp(x^{k}|\partial^{l})_{\bar{R},L}\overset{\partial|y}%
{\triangleright}\int\limits_{-\infty}^{y^{\overline{i}}}d_{L}\tilde
{y}^{\overline{i}}\,f(\tilde{y}^{j})\otimes1-1\otimes\exp(x^{k}|\partial
^{l})_{\bar{R},L}\overset{\partial|z}{\triangleright}\int\limits_{-\infty
}^{z^{\overline{i}}}d_{L}\tilde{z}^{\overline{i}}\,f(\bar{z}^{j})\nonumber\\
=\,  &  \int\limits_{-\infty}^{(x^{k}\oplus_{\bar{L}}y^{j})^{\overline{i}}%
}d_{L}\tilde{y}^{\overline{i}}\,f(\tilde{y}^{j})\otimes1-1\otimes
\int\limits_{-\infty}^{(x^{k}\oplus_{\bar{L}}z^{j})^{\overline{i}}}d_{L}%
\tilde{z}^{\overline{i}}\,f(\tilde{z}^{j})\nonumber\\
=\,  &  \int\limits_{(x^{k}\oplus_{\bar{L}}z^{j})^{\overline{i}}}%
^{(x^{k}\oplus_{\bar{L}}y^{j})^{\overline{i}}}d_{L}\tilde{y}^{\overline{i}%
}\,f(\tilde{y}^{j}).
\end{align}
With these identities at hand the considerations in (\ref{HerAddIntN}) now
lead to
\begin{align}
\int\limits_{(x^{k}\oplus_{\bar{L}}z^{j})^{\overline{i}}}^{(x^{k}\oplus
_{\bar{L}}y^{j})^{\overline{i}}}d_{L}\tilde{x}^{\overline{i}}\,f(\tilde{x}%
^{k})  &  =\int\limits_{-\infty}^{x^{\overline{i}}}d_{L}\tilde{x}%
^{\overline{i}}\,f(\tilde{x}^{k}\oplus_{\bar{L}}y^{j})-\int\limits_{-\infty
}^{x^{\overline{i}}}d_{L}\tilde{x}^{\overline{i}}\,f(\tilde{x}^{k}\oplus
_{\bar{L}}z^{j})\nonumber\\
&  =\int\limits_{-\infty}^{(x^{k}\oplus_{\bar{L}}y^{j})^{\overline{i}}}%
d_{L}\tilde{x}^{\overline{i}}\,f(\tilde{x}^{k})-\int\limits_{-\infty}%
^{(x^{k}\oplus_{\bar{L}}z^{j})^{\overline{i}}}d_{L}\tilde{x}^{\overline{i}%
}\,f(\tilde{x}^{k})\nonumber\\
&  =\int\limits_{-\infty}^{(x^{k}\oplus_{\bar{L}}y^{j})^{\overline{i}}}%
d_{L}\tilde{x}^{\overline{i}}f(\tilde{x}^{k})-\int\limits_{-\infty
}^{x^{\overline{i}}}d_{L}\tilde{x}^{\overline{i}}\,f(\tilde{x}^{k})\nonumber\\
&  -\int\limits_{-\infty}^{(x^{k}\oplus_{\bar{L}}z^{j})^{\overline{i}}}%
d_{L}\tilde{x}^{\overline{i}}\,f(\tilde{x}^{k})+\int\limits_{-\infty
}^{x^{\overline{i}}}d_{L}\tilde{x}^{\overline{i}}\,f(\tilde{x}^{k})\nonumber\\
&  =\int\limits_{x^{\overline{i}}}^{(x^{k}\oplus_{\bar{L}}y^{j})^{\overline
{i}}}d_{L}\tilde{x}^{\overline{i}}\,f(\tilde{x}^{k})-\int\limits_{x^{\overline
{i}}}^{(x^{k}\oplus_{\bar{L}}z^{j})^{\overline{i}}}d_{L}\tilde{x}%
^{\overline{i}}\,f(\tilde{x}^{k}),
\end{align}
being equivalent to%
\begin{equation}
\int\limits_{x^{\overline{i}}}^{(x^{k}\oplus_{\bar{L}}y^{j})^{\overline{i}}%
}d_{L}\tilde{x}^{\overline{i}}\,f(\tilde{x}^{k})=\int\limits_{(x^{k}%
\oplus_{\bar{L}}z^{j})^{\overline{i}}}^{(x^{k}\oplus_{\bar{L}}y^{j}%
)^{\overline{i}}}d_{L}\tilde{x}^{\overline{i}}\,f(\tilde{x}^{k})+\int
\limits_{x^{\overline{i}}}^{(x^{k}\oplus_{\bar{L}}z^{j})^{\overline{i}}}%
d_{L}\tilde{x}^{\overline{i}}\,f(\tilde{x}^{k}).
\end{equation}

Everything so far applies equally well to the other types of integrals, i.e.
we additionally have%
\begin{align}
\int\limits_{-\infty}^{(x^{k}\oplus_{L}y^{j})^{\overline{i}}}d_{\bar{L}}%
\tilde{x}^{\overline{i}}\,f(\tilde{x}^{j})  &  =\int\limits_{-\infty
}^{x^{\overline{i}}}d_{\bar{L}}\tilde{x}^{\overline{i}}\,f(\tilde{x}^{k}%
\oplus_{L}y^{j})\nonumber\\
\int\limits_{-\infty}^{(y^{j}\oplus_{\bar{R}}x^{k})^{\overline{i}}}d_{R}%
\tilde{x}^{\overline{i}}\,f(\tilde{x}^{j})  &  =\int\limits_{-\infty
}^{x^{\overline{i}}}d_{R}\tilde{x}^{\overline{i}}\,f(y^{j}\oplus_{\bar{R}%
}\tilde{x}^{k}),\nonumber\\
\int\limits_{-\infty}^{(y^{j}\oplus_{R}x^{k})^{\overline{i}}}d_{\bar{R}}%
\tilde{x}^{\overline{i}}\,f(\tilde{x}^{j})  &  =\int\limits_{-\infty
}^{x^{\overline{i}}}d_{\bar{R}}\tilde{x}^{\overline{i}}\,f(y^{j}\oplus
_{R}\tilde{x}^{k}), \label{TransIntegN}%
\end{align}
and%
\begin{align}
\int\limits_{x^{\overline{i}}}^{(x^{k}\oplus_{L}y^{j})^{\overline{i}}}%
d_{\bar{L}}\tilde{x}^{\overline{i}}\,f(\tilde{x}^{k})  &  =\int\limits_{(x^{k}%
\oplus_{L}z^{j})^{\overline{i}}}^{(x^{k}\oplus_{L}y^{j})^{\overline{i}}%
}d_{\bar{L}}\tilde{x}^{\overline{i}}\,f(\tilde{x}^{k})+\int
\limits_{x^{\overline{i}}}^{(x^{k}\oplus_{L}z^{j})^{\overline{i}}}d_{\bar{L}%
}\tilde{x}^{\overline{i}}\,f(\tilde{x}^{k}),\nonumber\\
\int\limits_{x^{\overline{i}}}^{(y^{j}\oplus_{\bar{R}}x^{k})^{\overline{i}}%
}d_{R}\tilde{x}^{\overline{i}}\,f(\tilde{x}^{k})  &  =\int\limits_{(z^{j}%
\oplus_{\bar{R}}x^{k})^{\overline{i}}}^{(y^{j}\oplus_{\bar{R}}x^{k}%
)^{\overline{i}}}d_{R}\tilde{x}^{\overline{i}}\,f(\tilde{x}^{k})+\int
\limits_{x^{\overline{i}}}^{(z^{j}\oplus_{\bar{R}}x^{k})^{\overline{i}}}%
d_{R}\tilde{x}^{\overline{i}}\,f(\tilde{x}^{k}),\nonumber\\
\int\limits_{x^{\overline{i}}}^{(y^{j}\oplus_{R}x^{k})^{\overline{i}}}%
d_{\bar{R}}\tilde{x}^{\overline{i}}\,f(\tilde{x}^{k})  &  =\int\limits_{(z^{j}%
\oplus_{R}x^{k})^{\overline{i}}}^{(y^{j}\oplus_{R}x^{k})^{\overline{i}}%
}d_{\bar{R}}\tilde{x}^{\overline{i}}\,f(\tilde{x}^{k})+\int
\limits_{x^{\overline{i}}}^{(z^{j}\oplus_{R}x^{k})^{\overline{i}}}d_{\bar{R}%
}\tilde{x}^{\overline{i}}\,f(\tilde{x}^{k}).
\end{align}
For a correct understanding of the relations in (\ref{TransIntegN}) one should
keep in mind that the set $\widetilde{\mathcal{A}}$ [cf. Eq. (\ref{RandBed})]
is invariant under translations, i.e. it holds%
\begin{equation}
f\in\widetilde{\mathcal{A}}\quad\Rightarrow\quad f(x^{i}\oplus_{L/\bar{L}%
}y^{j})\in\widetilde{\mathcal{A}},\quad f(\ominus_{L/\bar{L}}x^{i}%
)\in\widetilde{\mathcal{A}}.
\end{equation}
In other words, translations do not change the boundary conditions a function
shows at infinity. This assertion follows from a direct inspection of the
operator expressions for q-deformed translations (see for example Ref.
\cite{Wac04}).

\subsection{Integrals over the whole space and their translation invariance}

Up to now we have considered one-dimensional integrals only. We are now
looking for integrals over the entire space. In the work of Ref. \cite{Wac02}
such integrals were obtained by applying one-dimensional integrals in
succession, i.e.%
\begin{align}
\int\limits_{-\infty}^{+\infty}d_{L}^{n}x\,f  &  \equiv\int\limits_{-\infty
}^{+\infty}d_{L}x^{\bar{n}}\ldots\int\limits_{-\infty}^{+\infty}d_{L}%
x^{\bar{2}}\int\limits_{-\infty}^{+\infty}d_{L}x^{\bar{1}}\,f(x^{i}%
)\nonumber\\
&  =\lim_{x^{\bar{1}},\ldots,x^{\bar{n}}\rightarrow\infty}\big ((\partial
^{n})^{-1}\big |_{-\infty}^{x^{\bar{n}}}\ldots(\partial^{1})^{-1}%
\big |_{-\infty}^{x^{\bar{1}}}\triangleright f\big ),\nonumber\\[0.1in]
\int\limits_{-\infty}^{+\infty}d_{R}^{n}x\,f  &  \equiv\int\limits_{-\infty
}^{+\infty}d_{R}x^{1}\int\limits_{-\infty}^{+\infty}d_{R}x^{2}\ldots
\int\limits_{-\infty}^{+\infty}d_{R}x^{n}\,f(x^{i})\nonumber\\
&  =\lim_{x^{1},\ldots,x^{n}\rightarrow\infty}\big (f\triangleleft
(\hat{\partial}^{\bar{1}})^{-1}\big |_{-\infty}^{x^{1}}\ldots(\hat{\partial
}^{\bar{n}})^{-1}\big |_{-\infty}^{x^{n}}\big ). \label{VolIntN}%
\end{align}
There seems to be one difficulty with these definitions, since they can depend
on the order in which the one-dimensional integrals are performed. However, in
the above definitions we take for all the lower and upper limits of
integration $-\infty$ and $+\infty$, respectively. For this reason, the
one-dimensional integrals being of the form (\ref{IntegralE3N}) contribute to
the expressions in (\ref{VolIntN}) by their summands with $k=0$, only. This
observation relies on the fact that all summands with $k>0$ lead to surface
terms that can be neglected. This way, the expressions in (\ref{VolIntN})
should simplify to%
\begin{align}
\int\limits_{-\infty}^{+\infty}d_{L}^{n}x\,f(x^{i})  &  =(\partial_{\text{cl}%
}^{n}\left.  )^{-1}\right\vert _{-\infty}^{\infty}\ldots(\partial_{\text{cl}%
}^{1}\left.  )^{-1}\right\vert _{-\infty}^{\infty}\triangleright f,\nonumber\\
\int\limits_{-\infty}^{+\infty}d_{R}^{n}x\,f(x^{i})  &  =f\triangleleft
(\hat{\partial}_{\text{cl}}^{\bar{1}}\left.  )^{-1}\right\vert _{-\infty
}^{\infty}\ldots(\hat{\partial}_{\text{cl}}^{\bar{n}}\left.  )^{-1}\right\vert
_{-\infty}^{\infty}. \label{VerIntVoll}%
\end{align}
The point now is that the integral operators $(\partial_{\text{cl}}^{i})^{-1}%
$, $i=1,\ldots,n$, commute among each other up to certain powers of $q,$ as
can be seen from a direct inspection of their explicit form [cf.
(\ref{Int2dimN})]. Thus, we can conclude that the above definitions do not
really depend on the order of the one-dimensional integrals, since changing
their order leads to expressions that differ from the original ones by a
global factor, only.

Next, we wish to find the explicit form of the integrals over the whole space
for the cases we are interested in for physical reasons. This can be done as
sketched above. First, we identify the expressions for each $\partial
_{\text{cl}}^{i}.$ Taking the corresponding inverse and plugging it into the
formulae of (\ref{VerIntVoll}), we finally obtain

\begin{enumerate}
\item[(i)] (quantum plane)%
\begin{equation}
\int\limits_{-\infty}^{+\infty}d_{L}^{2}x\,f(x^{1},x^{2})\equiv-(D_{q^{2}}%
^{1})^{-1}\big |_{-\infty}^{\infty}q^{-2\hat{n}_{x^{2}}}(D_{q^{2}}^{2}%
)^{-1}\big |_{-\infty}^{\infty}q^{-\hat{n}_{x^{1}}}f, \label{ExpVolAnf}%
\end{equation}

\item[(ii)] (three-dimensional Euclidean space)%
\begin{gather}
\int\limits_{-\infty}^{+\infty}d_{L}^{3}x\,f(x^{+},x^{3},x^{-})\equiv
\nonumber\\
(D_{q^{4}}^{+})^{-1}\big |_{-\infty}^{\infty}(D_{q^{2}}^{3})^{-1}%
\big |_{-\infty}^{\infty}q^{2\hat{n}_{x^{+}}}(D_{q^{4}}^{-})^{-1}%
\big |_{-\infty}^{\infty}q^{2\hat{n}_{x^{3}}}f,
\end{gather}

\item[(iii)] (four-dimensional Euclidean space)%
\begin{gather}
\int\limits_{-\infty}^{+\infty}d_{L}^{4}x\,f(x^{4},x^{3},x^{2},x^{1}%
)\equiv\nonumber\\
(D_{q^{2}}^{1})^{-1}\big |_{-\infty}^{\infty}q^{\hat{n}_{x^{2}}+\,\hat
{n}_{x^{3}}}(D_{q^{2}}^{2})^{-1}q^{\hat{n}_{4}}\big |_{-\infty}^{\infty
}(D_{q^{2}}^{3})^{-1}\big |_{-\infty}^{\infty}q^{\hat{n}_{x^{4}}}(D_{q^{2}%
}^{4})^{-1}\big |_{-\infty}^{\infty}f,
\end{gather}

\item[(iv)] (q-deformed Minkowski space)%
\begin{gather}
\int\limits_{-\infty}^{+\infty}d_{L}^{4}x\,f(r^{2},x^{-},x^{3/0},x^{+}%
)\equiv\nonumber\\
-q\lambda_{+}^{-3}(D_{q^{-2}}^{r^{2}})^{-1}\big |_{-\infty}^{\infty}%
q^{2(\hat{n}_{x^{+}}+\,\hat{n}_{x^{3/0}}+\,\hat{n}_{x^{-}})}(D_{q^{-2}}%
^{+})^{-1}\big |_{-\infty}^{\infty}\nonumber\\
\times(D_{q^{-2}}^{3/0})^{-1}\big |_{-\infty}^{\infty}q^{2\hat{n}_{r^{2}}%
}(D_{q^{-2}}^{-})^{-1}\big |_{-\infty}^{\infty}q^{2\hat{n}_{r^{2}}}f.
\label{ExpVolEndeN}%
\end{gather}

\end{enumerate}

\noindent Notice that we have added Appendix \ref{MinRech} to explain in more
detail the reasonings leading to the expression in (\ref{ExpVolEndeN}).

Obviously, the above considerations apply to all types of one-dimensional
integrals without any difficulties. As we know, it suffices to concern
ourselves with one type of integrals, since the expressions for the
other\ versions are obtained most easily by the crossing symmetries graphed in
Fig. \ref{ConBild4}. Thus, we concretely have the correspondences%
\begin{align}
&  \int\limits_{-\infty}^{+\infty}d_{L}^{n}x\,f(x^{i})=(\partial^{n}%
)^{-1}\big |_{-\infty}^{\infty}\ldots(\partial^{1})^{-1}\big |_{-\infty
}^{\infty}\triangleright f\nonumber\\
\overset{%
\genfrac{}{}{0pt}{}{i\leftrightarrow\overline{i}}{q\leftrightarrow1/q}%
}{\longleftrightarrow}\hspace{0.15in}  &  \int\limits_{-\infty}^{+\infty
}d_{\bar{L}}^{n}x\,\tilde{f}(x^{i})=(\hat{\partial}^{\bar{n}})^{-1}%
\big |_{-\infty}^{\infty}\ldots(\hat{\partial}^{\bar{1}})^{-1}\big |_{-\infty
}^{\infty}\,\bar{\triangleright}\,\tilde{f},\label{CrosSymInt1N}\\[0.16in]
&  \int\limits_{-\infty}^{+\infty}d_{\bar{R}}^{n}x\,f(x^{i})=f\,\bar
{\triangleleft}\,(\partial^{1})^{-1}\big |_{-\infty}^{\infty}\ldots
(\partial^{n})^{-1}\big |_{-\infty}^{\infty}\nonumber\\
\overset{%
\genfrac{}{}{0pt}{}{i\leftrightarrow\overline{i}}{q\leftrightarrow1/q}%
}{\longleftrightarrow}\hspace{0.15in}  &  \int\limits_{-\infty}^{+\infty}%
d_{R}^{n}x\,\tilde{f}(x^{i})=\tilde{f}\triangleleft(\hat{\partial}^{\bar{1}%
})^{-1}\big |_{-\infty}^{\infty}\ldots(\hat{\partial}^{\bar{n}})^{-1}%
\big |_{-\infty}^{\infty},
\end{align}
and%
\begin{align}
\int\limits_{-\infty}^{+\infty}d_{R}^{n}x\,\tilde{f}(x^{i})  &  \overset
{i\leftrightarrow\overline{i}}{\longleftrightarrow}(-1)^{n}\int
\limits_{-\infty}^{+\infty}d_{\bar{L}}^{n}x\,\tilde{f}(x^{i}%
),\label{TransRulIn}\\
\int\limits_{-\infty}^{+\infty}d_{\bar{R}}^{n}x\,f(x^{i})  &  \overset
{i\leftrightarrow\overline{i}}{\longleftrightarrow}(-1)^{n}\int
\limits_{-\infty}^{+\infty}d_{L}^{n}x\,f(x^{i}),\nonumber
\end{align}
where $\overset{%
\genfrac{}{}{0pt}{}{i\leftrightarrow\overline{i}}{q\leftrightarrow1/q}%
}{\longleftrightarrow}$ and $\overset{i\leftrightarrow\overline{i}%
}{\longleftrightarrow}$ respectively indicate transitions via the
substitutions
\begin{equation}
(D_{q^{a}}^{i})^{-1}\rightarrow(D_{q^{-a}}^{\overline{i}})^{-1},\quad\hat
{n}_{x^{i}}\rightarrow\hat{n}_{x^{\overline{i}}},\quad x^{i}\rightarrow
x^{\overline{i}},\quad q\rightarrow q^{-1},
\end{equation}
and%
\begin{equation}
(D_{q^{a}}^{i})^{-1}\rightarrow(D_{q^{a}}^{\overline{i}})^{-1},\quad\hat
{n}_{x^{i}}\rightarrow\hat{n}_{x^{\overline{i}}},\quad x^{i}\rightarrow
x^{\overline{i}}.
\end{equation}
Furthermore, one should notice that the tilde on the symbol for the function
shall remind us of the fact that the corresponding expression refers to
reversed normal ordering. We would like to illustrate the transition rules in
(\ref{CrosSymInt1N})-(\ref{TransRulIn}) by the following example:%
\begin{align}
&  \int\limits_{-\infty}^{+\infty}d_{L}^{2}x\,f(x^{1},x^{2})\equiv-(D_{q^{2}%
}^{1})^{-1}\big |_{-\infty}^{\infty}q^{-2\hat{n}_{x^{2}}}(D_{q^{2}}^{2}%
)^{-1}\big |_{-\infty}^{\infty}q^{-\hat{n}_{x^{1}}}f\nonumber\\
\overset{%
\genfrac{}{}{0pt}{}{i\leftrightarrow\overline{i}}{q\leftrightarrow1/q}%
}{\longleftrightarrow}\hspace{0.15in}  &  \int\limits_{-\infty}^{+\infty
}d_{\bar{L}}^{2}x\,f(x^{2},x^{1})=-(D_{q^{-2}}^{2})^{-1}\big |_{-\infty
}^{\infty}q^{2\hat{n}_{x^{1}}}(D_{q^{-2}}^{1})^{-1}\big |_{-\infty}^{\infty
}q^{\hat{n}_{x^{2}}}f\nonumber\\
\overset{i\leftrightarrow\overline{i}}{\longleftrightarrow}\hspace{0.15in}  &
\int\limits_{-\infty}^{+\infty}d_{R}^{2}x\,f(x^{2},x^{1})=-(D_{q^{-2}}%
^{1})^{-1}\big |_{-\infty}^{\infty}q^{2\hat{n}_{x^{2}}}(D_{q^{-2}}^{2}%
)^{-1}\big |_{-\infty}^{\infty}q^{\hat{n}_{x^{1}}}f.
\end{align}

Remarkably, q-deformed integrals over the whole space can also be obtained in
a different manner, i.e. by seeking expressions being translationally
invariant. This way we begin by defining an integral over the whole space as a
mapping
\begin{equation}
\int:\widetilde{\mathcal{A}}\rightarrow\mathbb{C},
\end{equation}
being left and right invariant in the sense
\begin{equation}
\int\partial^{i}\triangleright f=\int f\triangleleft\partial^{i}%
=\varepsilon(\partial^{i})\int f=0. \label{StokThe}%
\end{equation}
Notice that this kind of invariance also implies invariance under the action
of generators of the quantum algebra describing the underlying symmetry, i.e.%
\begin{equation}
\int L^{ij}\triangleright f=\int f\triangleleft L^{ij}=\varepsilon(L^{ij})\int
f=0. \label{SymInv}%
\end{equation}
This can be seen rather easily from the fact that the generators of the
quantum algebra are realized within the differential calculus by
\begin{equation}
L^{ij}\equiv k(P_{A})^{ij}{}_{kl}\,\Lambda^{\alpha}X^{k}\partial^{l}%
=k^{\prime}(P_{A})^{ij}{}_{kl}\,\partial^{k}X^{l}\Lambda^{\alpha},
\end{equation}
where $k$ and $k^{\prime}$ stand for suitable constants, while $\Lambda
^{\alpha}$ and $P_{A}$ denote a scaling operator and a q-analog of an
antisymmetrizer, respectively.

That the integrals given by (\ref{VolIntN}) indeed satisfy (\ref{StokThe}) can
be readily checked by using the commutation relations involving inverse
partial derivatives [see for example (\ref{HauDifIntN})]. In this respect, we
should realize that each time we commute a partial derivative with an inverse
one a one-dimensional integral disappears and surface terms, which vanish at
infinity, arise.\ In the case of the two-dimensional quantum plane, for
example, we have%
\begin{align}
&  \lim_{x^{1},x^{2}\rightarrow\infty}(\partial^{2})^{-1}\big |_{-\infty
}^{x^{1}}(\partial^{1})^{-1}\big |_{-\infty}^{x^{2}}\triangleright
(\partial^{1}\triangleright f)\nonumber\\
&  =\lim_{x^{1},x^{2}\rightarrow\infty}(\partial^{2})^{-1}\big |_{-\infty
}^{x^{1}}\triangleright f=0,\\[0.16in]
&  \lim_{x^{1},x^{2}\rightarrow\infty}(\partial^{2})^{-1}\big |_{-\infty
}^{x^{1}}(\partial^{1})^{-1}\big |_{-\infty}^{x^{2}}\triangleright
(\partial^{2}\triangleright f)\nonumber\\
&  =q^{-1}\lim_{x^{1},x^{2}\rightarrow\infty}(\partial^{1})^{-1}%
\big |_{-\infty}^{x^{2}}\triangleright f=0,
\end{align}
where it is assumed that $f\in\widetilde{\mathcal{A}}$. It should be clear
that these reasonings require for the representations of the $\partial^{i}$ to
be of the same type as the representations of the $(\partial^{i})^{-1}.$
However, from (\ref{LinkRechtDarN}) and (\ref{RechtsLinksDarN}) we know that
the antipode and its inverse enable us to transform left and right actions
into each other. In this manner, we can proceed as follows:%
\begin{align}
&  \lim_{x^{\bar{1}},\ldots,x^{\bar{n}}\rightarrow\infty}\big [(\partial
^{n})^{-1}\big |_{-\infty}^{x^{\bar{n}}}\ldots(\partial^{1})^{-1}%
\big |_{-\infty}^{x^{\bar{1}}}\triangleright(f\triangleleft\hat{\partial}%
^{i})\big ]\nonumber\\
&  =\lim_{x^{\bar{1}},\ldots,x^{\bar{n}}\rightarrow\infty}\big [(\partial
^{n})^{-1}\big |_{-\infty}^{x^{\bar{n}}}\ldots(\partial^{1})^{-1}%
\big |_{-\infty}^{x^{\bar{1}}}\triangleright(-\hat{\partial}^{j}%
S^{-1}(\mathcal{L}_{\partial})_{j}^{i}\triangleright f)\big ]\nonumber\\
&  =0. \label{TranInvLR}%
\end{align}
The above considerations also apply for right integrals. Thus, we have%
\begin{align}
&  \lim_{x^{1},\ldots,x^{n}\rightarrow\infty}\big [(\partial^{i}\triangleright
f)\triangleleft(\hat{\partial}^{\bar{1}})^{-1}\big |_{-\infty}^{x^{1}}%
\ldots(\hat{\partial}^{\bar{n}})^{-1}\big |_{-\infty}^{x^{n}}\big ]\nonumber\\
&  =\lim_{x^{1},\ldots,x^{n}\rightarrow\infty}\big [(f\triangleleft
\hat{\partial}^{i})\triangleleft(\hat{\partial}^{\bar{1}})^{-1}\big |_{-\infty
}^{x^{1}}\ldots(\hat{\partial}^{\bar{n}})^{-1}\big |_{-\infty}^{x^{n}%
}\big ]\nonumber\\
&  =0. \label{TranInvKUk}%
\end{align}
Moreover, conjugate actions of partial derivatives can be expressed by
unconjugate ones and vice versa, since we have \cite{LWW97, Oca96}%
\begin{align}
\hat{\partial}^{i}\,\bar{\triangleright}\,f  &  =k_{1}\partial^{i}%
\triangleright(\Lambda\triangleright f)+k_{2}(g_{ij}\partial^{i}\partial
^{i})\triangleright(\Lambda\triangleright(x^{i}\circledast f)),\nonumber\\
\partial^{i}\triangleright f  &  =\tilde{k}_{1}\hat{\partial}^{i}%
\,\bar{\triangleright}\,(\tilde{\Lambda}\triangleright f)+\tilde{k}_{2}%
(g_{ij}\hat{\partial}^{i}\hat{\partial}^{i})\,\bar{\triangleright}%
\,(\tilde{\Lambda}\triangleright(x^{i}\circledast f)),
\end{align}
and
\begin{align}
f\triangleleft\hat{\partial}^{i}  &  =k_{3}(f\triangleleft\Lambda^{-1}%
)\,\bar{\triangleleft}\,\partial^{i}+k_{4}((f\circledast x^{i})\triangleleft
\Lambda^{-1})\,\bar{\triangleleft}\,(\partial^{i}\partial^{i}g_{ij}%
),\nonumber\\
f\,\bar{\triangleleft}\,\partial^{i}  &  =\tilde{k}_{3}(f\triangleleft
\tilde{\Lambda}^{-1})\,\bar{\triangleleft}\,\hat{\partial}^{i}+\tilde{k}%
_{4}((f\circledast x^{i})\triangleleft\tilde{\Lambda}^{-1})\triangleleft
(\hat{\partial}^{i}\hat{\partial}^{i}g_{ij}),
\end{align}
with suitable constants $k_{i},\tilde{k}_{i}\in\mathbb{R}$ and scaling
operators $\Lambda$ and $\tilde{\Lambda}.$ Together with (\ref{TranInvLR}) and
(\ref{TranInvKUk}) these relations give rise to the identities%
\begin{align}
&  \int d_{A}^{n}x(\partial^{i}\triangleright f)=\int d_{A}^{n}x(\hat
{\partial}^{i}\,\bar{\triangleright}\,f)\nonumber\\
=  &  \int d_{A}^{n}x(f\triangleleft\hat{\partial}^{i})=\int d_{A}%
^{n}x(f\,\bar{\triangleleft}\,\partial^{i})=0, \label{TranProp}%
\end{align}
where $A\in\{L,\bar{L},R,\bar{R}\}.$

Translation invariance of integrals can help to simplify the rules for
integration by parts. This becomes quite clear from the calculation%
\begin{align}
\int\limits_{-\infty}^{+\infty}d_{L/R}x^{n}\,f\circledast(\partial
^{i}\triangleright g)  &  =\int\limits_{-\infty}^{+\infty}d_{L/R}%
x^{n}\,(f\triangleleft\partial^{i})\circledast g\nonumber\\
&  +\int\limits_{-\infty}^{+\infty}d_{L/R}x^{n}\,(\mathcal{L}_{\partial}%
)_{j}^{i}\triangleright\left[  (f\triangleleft\partial^{j})\circledast
g\right] \nonumber\\
&  =\int\limits_{-\infty}^{+\infty}d_{L/R}x^{n}\,(f\triangleleft\partial
^{i})\circledast g. \label{SimParN}%
\end{align}
The first step is an application of the coproduct for\ partial derivatives.
For the second step we have to recall that the entries of the L-matrix are
elements of the symmetry algebra. [To be more precise, we should mention that
the entries of the L-matrix depend on a scaling operator. But the invariance
of our integrals under the action of this scaling operator is a rather subtle
point. See the discussion of (\ref{PerJackN})]. Applying (\ref{SymInv})
together with $\varepsilon(\mathcal{L}_{j}^{i})=\delta_{j}^{i}$ then yields
the final result. By a slight modification we can repeat these arguments for
the symmetry generators $L^{ij}.$ In doing so, we get%
\begin{align}
&  \int\limits_{-\infty}^{+\infty}d_{L/R}x^{n}\,f\circledast(L^{ij}%
\triangleright g)\nonumber\\
&  =\int\limits_{-\infty}^{+\infty}d_{L/R}x^{n}\,(L^{ij})_{(2)}\triangleright
\left(  \left[  S^{-1}((L^{ij})_{(1)})\triangleright f\right]  \circledast
g\right) \nonumber\\
&  =\int\limits_{-\infty}^{+\infty}d_{L/R}x^{n}\,\varepsilon((L^{ij}%
)_{(2)})\left[  \left(  f\triangleleft(L^{ij})_{(1)}\right)  \circledast
g\right] \nonumber\\
&  =\int\limits_{-\infty}^{+\infty}d_{L/R}x^{n}\,\left(  f\triangleleft
L^{ij}\right)  \circledast g,
\end{align}
where the second step again makes use of (\ref{SymInv}).

Next, we would like to rewrite the condition of translation invariance in
terms of q-translations. To this end, we first reformulate the properties in
(\ref{TranProp}) by inserting the expressions for the coregular actions [cf.
(\ref{CorWirkAnfN}) and (\ref{CorWirkEndN})]. The non-degeneracy of the
pairings then\ implies
\begin{align}
&  \big \langle\partial^{i},f_{(\bar{L},1)}\big \rangle_{L,\bar{R}}\,\int
d_{A}^{n}x\,f_{(\bar{L},2)}=\big \langle\partial^{i},1\big \rangle_{L,\bar{R}%
}\,\int d_{A}^{n}x\,f,\nonumber\\
\Rightarrow\hspace{0.15in}  &  f_{(\bar{L},1)}\otimes\int d_{A}^{n}%
x\,f_{(\bar{L},2)}=1\otimes\int d_{A}^{n}x\,f,\\[0.16in]
&  \big \langle\hat{\partial}^{i},f_{(L,1)}\big \rangle_{\bar{L},R}\,\int
d_{A}^{n}x\,f_{(L,2)}=\big \langle\hat{\partial}^{i},1\big \rangle_{\bar{L}%
,R}\,\int d_{A}^{n}x\,f,\nonumber\\
\Rightarrow\hspace{0.15in}  &  f_{(L,1)}\otimes\int d_{A}^{n}x\,f_{(L,2)}%
=1\otimes\int d_{A}^{n}x\,f,
\end{align}
and%
\begin{align}
&  \int d_{A}^{n}x\,f_{(\bar{R},1)}\,\big \langle f_{(\bar{R},2)},\partial
^{i}\big \rangle_{\bar{L},R}=\int d_{A}^{n}x\,f\,\big \langle1,\partial
^{i}\big \rangle_{\bar{L},R},\nonumber\\
\Rightarrow\hspace{0.15in}  &  \int d_{A}^{n}x\,f_{(\bar{R},1)}\otimes
f_{(\bar{R},2)}=\int d_{A}^{n}x\,f\otimes1,\\[0.16in]
&  \int d_{A}^{n}x\,f_{(R,1)}\,\big \langle f_{(R,2)},\hat{\partial}%
^{i}\big \rangle_{L,\bar{R}}=\int d_{A}^{n}x\,f\,\big \langle1,\hat{\partial
}^{i}\big \rangle_{L,\bar{R}},\nonumber\\
\Rightarrow\hspace{0.15in}  &  \int d_{A}^{n}x\,f_{(R,1)}\otimes
f_{(R,2)}=\int d_{A}^{n}x\,f\otimes1.
\end{align}
To sum up, we have%
\begin{align}
\int\limits_{-\infty}^{+\infty}d_{A}x^{n}\,f(x^{i})  &  =\int\limits_{-\infty
}^{+\infty}d_{A}x^{n}\,f(y^{j}\oplus_{\bar{L}}x^{i})=\int\limits_{-\infty
}^{+\infty}d_{A}x^{n}\,f(x^{i}\oplus_{\bar{R}}y^{j})\nonumber\\
&  =\int\limits_{-\infty}^{+\infty}d_{A}x^{n}\,f(y^{j}\oplus_{L}x^{i}%
)=\int\limits_{-\infty}^{+\infty}d_{A}x^{n}\,f(x^{i}\oplus_{R}y^{j}),
\end{align}
with $A\in\{L,\bar{L},R,\bar{R}\}$.

Let us return to the property in (\ref{StokThe}). It tells us that an integral
over the whole space is nothing other than a functional which is invariant
under translations. For this reason, the problem of constructing an integral
over the whole space is equivalent to that of seeking a functional being
invariant under translations. To find such a functional it is helpful to
realize that all functions subject to%
\begin{equation}
\partial^{i}\triangleright F(x^{j})=F(x^{j})\triangleleft\partial^{i}%
=0,\quad\text{for all }i\text{,}\label{VerPar}%
\end{equation}
have to be periodic in the sense

\begin{enumerate}
\item[(i)] (quantum plane)%
\begin{equation}
F(q^{\pm2}x^{1})=F(q^{\pm2}x^{2})=F, \label{PerAnf}%
\end{equation}

\item[(ii)] (three-dimensional Euclidean space)%
\begin{equation}
F(q^{\pm4}x^{+})=F(q^{\pm2}x^{3})=F(q^{\pm4}x^{-})=F,
\end{equation}

\item[(iii)] (four-dimensional Euclidean space)%
\begin{equation}
F(q^{\pm2}x^{1})=F(q^{\pm2}x^{2})=F(q^{\pm2}x^{3})=F(q^{\pm2}x^{4})=F,
\end{equation}

\item[(iv)] (q-deformed Minkowski space)%
\begin{equation}
F(q^{\pm2}r^{2})=F(q^{\pm2}x^{+})=F(q^{\pm2}x^{3/0})=F(q^{\pm2}x^{-})=F.
\label{PerEnd}%
\end{equation}

\end{enumerate}

\noindent These conditions follow from a direct inspection of the results in
Ref. \cite{Wac04}. In other words (\ref{PerAnf})-(\ref{PerEnd}) characterize
functions being constant from the point of view provided by q-deformation. For
the sake of completeness we also note that due to (\ref{q-TayN}) and
(\ref{q-TayRecN}) the requirement in (\ref{VerPar}) is equivalent to
\begin{equation}
F(y^{j}\oplus_{L/\bar{L}}x^{i})=F(x^{i}),\quad F(x^{i}\oplus_{R/\bar{R}}%
y^{j})=F(x^{i}).
\end{equation}

Now, we are ready to solve the problem of constructing translationally
invariant functionals. Functions satisfying the periodicity conditions in
(\ref{PerAnf})-(\ref{PerEnd}) are obtained most easily by means of the Jackson
integrals $(a>0,$ $q>1)$%
\begin{align}
F  &  =(D_{q^{\pm a}}^{i})^{-1}\big |_{0}^{\infty}f=\mp(1-q^{\pm a}%
)\sum_{k=-\infty}^{\infty}(q^{ak}x^{i})f(q^{ak}x^{i}),\quad x^{i}%
>0,\nonumber\\
F  &  =(D_{q^{\pm a}}^{i})^{-1}\big |_{-\infty}^{0}f=\pm(1-q^{\pm a}%
)\sum_{k=-\infty}^{\infty}(q^{ak}x^{i})f(q^{ak}x^{i}),\quad x^{i}<0,
\label{PerJackN}%
\end{align}
since it holds%
\begin{align}
\big ((D_{q^{a}}^{i})^{-1}\big |_{0}^{\infty}f\big )(q^{\pm a}x^{i})  &
=(D_{q^{a}}^{i})^{-1}\big |_{0}^{\infty}f,\nonumber\\
\big ((D_{q^{a}}^{i})^{-1}\big |_{-\infty}^{0}f\big )(q^{\pm a}x^{i})  &
=(D_{q^{a}}^{i})^{-1}\big |_{-\infty}^{0}f.
\end{align}
A short glance at (\ref{ExpVolAnf})-(\ref{ExpVolEndeN}) tells us that our
integrals over the whole space are built up out of Jackson integrals. From
this observation we see once again that the integrals in (\ref{ExpVolAnf}%
)-(\ref{ExpVolEndeN}) are indeed invariant under translations.

However, there is a point we have to discuss more carefully. Remarkably, the
Jackson integrals over the intervals $[\,0,-\infty)$ and $(-\infty,0\,]$ still
depend on the integration variable. Thus, the value of the Jackson integral in
(\ref{PerJackN}) is not completely determined. This ambiguity can be removed
by setting the value of the integration variable equal to $1$ or $-1$, i.e.%
\begin{equation}
(D_{q^{\pm a}}^{i})^{-1}\big |_{-\infty}^{\infty}f\equiv\mp(1-q^{\pm a}%
)\sum_{k=-\infty}^{\infty}q^{ak}\big (f(q^{ak})+f(-q^{ak})\big ).
\label{JackDef}%
\end{equation}
Nevertheless, it arises the question where this ambiguity comes from and what
it physically means. First of all, we can say that due to the periodicity of
the Jackson integrals in (\ref{PerJackN}) their values can vary within a
certain range. So to speak, an infinite Jackson integral cannot be
characterized by a certain value, only, but an interval of values. We can
believe that this feature expresses the impossibility of making sharp measurements.

There is another very important point which should be clarified. It concerns
the behavior of integrals under the action of scaling operators. Up to now we
know that the values of infinite Jackson integrals are not uniquely
determined, since they depend on a parameter. Important for us is the fact
that scaling operators of the form $q^{\alpha\hat{n}_{i}}$ have an influence
on this parameter, when they act on infinite Jackson integrals. This can be
seen rather easily by the following calculation:
\begin{align}
q^{\alpha\hat{n}_{i}}\triangleright(D_{q^{\pm a}}^{i})^{-1}\big |_{0}^{\infty
}f  &  =\mp(1-q^{\pm a})\sum_{k=-\infty}^{\infty}q^{ak}(q^{\alpha}%
x^{i})f(q^{ak}(q^{\alpha}x^{i}))\nonumber\\
&  =\big ((D_{q^{\pm a}}^{i})^{-1}\big |_{0}^{\infty}f\big )(q^{\alpha}x^{i}).
\end{align}
However, if we keep in mind that the parameters of infinite Jackson integrals
can take on arbitrary values, we are always allowed to drop scaling operators
acting on infinite Jackson integrals.

Now, we can go one step further. The considerations so far show us that right
and left invariant functionals are obtained by products of Jackson integrals
of the form (\ref{JackDef}). The integrals in (\ref{ExpVolAnf}%
)-(\ref{ExpVolEndeN}) give examples for such functionals. Due to the
identities%
\begin{align}
(D_{q^{a}}^{i})^{-1}\big |_{-\infty}^{\infty}f  &  =q^{a}(D_{q^{-a}}^{i}%
)^{-1}\big |_{-\infty}^{\infty}f,\nonumber\\[0.06in]
(D_{q^{a}}^{i})^{-1}\big |_{-\infty}^{\infty}q^{c\hat{n}_{i}}f(x^{i})  &
=q^{-c}q^{c\hat{n}_{i}}(D_{q^{a}}^{i})^{-1}\big |_{-\infty}^{\infty}f,
\end{align}
which are a direct consequence of the definitions in (\ref{PerJackN}), these
integrals are unique up to a global factor if we neglect the parametrization
of infinite Jackson integrals and if we do not take into account the
appearance of scaling operators.

Up to now we know that integrals over the whole space are invariant under
translations and behave like scalars under the action of the algebra
describing the quantum symmetry. This way, integrals over the whole space
should have trivial braiding. However, a short glance at the L-matrices [cf.
(\ref{LMa2dimN}) and (\ref{LMa2dimKonNN})] shows us that the braiding is not
only determined by the action of symmetry generators but also by that of
certain scaling operators. From the discussion above we know that a scaling
operator $\Lambda$ with $\Lambda\triangleright x^{i}=q^{\alpha}x^{i}$ does not
really change a Jackson integral over the intervals $[\,0,-\infty)$ and
$(-\infty,0\,]$. This, in turn, implies that integrals over the whole space
remain unchanged under the action of scaling operators. Hence, integrals over
the whole space can indeed be treated as objects having trivial braiding.

Lastly, let us remind of the fact that the explicit form of our integrals
depends on the choice for the normal ordering of quantum space coordinates. On
this ground, it would be interesting to understand how a change of the
underlying normal ordering affects the evaluation of integrals over the whole
space. As we know from the discussion in Section \ref{Sec1}, there are
operators transforming functions to those representing the same quantum space
element but for reversed normal ordering. With these operators at hand we show
in Appendix \ref{MinRech} how the expressions for integrals over q-deformed
Minkowski space change under reversing the normal ordering. The same method
applies to the other quantum spaces we are considering in this article.\ In
this manner we get

\begin{enumerate}
\item[(i)] (quantum plane)%
\begin{align}
\int\limits_{-\infty}^{+\infty}d_{L/\bar{R}}^{2}x\,f(x^{1},x^{2})  &
=\int\limits_{-\infty}^{+\infty}d_{L/\bar{R}}^{2}x\,q^{-\hat{n}_{x^{2}}\hat
{n}_{x^{1}}}\tilde{f}(x^{2},x^{1}),\\
\int\limits_{-\infty}^{+\infty}d_{\bar{L}/R}^{2}x\,\tilde{f}(x^{2},x^{1})  &
=\int\limits_{-\infty}^{+\infty}d_{\bar{L}/R}^{2}x\,q^{\hat{n}_{x^{2}}\hat
{n}_{x^{1}}}f(x^{1},x^{2}),
\end{align}

\item[(ii)] (three-dimensional Euclidean space)%
\begin{align}
&  \int\limits_{-\infty}^{+\infty}d_{L/\bar{R}}^{3}x\,f(x^{+},x^{3}%
,x^{-})=\int\limits_{-\infty}^{+\infty}d_{L/\bar{R}}^{3}x\,q^{2\hat{n}_{x^{3}%
}(\hat{n}_{x^{+}}+\,\hat{n}_{x^{-}})}\tilde{f}(x^{-},x^{3},x^{+}),\\[0.16in]
&  \int\limits_{-\infty}^{+\infty}d_{\bar{L}/R}^{3}x\,\tilde{f}(x^{-}%
,x^{3},x^{+})=\int\limits_{-\infty}^{+\infty}d_{\bar{L}/R}^{3}x\,q^{-2\hat
{n}_{x^{3}}(\hat{n}_{x^{+}}+\,\hat{n}_{x^{-}})}f(x^{+},x^{3},x^{-}),
\end{align}

\item[(iii)] (four-dimensional Euclidean space)%
\begin{gather}
\int\limits_{-\infty}^{+\infty}d_{\bar{L}/R}^{4}x\,f(x^{1},x^{2},x^{3}%
,x^{4})=\nonumber\\
\int\limits_{-\infty}^{+\infty}d_{\bar{L}/R}^{4}x\,q^{-(\hat{n}_{x^{2}}%
+\,\hat{n}_{x^{3}})(\hat{n}_{x^{1}}+\,\hat{n}_{x^{4}})}\tilde{f}(x^{4}%
,x^{2},x^{3},x^{1}),\\[0.16in]
\int\limits_{-\infty}^{+\infty}d_{L/\bar{R}}^{4}x\,\tilde{f}(x^{4},x^{3}%
,x^{2},x^{1})=\nonumber\\
\int\limits_{-\infty}^{+\infty}d_{L/\bar{R}}^{4}x\,q^{(\hat{n}_{x^{2}}%
+\,\hat{n}_{x^{3}})(\hat{n}_{x^{1}}+\,\hat{n}_{x^{4}})}f(x^{1},x^{2}%
,x^{3},x^{4}),
\end{gather}

\item[(iv)] (q-deformed Minkowski space)%
\begin{gather}
\int\limits_{-\infty}^{+\infty}d_{\bar{L}/R}^{4}x\,f(r^{2},x^{+},x^{3/0}%
,x^{-})=\nonumber\\
\int\limits_{-\infty}^{+\infty}d_{\bar{L}/R}^{4}x\,q^{2\hat{n}_{x^{+}}\hat
{n}_{x^{-}}+\,2\hat{n}_{x^{3}}(\hat{n}_{x^{+}}+\,\hat{n}_{x^{-}})}\tilde
{f}(x^{-},x^{3/0},x^{+},r^{2}),\\[0.16in]
\int\limits_{-\infty}^{+\infty}d_{\bar{L}/R}^{4}x\,\tilde{f}(x^{-}%
,x^{3/0},x^{+},r^{2})=\nonumber\\
\int\limits_{-\infty}^{+\infty}d_{\bar{L}/R}^{4}x\,q^{-2\hat{n}_{x^{+}}\hat
{n}_{x^{-}}-\,2\hat{n}_{x^{3}}(\hat{n}_{x^{+}}+\,\hat{n}_{x^{-}})}%
f(r^{2},x^{+},x^{3/0},x^{-}),
\end{gather}

\end{enumerate}

\noindent where $f$ and $\tilde{f}$ represent the same quantum space element
but for different normal orderings.

\subsection{Conjugation properties of q-deformed integrals}

Now, we turn to the conjugation properties of our integrals. To begin with,
let us mention that from an algebraic point of view a consistent conjugation
is introduced by%
\begin{equation}
\overline{(\partial^{i})^{-1}}=(\overline{\partial^{i}})^{-1}. \label{KonInt}%
\end{equation}
With this convention the properties in (\ref{RegConAblN})\ imply
for\ representations of inverse partial derivatives the following identities%
\begin{align}
\overline{\big ((\partial^{i})^{-1}\triangleright f(x^{j})\big )}  &
=\overline{f(x^{j})}\,\bar{\triangleleft}\,(\overline{\partial^{i}})^{-1}, &
\overline{\big (f(x^{j})\,\bar{\triangleleft}\,(\partial^{i})^{-1}\big )}  &
=(\overline{\partial^{i}})^{-1}\triangleright\overline{f(x^{j})},\nonumber\\
\overline{\big ((\hat{\partial}^{i})^{-1}\,\bar{\triangleright}\,f(x^{j}%
)\big )}  &  =\overline{f(x^{j})}\triangleleft(\overline{\hat{\partial}^{i}%
})^{-1}, & \overline{\big (f(x^{j})\triangleleft(\hat{\partial}^{i}%
)^{-1}\big )}  &  =(\overline{\hat{\partial}^{i}})^{-1}\,\bar{\triangleright
}\,\overline{f(x^{j})}.
\end{align}
Moreover, from (\ref{KonInt}) it follows that inverse partial derivatives are
imaginary, i.e.
\begin{equation}
\overline{(\partial^{i})^{-1}}=-(\overline{\partial}_{i})^{-1}=-(\partial
_{i})^{-1}=-g_{ij}(\partial^{j})^{-1}. \label{ConInvAbl}%
\end{equation}
However, for this to be true in general we have to deal with $-$%
i$(\partial^{0})^{-1}$ instead of $(\partial^{0})^{-1}$ in the case of
q-deformed Minkowski space.

In addition to this, we can ask about the conjugation properties of integrals
over the whole space. Applying the above results to the expressions in
(\ref{VolIntN}) yields%
\begin{align}
\overline{\int\limits_{-\infty}^{+\infty}d_{L}^{n}x\,f}  &  =\lim_{x^{\bar{1}%
},\ldots,x^{\bar{n}}\rightarrow\infty}\overline{(\partial^{n})^{-1}%
\big |_{-\infty}^{x^{\bar{n}}}\ldots(\partial^{1})^{-1}\big |_{-\infty
}^{x^{\bar{1}}}\triangleright f}\nonumber\\
&  =(-1)^{n}\lim_{x^{1},\ldots,x^{n}\rightarrow\infty}\bar{f}\,\bar
{\triangleleft}\,(\partial^{\bar{1}})^{-1}\big |_{-\infty}^{x^{1}}%
\ldots(\partial^{\bar{n}})^{-1}\big |_{-\infty}^{x^{n}}\nonumber\\
&  =(-1)^{n}\int\limits_{-\infty}^{+\infty}d_{\bar{R}}^{n}x\,\bar
{f},\label{ConjIntAnfN}\\[0.16in]
\overline{\int\limits_{-\infty}^{+\infty}d_{R}^{n}x\,f}  &  =\lim
_{x^{1},\ldots,x^{n}\rightarrow\infty}\overline{f\triangleleft(\hat{\partial
}^{\bar{1}})^{-1}\big |_{-\infty}^{x^{1}}\ldots(\hat{\partial}^{\bar{n}}%
)^{-1}\big |_{-\infty}^{x^{n}}}\nonumber\\
&  =(-1)^{n}\lim_{x^{\bar{1}},\ldots,x^{\bar{n}}\rightarrow\infty}%
(\hat{\partial}^{n})^{-1}\big |_{-\infty}^{x^{\bar{n}}}\ldots(\hat{\partial
}^{1})^{-1}\big |_{-\infty}^{x^{\bar{1}}}\,\bar{\triangleright}\,\bar
{f}\nonumber\\
&  =(-1)^{n}\int\limits_{-\infty}^{+\infty}d_{\bar{L}}^{n}x\,\bar{f},
\end{align}
where we used the fact that the metrical factors stemming from
(\ref{ConInvAbl}) cancel against each other. In very much the same way, we
obtain%
\begin{align}
\overline{\int\limits_{-\infty}^{+\infty}d_{\bar{L}}^{n}x\,f}  &
=(-1)^{n}\int\limits_{-\infty}^{+\infty}d_{R}^{n}x\,\bar{f},\nonumber\\
\overline{\int\limits_{-\infty}^{+\infty}d_{\bar{R}}^{n}x\,f}  &
=(-1)^{n}\int\limits_{-\infty}^{+\infty}d_{L}^{n}x\,\bar{f}.
\label{ConjIntEndN}%
\end{align}
It is easily seen that under conjugation left integrals are transformed into
right ones and vice versa.

If we want to have an integral with%
\begin{equation}
\overline{\int f}=\int\bar{f}, \label{RealInt}%
\end{equation}
it should be clear that we can take the combinations%
\begin{align}
&  \text{i}^{n}\int\limits_{-\infty}^{+\infty}d_{\bar{L}}^{n}x\,f+\text{i}%
^{n}\int\limits_{-\infty}^{+\infty}d_{R}^{n}x\,f,\nonumber\\
&  \text{i}^{n}\int\limits_{-\infty}^{+\infty}d_{L}^{n}x\,f+\text{i}^{n}%
\int\limits_{-\infty}^{+\infty}d_{\bar{R}}^{n}x\,f.
\end{align}
Notice that for q-deformed Minkowski space integration over the time
coordinate requires an additional imaginary factor in the above expressions.

\section{Conclusion\label{SecCon}}

Let us end with some comments on what we have done so far. The aim of our
program was to provide us with a q-deformed version of analysis on quantum
spaces of physical interest, i.e. Manin plane, q-deformed Euclidean space in
three or four dimensions, and q-deformed Minkowski space. In our previous work
attention was focused on explicit formulae realizing the elements of
q-analysis on commutative coordinate algebras. This way, we got
multi-dimensional versions of the well-known q-calculus with Jackson integrals
and Jackson derivatives \cite{Kac00}. In the present article we continued
these considerations, but stress was taken on a presentation which reveals the
properties of the elements of q-deformed analysis. Remarkably, our reasonings
are in complete analogy to the classical situation, which we regain in the
undeformed limit as $q\rightarrow1$.

It should be mentioned that in some sense our work is inspired by the
so-called star-product formalism. Nevertheless we do not explicitly consider
expansions in terms of a deformation parameter, since we try to work with
closed expressions, only. In the former literature the problem of formulating
physical theories is often attacked by the construction of Hilbert space
representations \cite{Fio93, Zip95, CW98, FBM03}. Although this approach is
rather rigorous from a mathematical point of view it is not so easy to
understand the physical meaning of its results. As was pointed out above, in
our framework, however, the relationship to the classical limit is very clear.
This observation nourishes the hope that our concept will enable us to develop
a q-deformed field theory along the same line of reasonings\ as its undeformed
counterpart.\vspace{0.16in}

\noindent\textbf{Acknowledgement}

First of all I am very grateful to Eberhard Zeidler for very interesting and
useful discussions, special interest in my work, and financial support. Also I
wish to express my gratitude to Julius Wess for his efforts, suggestions, and
discussions. Furthermore I would like to thank Alexander Schmidt, Fabian
Bachmaier, and Florian Koch for useful discussions and their steady support.
Finally, I thank Dieter L\"{u}st for kind hospitality.

\appendix

\section{Quantum spaces\label{AppQuan}}

The coordinates of the two-dimensional q-deformed quantum plane fulfill the
relation \cite{Man88,SS90}
\begin{equation}
X^{1}X^{2}=qX^{2}X^{1}, \label{2dimQuan}%
\end{equation}
whereas the quantum metric is given by a matrix $\varepsilon^{ij}$ with
non-vanishing elements
\begin{equation}
\varepsilon^{12}=q^{-1/2},\quad\varepsilon^{21}=-q^{1/2}.
\end{equation}
The inverse of $\varepsilon^{ij}$ is given by
\begin{equation}
(\varepsilon^{-1})^{ij}=\varepsilon_{ij}=-\varepsilon^{ij}.
\end{equation}

In the case of the q-deformed Euclidean space in three dimensions the
commutation relations read \cite{LWW97}
\begin{align}
X^{3}X^{+}  &  =q^{2}X^{+}X^{3},\nonumber\\
X^{-}X^{3}  &  =q^{2}X^{3}X^{-},\nonumber\\
X^{-}X^{+}  &  =X^{+}X^{-}+\lambda X^{3}X^{3}. \label{Koord3dimN}%
\end{align}
The non-vanishing elements of the quantum metric are
\begin{equation}
g^{+-}=-q,\quad g^{33}=1,\quad g^{-+}=-q^{-1}.
\end{equation}
Its inverse is determined by
\begin{equation}
(g^{-1})^{AB}=g_{AB}=g^{AB}.
\end{equation}

For the four-dimensional q-deformed\ Euclidean space we have the relations
\cite{CSSW90, Oca96}
\begin{align}
X^{1}X^{2}  &  =qX^{2}X^{1},\nonumber\\
X^{1}X^{3}  &  =qX^{3}X^{1},\nonumber\\
X^{3}X^{4}  &  =qX^{4}X^{3},\nonumber\\
X^{2}X^{4}  &  =qX^{4}X^{2},\nonumber\\
X^{2}X^{3}  &  =X^{3}X^{2},\nonumber\\
X^{4}X^{1}  &  =X^{1}X^{4}+\lambda X^{2}X^{3}. \label{Algebra4N}%
\end{align}
The metric has as non-vanishing components
\begin{equation}
g^{14}=q^{-1},\quad g^{23}=g^{32}=1,\quad g^{41}=q,
\end{equation}
and it holds
\begin{equation}
(g^{-1})^{ij}=g_{ij}=g^{ij}. \label{Metrik}%
\end{equation}

In the case of the q-deformed Minkowski space the coordinates obey the
relations \cite{CSSW90}
\begin{align}
X^{\mu}X^{0} &  =X^{0}X^{\mu},\quad\mu\in\{0,+,-,3\},\nonumber\\
X^{-}X^{3}-q^{2}X^{3}X^{-} &  =-q\lambda X^{0}X^{-},\nonumber\\
X^{3}X^{+}-q^{2}X^{+}X^{3} &  =-q\lambda X^{0}X^{+},\nonumber\\
X^{-}X^{+}-X^{+}X^{-} &  =\lambda(X^{3}X^{3}-X^{0}X^{3}),\label{MinrelN}%
\end{align}
and the non-vanishing components of the metric read
\begin{equation}
\eta^{00}=-1,\quad\eta^{33}=1,\quad\eta^{+-}=-q,\quad\eta^{-+}=-q^{-1}.
\end{equation}
Again, we have
\begin{equation}
(\eta^{-1})^{\mu\nu}=\eta_{\mu\nu}=\eta_{\mu\nu}.
\end{equation}
For other deformations of Minkowski spacetime we refer to Refs. \cite{Lu92,
Cas93, Dob94, ChDe95, DFR95, ChKu04, Koch04}.

\section{\label{MinRech}Integrals for q-deformed Minkowski space}

In this appendix we would like to deal with integrals for q-deformed Minkowski
space, which from a physical point of view is the most interesting case among
q-deformed quantum spaces. Especially, we derive expressions for
one-dimensional integrals as well as for integrals over the whole space. In
the latter case, we additionally explain how these expressions depend on the
choice for the normal ordering.

To begin with, we recall the explicit form of conjugate left representations
of partial derivatives \cite{Wac04}:%
\begin{align}
&  \hat{\partial}^{3}\,\bar{\triangleright}\,f(r^{2},x^{+},x^{3/0}%
,x^{-})=\underline{D_{q^{2}}^{3/0}f(q^{2}r^{2})}\nonumber\\
&  \quad+q^{-1}\lambda_{+}x^{3/0}D_{q^{2}}^{r^{2}}f(q^{2}x^{+})\nonumber\\
&  \quad+\big (-q^{-2}r^{2}(x^{3/0})^{-1}-q^{-4}x^{3/0}\big )D_{q^{2}}^{r^{2}%
}f\nonumber\\
&  \quad+q^{-2}\lambda_{+}x^{+}x^{-}(x^{3/0})^{-1}D_{q^{2}}^{r^{2}}%
f(q^{2}x^{3/0})\nonumber\\
&  \quad-\,q\lambda_{+}^{-1}\lambda x^{3/0}D_{q^{2}}^{+}D_{q^{2}}^{-}%
f(q^{2}r^{2}),\label{IntMinAnf}\\[0.16in]
&  \hat{\partial}^{+}\,\bar{\triangleright}\,f(r^{2},x^{+},x^{3/0}%
,x^{-})=\underline{-qD_{q^{2}}^{-}f(q^{2}r^{2})}\nonumber\\
&  \quad+q^{-1}\lambda_{+}x^{+}D_{q^{2}}^{r^{2}}f,\\[0.16in]
&  \hat{\partial}^{-}\,\bar{\triangleright}\,f(r^{2},x^{+},x^{3/0}%
,x^{-})=\underline{-q^{-1}D_{q^{2}}^{+}f}\nonumber\\
&  \quad+q^{-1}\lambda_{+}x^{-}D_{q^{2}}^{r^{2}}f(q^{2}x^{+},q^{2}%
x^{3/0})\nonumber\\
&  \quad+\,q^{-2}\lambda(x^{3/0})^{2}D_{q^{2}}^{+}D_{q^{2}}^{r^{2}}%
f(q^{2}x^{+},q^{2}x^{-})\nonumber\\
&  \quad-\,q^{-1}\lambda^{2}(x^{3/0})^{2}x^{-}D_{q^{2}}^{+}D_{q^{2}}%
^{-}D_{q^{2}}^{r^{2}}f(q^{2}x^{+}),\\[0.16in]
&  \hat{\partial}^{2}\,\bar{\triangleright}\,f(r^{2},x^{+},x^{3/0}%
,x^{-})=\underline{q\lambda_{+}^{3}D_{q^{2}}^{r^{2}}f(q^{2}x^{+},q^{2}%
x^{3/0},q^{2}x^{-})}\nonumber\\
&  \quad+q^{3}\lambda_{+}^{3}r^{2}(D_{q^{2}}^{r^{2}})^{2}f(q^{2}x^{+}%
,q^{2}x^{3/0},q^{2}x^{-})\nonumber\\
&  \quad+\,(q^{-1}\lambda_{+})^{2}x^{-}D_{q^{2}}^{-}D_{q^{2}}^{r^{2}}%
f(q^{2}x^{+},q^{2}x^{3/0})\nonumber\\
&  \quad+\,(q^{-1}\lambda_{+})^{2}x^{3/0}D_{q^{2}}^{3/0}D_{q^{2}}^{r^{2}%
}f(q^{2}x^{+})\nonumber\\
&  \quad-\,\lambda_{+}D_{q^{2}}^{+}D_{q^{2}}^{-}f(q^{2}r^{2})+(q^{-1}%
\lambda_{+})^{2}x^{+}D_{q^{2}}^{+}D_{q^{2}}^{r^{2}}f, \label{IntMinEnd}%
\end{align}
where $\lambda_{+}\equiv q+q^{-1}.$

Next, we wish to apply the result of (\ref{IntegralE3N}) to the above
representations. To this end, we first have to identify the expressions for
the $\hat{\partial}_{\text{cl}}^{i}.$ They have to be invertible operators
that do not vanish in the classical limit. In this manner, we choose for the
$\hat{\partial}_{\text{cl}}^{i}$ the expressions we have underlined in
(\ref{IntMinAnf})-(\ref{IntMinEnd}). We find as corresponding inverse
operators%
\begin{align}
(\hat{\partial}_{\text{cl}}^{3})^{-1}\,\bar{\triangleright}\,f  &  =(D_{q^{2}%
}^{3/0})^{-1}f(q^{-2}r^{2}),\nonumber\\
(\hat{\partial}_{\text{cl}}^{+})^{-1}\,\bar{\triangleright}\,f  &
=-q^{-1}(D_{q^{2}}^{-})^{-1}f(q^{-2}r^{2}),\nonumber\\
(\hat{\partial}_{\text{cl}}^{-})^{-1}\,\bar{\triangleright}\,f  &
=-q(D_{q^{2}}^{+})^{-1}f,\nonumber\\
(\hat{\partial}_{\text{cl}}^{r^{2}})^{-1}\,\bar{\triangleright}\,f  &
=q^{-1}\lambda_{+}^{-3}(D_{q^{2}}^{r^{2}})^{-1}f(q^{-2}x^{+},q^{-2}%
x^{3/0},q^{-2}x^{-}).
\end{align}
Consequently, the terms being not underlined contribute to $\hat{\partial
}_{\text{cor}}^{i}$. Inserting the explicit form of $(\hat{\partial
}_{\text{cl}}^{i})^{-1}$ and $\hat{\partial}_{\text{cor}}^{i}$ into
(\ref{IntegralE3N}) then yields the wanted expressions for the integral
operator $(\hat{\partial}^{i})^{-1}.$ We leave the details to the reader and
write down the result for $(\hat{\partial}^{+})^{-1}$ as an example:%
\begin{align}
(\hat{\partial}^{+})^{-1}\,\bar{\triangleright}\,f  &  =-q^{-1}\sum
_{k=0}^{\infty}(q^{-2}\lambda_{+})^{k}q^{k(k+1)}(x^{+})^{k}\nonumber\\
&  \quad\times(D_{q^{2}}^{r^{2}})^{k}(D_{q^{2}}^{-})^{-(k+1)}f(q^{-2(k+1)}%
r^{2}).
\end{align}

Now, we come to integrals over the whole Minkowski-space. They can be
introduced by%
\begin{equation}
\int d_{\bar{L}}^{4}x\,f(x^{i})\equiv(\hat{\partial}^{2})^{-1}\big |_{-\infty
}^{\infty}(\hat{\partial}^{+})^{-1}\big |_{-\infty}^{\infty}(\hat{\partial
}^{3})^{-1}\big |_{-\infty}^{\infty}(\hat{\partial}^{-})^{-1}\big |_{-\infty
}^{\infty}\,\bar{\triangleright}\,f.
\end{equation}
It is our aim to rewrite the last expression in terms of Jackson integrals. To
this end we first try to simplify the formulae for the one-dimensional
integrals. In the case of $(\hat{\partial}^{-})^{-1},$ for example, we can
proceed as follows:%
\begin{align}
&  (\hat{\partial}^{-})^{-1}\big |_{-\infty}^{\infty}\,\bar{\triangleright
}\,f=\sum_{k=0}^{\infty}\left(  -1\right)  ^{k}\left[  (\hat{\partial
}_{\text{cl}}^{-})^{-1}\big |_{-\infty}^{\infty}\hat{\partial}_{\text{cor}%
}^{-}\right]  ^{k}(\hat{\partial}_{\text{cl}}^{-})^{-1}\big |_{-\infty
}^{\infty}f\nonumber\\
&  \quad=-q\sum_{k=0}^{\infty}\Big [q(D_{q^{2}}^{+})^{-1}\big |_{-\infty
}^{\infty}\Big (q^{-1}\lambda_{+}x^{-}D_{q^{2}}^{r^{2}}q^{2(\hat{n}_{x^{+}%
}+\,\hat{n}_{x^{3/0}})}\nonumber\\
&  \qquad\qquad+\,q^{-2}\lambda(x^{3/0})^{2}D_{q^{2}}^{+}D_{q^{2}}^{r^{2}%
}q^{2(\hat{n}_{x^{+}}+\,\hat{n}_{x^{-}})}\nonumber\\
&  \qquad\qquad-\,q^{-1}\lambda^{2}(x^{3/0})^{2}x^{-}D_{q^{2}}^{+}D_{q^{2}%
}^{-}D_{q^{2}}^{r^{2}}q^{2\hat{n}_{x^{+}}}\Big )\Big ]^{k}(D_{q^{2}}^{+}%
)^{-1}\big |_{-\infty}^{\infty}f\nonumber\\
&  \quad=-q\sum_{k=0}^{\infty}\Big [\lambda_{+}x^{-}(D_{q^{2}}^{+}%
)^{-1}\big |_{-\infty}^{\infty}D_{q^{2}}^{r^{2}}q^{2(\hat{n}_{x^{+}}+\,\hat
{n}_{x^{3/0}})}\Big ]^{k}(D_{q^{2}}^{+})^{-1}\big |_{-\infty}^{\infty
}f\nonumber\\
&  \quad=-q\sum_{k=0}^{\infty}q^{k(k+1)}(\lambda_{+}x^{-}D_{q^{2}}^{r^{2}%
})^{k}(D_{q^{2}}^{+})^{-(k+1)}\big |_{-\infty}^{\infty}f(q^{2k}x^{+}%
,q^{2k}x^{3/0}).
\end{align}
The first equality is the formula from (\ref{IntegralE3N}), for the second
equality we insert the explicit form of $(\hat{\partial}_{\text{cl}}^{-}%
)^{-1}$ and $\hat{\partial}_{\text{cor}}^{-}$, while the third equality uses%
\begin{align}
D_{q^{a}}^{i}(D_{q^{a}}^{i})^{-1}\big |_{-\infty}^{\infty}f  &  =\frac
{(D_{q^{a}}^{i})^{-1}\big |_{-\infty}^{\infty}f-\big ((D_{q^{a}}^{i}%
)^{-1}\big |_{-\infty}^{\infty}f\big )(q^{a}x^{i})}{(1-q^{a})x^{i}}\nonumber\\
&  =\frac{(D_{q^{a}}^{i})^{-1}\big |_{-\infty}^{\infty}f-(D_{q^{a}}^{i}%
)^{-1}\big |_{-\infty}^{\infty}f}{(1-q^{a})x^{i}}\nonumber\\
&  =0. \label{IntAbl1N}%
\end{align}
In very much the same way we can show that%
\begin{align}
&  (\hat{\partial}^{2})^{-1}\big |_{-\infty}^{\infty}\,\bar{\triangleright
}\,f=\nonumber\\
&  \hspace{0.1in}\sum_{k=0}^{\infty}q^{(\frac{3}{2}k-1)(k+1)}\lambda
_{+}^{-2k-3}(D_{q^{2}}^{+}D_{q^{2}}^{-})^{k}\nonumber\\
&  \hspace{0.1in}\times(D_{q^{2}}^{r^{2}})^{-(k+1)}\big |_{-\infty}^{\infty
}f(q^{2k}r^{2},q^{-2(k+1)}x^{+},q^{-2(k+1)}x^{3/0},q^{-2(k+1)}x^{-}).
\end{align}
Putting everything together we obtain%
\begin{align}
&  (\hat{\partial}^{+})^{-1}\big |_{-\infty}^{\infty}(\hat{\partial}^{3}%
)^{-1}\big |_{-\infty}^{\infty}(\hat{\partial}^{-})^{-1}\big |_{-\infty
}^{\infty}(\hat{\partial}^{2})^{-1}\big |_{-\infty}^{\infty}\,\bar
{\triangleright}\,f=\nonumber\\
&  \qquad\sum_{j=0}^{\infty}(q^{-2}\lambda_{+})^{j}q^{j(j+1)}(x^{+}%
)^{j}(D_{q^{2}}^{r^{2}})^{j}(D_{q^{2}}^{-})^{-(j+1)}\big |_{-\infty}^{\infty
}q^{-2(j+1)\hat{n}_{r^{2}}}\nonumber\\
&  \qquad\times\sum_{k=0}^{\infty}(-1)^{k}\Big [(D_{q^{2}}^{3/0}%
)^{-1}\big |_{-\infty}^{\infty}q^{-2\hat{n}_{r^{2}}}\Big (q^{-1}\lambda
_{+}x^{3/0}D_{q^{2}}^{r^{2}}q^{2\hat{n}_{x^{+}}}\nonumber\\
&  \qquad\qquad+\big (-q^{-2}r^{2}(x^{3/0})^{-1}-q^{-4}x^{3/0}\big )D_{q^{2}%
}^{r^{2}}\nonumber\\
&  \qquad\qquad+q^{-2}\lambda_{+}x^{+}x^{-}(x^{3/0})^{-1}D_{q^{2}}^{r^{2}%
}q^{-2\hat{n}_{x^{3/0}}}\\
&  \qquad\qquad-\,q\lambda_{+}^{-1}\lambda x^{3/0}D_{q^{2}}^{+}D_{q^{2}}%
^{-}q^{2\hat{n}_{r^{2}}}\Big )\Big ]^{k}(D_{q^{2}}^{3/0})^{-1}\big |_{-\infty
}^{\infty}q^{-2\hat{n}_{r^{2}}}\nonumber\\
&  \qquad\times\sum_{l=0}^{\infty}q^{l(l+1)}(\lambda_{+}x^{-}D_{q^{2}}^{r^{2}%
})^{l}(D_{q^{2}}^{+})^{-(l+1)}\big |_{-\infty}^{\infty}q^{2l(\hat{n}_{x^{+}%
}+\,\,\hat{n}_{x^{3/0}})}f\nonumber\\
&  \qquad\times\sum_{m=0}^{\infty}q^{(\frac{3}{2}m-1)(m+1)}\lambda_{+}%
^{-2m-3}(D_{q^{2}}^{+}D_{q^{2}}^{-})^{m}\nonumber\\
&  \qquad\times(D_{q^{2}}^{r^{2}})^{-(m+1)}\big |_{-\infty}^{\infty}%
f(q^{2m}r^{2},q^{-2(m+1)}x^{+},q^{-2(m+1)}x^{3/0},q^{-2(m+1)}x^{-})\nonumber\\
&  =-q^{-1}\lambda_{+}^{-3}(D_{q^{2}}^{-})^{-1}\big |_{-\infty}^{\infty
}q^{-2\hat{n}_{r^{2}}}(D_{q^{2}}^{3/0})^{-1}\big |_{-\infty}^{\infty}%
q^{-2\hat{n}_{r^{2}}}\nonumber\\
&  \qquad\times(D_{q^{2}}^{+})^{-1}\big |_{-\infty}^{\infty}(D_{q^{2}}^{r^{2}%
})^{-1}\big |_{-\infty}^{\infty}q^{-2(\hat{n}_{x^{+}}+\,\hat{n}_{x^{3/0}}%
+\hat{n}_{x^{-}})}f. \label{MinVolIntN}%
\end{align}
For the first identity the expressions we have derived so far are substituted
for the one-dimensional integrals. The second equality makes use of
(\ref{IntAbl1N}) and
\begin{equation}
(D_{q^{a}}^{i})^{-1}\big |_{-\infty}^{\infty}(D_{q^{a}}^{i})^{n}f=\lim
_{x^{i}\rightarrow\infty}(D_{q^{a}}^{i})^{n-1}f-\lim_{x^{i}\rightarrow-\infty
}(D_{q^{a}}^{i})^{n-1}f=0. \label{IntAbl2}%
\end{equation}
Notice that the last result only holds if we assume $f\in\widetilde
{\mathcal{A}}$ [cf. (\ref{RandBed})].

Lastly, we would like to discuss how the expression in (\ref{MinVolIntN})
changes under reversing the normal ordering. To this end we need an operator
that transforms a function representing an element of q-deformed Minkowski
space for ordering $X^{-}X^{3/0}X^{+}\hat{r}^{2}$ to a function representing
the same quantum space element but now for reversed ordering $\hat{r}^{2}%
X^{+}X^{3/0}X^{-}.$ Its explicit form reads \cite{BW01}%
\begin{align}
f(r^{2},x^{+},x^{3/0},x^{-})=\,  &  \hat{U}^{-1}\big (f(x^{-},x^{3/0}%
,x^{+},r^{2})\big )\nonumber\\
=\,  &  \sum_{i,j=0}^{\infty}(\lambda\lambda_{+}^{-1})^{i}\frac{(-1)^{j}%
q^{-(i-k)-k^{2}}}{[[k]]_{q^{2}}![[i-k]]_{q^{2}}!}\nonumber\\
\,  &  \quad\times r^{2(i-k)}(x^{3/0})^{2k}\,q^{2\hat{n}_{x^{+}}\hat{n}%
_{x^{-}}+\,(\hat{n}_{x^{+}}+\,\hat{n}_{x^{-}})(2\hat{n}_{x^{3/0}}+i)+2\hat
{n}_{x^{3/0}}i}\nonumber\\
\,  &  \quad\times\big ((D_{q^{2}}^{+}D_{q^{2}}^{-})^{i}\tilde{f}%
\big )(q^{i-2k}x^{-},x^{3/0},q^{i-2k}x^{+},r^{2}). \label{UmMin2N}%
\end{align}
As a next step we take the integral of $f=\hat{U}^{-1}\tilde{f}$ over the
whole Minkowski space. By virtue of (\ref{MinVolIntN}) and (\ref{IntAbl2}) we
finally arrive at%
\begin{equation}
\int d_{\bar{L}}^{4}x\,f=\int d_{\bar{L}}^{4}x\,q^{2\hat{n}_{x^{+}}\hat
{n}_{x^{-}}+\,2(\hat{n}_{x^{+}}+\,\hat{n}_{x^{-}})\hat{n}_{x^{3/0}}}\tilde{f},
\end{equation}
i.e. the terms with $i>0$ in (\ref{UmMin2N}) do not contribute to the integral.

\end{document}